\documentclass[%
reprint,
superscriptaddress,
 amsmath,amssymb,
 aps,
]{revtex4-1}

\usepackage{mathtools}
\usepackage{graphicx}
\usepackage{dcolumn}
\usepackage{bm}
\usepackage{physics}

\usepackage{amsmath}
\usepackage{amssymb}
\usepackage{amsthm}
\usepackage{graphicx}
\usepackage{algorithmic}
\usepackage{mathrsfs}
\usepackage{braket}
\usepackage{xcolor}
\usepackage{tikz}
\usepackage[roman]{complexity}
\usepackage[caption=false]{subfig}

\usepackage{pdfpages}
\usepackage{pgffor}
\makeatletter
\AtBeginDocument{\let\LS@rot\@undefined}
\makeatother

\begin{document}

\title{Analytic theory for the dynamics of wide quantum neural networks}

\author{Junyu Liu}
\email{junyuliu@uchicago.edu}
\affiliation{Pritzker School of Molecular Engineering, The University of Chicago, Chicago, IL 60637, USA}
\affiliation{Chicago Quantum Exchange, Chicago, IL 60637, USA}
\affiliation{Kadanoff Center for Theoretical Physics, The University of Chicago, Chicago, IL 60637, USA}

\author{Khadijeh Najafi}
\email{knajafi@ibm.com}
\affiliation{IBM Quantum, IBM T. J. Watson Research Center, Yorktown Heights, NY 10598, USA}

\author{Kunal Sharma}
\email{kunals@ibm.com}
\affiliation{IBM Quantum, IBM T. J. Watson Research Center, Yorktown Heights, NY 10598, USA}
\affiliation{Joint Center for Quantum Information and Computer Science, University of Maryland, College Park, Maryland 20742, USA}

\author{\mbox{Francesco Tacchino}}
\email{fta@zurich.ibm.com}
\affiliation{IBM Quantum, IBM Research, Zurich, 8803 R\"{u}schlikon, Switzerland}

\author{Liang Jiang}
\email{liang.jiang@uchicago.edu}
\affiliation{Pritzker School of Molecular Engineering, The University of Chicago, Chicago, IL 60637, USA}
\affiliation{Chicago Quantum Exchange, Chicago, IL 60637, USA}

\author{Antonio Mezzacapo}
\email{mezzacapo@ibm.com}
\affiliation{IBM Quantum, IBM T. J. Watson Research Center, Yorktown Heights, NY 10598, USA}

\date{\today}

\begin{abstract}
Parameterized quantum circuits can be used as quantum neural networks and have the potential to outperform their classical counterparts when trained for addressing learning problems. To date, much of the results on their performance on practical problems are heuristic in nature. In particular, the convergence rate for the training of quantum neural networks is not fully understood. 
Here, we analyze the dynamics of gradient descent for the training error of a class of variational quantum machine learning models. We define wide quantum neural networks as parameterized quantum circuits in the limit of a large number of qubits and variational parameters. We then find a simple analytic formula that captures the average behavior of their loss function and discuss the consequences of our findings. For example, for random quantum circuits, we predict and characterize an exponential decay of the residual training error as a function of the parameters of the system. 
We finally validate our analytic results with numerical experiments.
\end{abstract}

\maketitle

Machine learning has revolutionized data processing for several practical applications. With the abundance of data and computational resources, heuristic deep learning algorithms have been successfully employed for several applications, including speech recognition, translation, drug discovery, genomics, and self-driving cars \cite{lecun2015deep}. The theory of deep learning owes part of its successes to analytic insights that facilitate the design of learning algorithms \cite{lee2017deep,jacot2018neural,lee2019wide,arora2019exact,sohl2020infinite,yang2020feature,yaida2020non}. 

Recent experimental progress on quantum hardware and algorithms has generated great excitement in trying to identify applications that can lead to a quantum advantage over classical devices~\cite{huang2020power,abbas2021power,liu2021rigorous,aharonov2022quantum}. One such application is quantum machine learning (QML) which employs parameterized quantum circuits to analyze either classical or quantum data~\cite{schuld2015introduction,biamonte2017quantum,dunjko2018machine}.  
Contrary to the classical setting, where experiments are routinely performed on large-scale problems, current quantum processors are limited both in number of qubits and by decoherence noise~\cite{IBM_blog_post}, which makes it challenging to test QML algorithms in practice. Analytic tools are currently among the best resources that can help us quantify the performance of QML models and design new algorithms~\cite{schuld2022quantum}. 

Similar to classical neural networks, several models for quantum neural networks (QNNs) have been proposed \cite{farhi2018classification,cong2019quantum,bausch2020recurrent,beer2020training,Mangini_2021}. The success of QNNs relies on several factors: trainability, expressivity, generalization, and convergence rate. Although trainability \cite{mcclean2018barren,cerezo2021cost,pesah2021absence,sharma2020trainability,liu2021presence,kieferova2021quantum,holmes2021connecting,larocca2021diagnosing,patti2021entanglement,zhao2021analyzing,wang2021noise,thanasilp2021subtleties}, expressivity \cite{sim2019expressibility,nakaji2021expressibility,holmes2021connecting, du2021efficient,sharma2022reformulation}, and generalization capabilities \cite{caro2021generalization,du2021efficient,huang2021information,abbas2021power,huang2011power} of QNNs have been extensively studied, analytic understanding of the convergence rate of the training error is still lacking.

In this article, we present an analytic theory for the dynamics of a wide QNN trained with gradient descent, in the limit of large number of parameters. Our results are 
based on the framework of the quantum neural tangent kernel (QNTK), recently developed in Refs.~\cite{liu:2021wqr, shirai2021quantum}. Of particular interest here is the lazy training regime where the QNTK becomes constant - or frozen, see Ref.~\cite{liu:2021wqr}. 

Using the QNTK framework, we calculate the behavior of the residual training error for random parameterized quantum circuits. In the high-dimensional limit and for a sufficiently large number of variational parameters, we find an analytic solution characterizing the convergence of the residual training error. We denote the rate of convergence as $\gamma$, and show its dependence on the number of variational parameters $L$, the dimension of the Hilbert space $D$, learning rate $\eta$, and $\Tr(O^2)$ for an observable $O$. High values of $\gamma$ imply that a QNN is in the \textit{overparameterized} regime, leading to an exponential convergence of the residual error on average. We note that prior to our analytic results, overparameterization in QNNs was only numerically investigated for some systems in \cite{kiani2020learning,wiersema2020exploring} and connected to the dimension of the dynamical Lie algebra associated with periodic structure ansatzes \cite{larocca2021theory}. 

The paper is organized as follows: we first review the QNTK theory. Using the QNTK, we derive an analytic solution characterizing the convergence of the residual training error. We then derive conditions on the parameters of the system, such that the residual error decays exponentially, {}{and extend our results to general supervised learning problems}. Finally, we provide numerics verifying our results. We conclude with a brief summary and discuss the implications of our results. In our Supplemental Material we provide detailed proofs of our results.

{\it Quantum Neural Tangent Kernel.---}We begin by reviewing the QNTK theory as described in Ref.~\cite{liu:2021wqr}. Let $D$ denote the dimension of a Hilbert space $\mathcal{H}$. We consider a general class of parameterized quantum circuits on $\log(D)$ qubits, expressed as follows:
\begin{equation}\label{eq:ansatz}
U(\vec{\theta}) =  {\prod\limits_{\ell  = 1}^L {{W_\ell }} \exp \left( {i{\theta _\ell }{X_\ell }} \right)}  \equiv  {\prod\limits_{\ell  = 1}^L {{W_\ell }} U_{\ell}},
\end{equation}
where $\vec{\theta} = \{\theta_\ell\}_{\ell=1}^L$ is a set of continuous parameters, $W_\ell$ denote unparameterized gates, and $X_\ell$ are Hermitian operators. Here, $\vec{\theta}$ are optimized to minimize a loss function that can be expressed as the expectation value of an observable~$O$: 
\begin{align}\label{opt_problem}
{\cal L}(\vec{\theta}) \equiv \frac{1}{2}{\left( {\left\langle {{\Psi _0}\left| {{U^\dag }(\vec{\theta} )OU(\vec{\theta} )} \right|{\Psi _0}} \right\rangle  - {O_0}} \right)^2} \equiv \frac{1}{2}{\varepsilon ^2}~,
\end{align}
where $\ket{\psi_0}$ is an input state, $O_0$ denotes the target value, and $\varepsilon$  denotes the residual error \cite{liu:2021wqr}. 

We note that in Eq.~\eqref{opt_problem}, we {}{start with} a simpler problem than a general supervised learning task where one has access to a labeled dataset. In general, the loss function for a general supervised learning task is given by
\begin{equation}\label{opt_problem-gen}
   {\cal L}_{\mathcal{A}}(\vec \theta ) = \sum\limits_{i,\tilde \alpha  \in \mathcal{A}} {\frac{1}{2}{{\left( {{z_i}(\vec \theta ,{{\bf{x}}_{\tilde \alpha }}) - {y_{\tilde \alpha ,i}}} \right)}^2}}  \equiv \sum\limits_{i,\tilde \alpha  \in \mathcal{A}} {\frac{1}{2}\varepsilon _{\tilde \alpha ,i}^2(\vec \theta )} ~,
\end{equation}
    where $\tilde{\alpha}$ labels the elements from the training set~$\mathcal{A}$, $\mathbf{x}_{\tilde{\alpha}}$ and $y_{\tilde{\alpha},i}$ form the data inputs and outputs respectively in the training set, where the output dimension has the index $i$. $z_i (\mathbf{x}_{\tilde{\alpha}})=\left\langle {\Psi ({{\bf{x}}_{\tilde \alpha }})\left| {{U^\dag }(\vec \theta ){O_i}U(\vec \theta )} \right|\Psi ({{\bf{x}}_{\tilde \alpha }})} \right\rangle $ is the model output with the embedding map $\ket{\Psi (\mathbf{x}_{\tilde{\alpha}})}$, and ${\varepsilon _{\tilde \alpha ,i}} = {z_i}(\vec \theta ,{{\bf{x}}_{\tilde \alpha }}) - {y_{\tilde \alpha ,i}}$ is the residual training error. 

{}{Below we provide a detailed summary of our results for Eq.~\eqref{opt_problem} and briefly discuss our results for Eq.~\eqref{opt_problem-gen}. We provide detailed proofs for both cases in Supplemental Material \cite{SM}.}
Based on the gradient of the loss function in Eq.~\eqref{opt_problem}, the gradient descent algorithm updates the variational parameters as 
\begin{equation}
     \delta\theta_\ell \equiv \theta_{\ell}(t+1) - \theta_{\ell}(t) = -\eta \varepsilon\frac{\partial \varepsilon}{\partial \theta_\ell},
\end{equation} 
where $\eta$ is the learning rate and $t$ refers to the time step of the gradient descent dynamics. Similarly, we define the change in the residual training error as $\delta \varepsilon \equiv \varepsilon(t+1) - \varepsilon(t)$. When the learning rate $\eta$ is small, from the Taylor expansion of $\delta \varepsilon$ we get
\begin{align}\label{taylor}
\delta \varepsilon  \approx \sum\limits_\ell  {\frac{{\partial \varepsilon }}{{\partial {\theta _\ell }}}} \delta {\theta _\ell } =  - \eta \sum\limits_\ell  {\frac{{\partial \varepsilon }}{{\partial {\theta _\ell }}}} \frac{{\partial \varepsilon }}{{\partial {\theta _\ell }}}\varepsilon  =  - \eta K\varepsilon ~,
\end{align}
where the quantity 
\begin{align}
K \equiv \sum\limits_\ell  {\frac{{\partial \varepsilon }}{{\partial {\theta _\ell }}}} \frac{{\partial \varepsilon }}{{\partial {\theta _\ell }}}    
\end{align} 
is called the Quantum Neural Tangent Kernel (QNTK) \cite{liu:2021wqr}, which is a non-negative number. In a general supervised learning setting, as defined in Eq.~\eqref{opt_problem-gen}, $K$ is a symmetric positive-semidefinite matrix.

In the regime of lazy training -- where variational angles do not change much -- QNTK becomes constant (frozen) \cite{liu:2021wqr}. 
For a frozen QNTK, at the gradient descent step $t$,  the residual error decays as follows \cite{liu:2021wqr}:
\begin{align}\label{varep}
\varepsilon (t) \approx {(1 - \eta K)^t}\varepsilon (0)~,
\end{align}
where $\varepsilon(0)$ denotes the residual error at $t=0$. Thus for a small learning rate $\eta$ and a frozen QNTK, the residual error $\varepsilon$ decays exponentially.  

We first analyze Eq.~\eqref{varep} for the case when $K \approx \mathbb{E}(K) \equiv \overline{K}$,
where the average of $K$ is over $\vec{\theta}$. We later derive conditions under which Eq.~\eqref{varep} is valid. Note that an average of $K$ over $\vec{\theta}$ depends on the choice of the ansatz, as defined in Eq.~\eqref{eq:ansatz}. For such ansatzes,  $\partial\varepsilon/\partial\theta_l$ can be expressed as 
\begin{equation}
\partial\varepsilon/\partial\theta_l = -i\langle\Psi_{0}|U_{+, \ell}^{\dagger}[X_{\ell},  U_{-, \ell}^{\dagger} O U_{-, \ell}] U_{+, \ell}| \Psi_{0}\rangle,    
\end{equation}
where ${U_{ - ,\ell }} \equiv \prod\limits_{k = 1}^{\ell  } {{W_{k}}} {U_{k}}$ and ${U_{ + ,\ell }} \equiv \prod\limits_{k = \ell  + 1}^L {{W_{k}}} {U_{k}}$. Let $\Tr(X_l^2) = c N$, where $c$ is a constant.

We now derive our results on the residual training error of random parameterized quantum circuits. Note that our results can be generalized to other ansatzes, and later we discuss the relevance of our results for periodic structure ansatzes \cite{larocca2021diagnosing, larocca2021theory}. Suppose that $U(\vec{\theta})$ is sufficiently random, such that for each $l$, both $U_{-, l}$ and $U_{+, l}$ are independent and match the Haar distribution up to the second moment. Then we get the following averaged value of $K$ in the large $D$ limit~\cite{SM}:
\begin{align}\label{fullqntke}
\overline{K} \approx \frac{{L{\mathop{\rm Tr}\nolimits} \left( {{O^2}} \right)}}{{{D^2}}}~,
\end{align}
which implies that on average, the residual training error decays as 
\begin{align}\label{eq:exp-decay}
    \varepsilon(t) \approx e^{- \gamma t} \varepsilon(0)~,
\end{align}
where the decay rate is given by 
\begin{align}\label{eq:decay-rate}
    \gamma \equiv \eta \overline{K} =  (\eta L \Tr(O^2))/D^2~.
\end{align}

Note that Eq.~\eqref{eq:exp-decay} holds for $\eta \ll 1$, as we have discarded higher-order terms in Eq.~\eqref{taylor}. We later show that in the large-$D$ limit, the second order term can be discarded even for high values of $\eta$.

The average result Eq.~\eqref{fullqntke} follows for 2-design random circuits which can be implemented using one-dimensional $\mathcal{O}(\log^2 (D))$ gates \cite{dankert2009exact}. Thus, for such efficiently-implementable circuits, Eq.~\eqref{eq:exp-decay} provides an analytic solution to the average behavior of the residual training error. For circuits that approximate a 4-design, fluctuations in K from $\overline{K}$ are also small.

In general, the decay rate $\gamma$ is small because of the $1/D^2$ dependence on the dimension of the Hilbert space. By setting $L \approx D^2/(\eta \Tr(O^2))$, the residual training error decays exponentially with the decay with rate $\gamma = \mathcal{O}(1)$. This leads us to define the \textit{overparameterized} regime for a quantum neural network (QNN): A QNN is overparameterized if the number of parameters of the system are sufficiently large such that $\gamma = \mathcal{O}(1)$.

We now discuss two cases for overparameterized random quantum circuits.
For $\Tr(O^2) \in \mathcal{O}(D)$, which holds for physical Hamiltonians that can be expressed as a linear combination of Pauli operators on $\log D$ qubits, $L\sim D$ is sufficient for making the corresponding QNN overparameterized. Similarly, for low-rank observables, $\Tr(O^2) \in \mathcal{O}(\log(D))$ which implies that $L \sim D^2$ make the corresponding QNN overparameterzied. 

Prior to our work, the overparameterization of a QNN was first numerically observed in Ref.~\cite{kiani2020learning}, where the authors investigated the task of learning Haar random unitaries $U(D)$ using parameterized alternating operator sequences. 
In particular, they numerically observed that when the number of parameters in the sequence was greater than or equal to $D^2$, the gradient descent always finds the target unitary. 
On the other hand, in Ref.~\cite{wiersema2020exploring}, overparameterization phenomena were studied in the context of estimating the ground state energies of transverse-field Ising and XXZ models. 
In particular, they employed the Hamiltonian variational ansatz from Ref.~\cite{wecker2015progress} for these problems and numerically observed that the computational phase transition (or overparameterization) takes place when the number of parameters is much less than $D^2$.

Furthermore, in Ref.~\cite{larocca2021theory}, the overparameterization was defined using the rank of the quantum Fisher information matrix associated with a QNN. They particularly focused on periodic structure ansatzes (PSA) and argued that a PSA is overparameterized if the number of parameters scale as the dimension of the dynamical Lie algebra associated with a periodic structure ansatz~\cite{larocca2021theory}. The results provided here show a deeper understanding of the training error dynamics, which cannot be obtained with algebraic arguments alone.

Although the overparameterized regime for random quantum circuits provides an analytic understanding of exponential convergence of the training error, it is not amenable to practical implementations as the number of parameters are required to scale as the dimension of the Hilbert space. {}{However, our result should not be interpreted as a no-go theorem. In fact, our model simply cannot make predictions if the number of parameters is not large enough.} On the other hand, Eqs.~\eqref{eq:exp-decay}--\eqref{eq:decay-rate} are in general valid for $k$-design circuits, which can be efficiently implemented, but there are two major issues for such circuits: 1) the decay rate is small due to $1/D^2$ dependence, and 2) these circuits suffer from barren plateaus \cite{mcclean2018barren}. Thus, a practical challenge is to identify ansatzes that are trainable and have fast exponential convergence of the training error. 
In this regard, our analytic results could be applied to the case when the dynamical Lie algebra associated with the generators of a periodic structure ansatz share a symmetry \cite{larocca2021diagnosing}. As discussed in Ref.~\cite{larocca2021diagnosing}, a symmetry can cause the state space to break into invariant subspaces, and the system may become reducible. Let us assume that such a reducible system is controllable on some or all of the invariant subspaces.  
More concretely, let $\mathcal{H} = \bigoplus_k \mathcal{H}_k$, and let the system be controllable on a subspace $\mathcal{H}_k$ of dimension $D_k$. If the initial state $\psi \in \mathcal{H}_k$, then under the condition that $D_k \in \mathcal{O}(\text{poly}(\log D))$, it is possible to get the decay rate $\gamma = \mathcal{O}(1)$ for $L\in \mathcal{O}(\text{poly}(\log D))$ number of parameters. Thus for such ansatzes, it is possible to get trainability guarantees along with an exponential decay of the residual error.  \

{\it Concentration of the QNTK and validity of the analytic regime.---}We proved Eq.~\eqref{eq:exp-decay} under two assumptions: 1) For small learning rate $\eta$, in Eq.~\eqref{taylor}, we expanded $\delta \varepsilon$ up to the first order in $\eta$, 2) we considered $K \approx \overline{K}$ in Eq.~\eqref{varep}. We now derive conditions in support of these assumptions. First, we derive conditions under which $K$ does not fluctuate much around $\overline{K}$. In particular, under the assumption that the ansatz in Eq.~\eqref{eq:ansatz} 
forms a $4$-design and in the large $D$ limit, we get that~$\Delta K = \sqrt {{\mathbb{E}}[ {{{(K - \overline{K})}^2}} ]}$ scales as~\cite{SM}
\begin{align}\label{deltak}
\Delta K  \approx \frac{{\sqrt L }}{{{D^2}}}\sqrt {\left( {8{{{\mathop{\rm Tr}\nolimits} }^2}\left( {{O^2}} \right) + 12{\mathop{\rm Tr}\nolimits} \left( {{O^4}} \right)} \right)} ~.
\end{align}

Similarly, we analyze the higher order corrections to the Taylor expansion of $\delta\varepsilon$ in Eq.~\eqref{taylor}, which was called the quantum meta-kernel (dQNTK) in Ref.~\cite{liu:2021wqr}. In particular, 
\begin{equation}
\delta \varepsilon = - \eta K \varepsilon + (1/2)\eta^2 \varepsilon^2 \mu~,
\end{equation}
where 
\begin{align}
\mu=\sum_{\ell_{1}, \ell_{2}} \frac{\partial^{2} \varepsilon}{\partial \theta_{\ell_{1}} \partial \theta_{\ell_{2}}} \frac{\partial \varepsilon}{\partial \theta_{\ell_{1}}} \frac{\partial \varepsilon}{\partial \theta_{\ell_{2}}}~.
\end{align}
Under the assumption that the ansatz in \eqref{eq:ansatz} forms a $6$-design, we show that $\mathbb{E}(\mu) =0$. This condition is similar to the classical counterpart of deep neural networks \cite{roberts2021principles}. Moreover, in the large-$D$ limit $ \Delta \mu \equiv \sqrt {{\mathbb{E}}\left( {{\mu ^2}} \right)}$ scales as~\cite{SM}
\begin{align}\label{deltamu}
\Delta \mu  \approx \frac{{\sqrt {32} \eta L}}{{{D^3}}}{{\mathop{\rm Tr}\nolimits} ^{3/2}}\left( {{O^2}} \right)~.
\end{align}

Therefore, assumptions made in deriving Eq.~\eqref{eq:exp-decay} are valid as long as 
\begin{align}\label{eq:fluc}
    \frac{\Delta K}{\overline{K}} \approx \frac{1}{\sqrt{L}} \ll 1, \quad \frac{\Delta \mu}{\overline{K}} \approx \frac{\left(\eta\sqrt{\Tr(O^2)}\right)}{D} \ll 1
\end{align} 

We refer to these conditions as the \emph{concentration conditions}. Note that $\text{poly(t)}\cdot\log^2(D)$-depth local random circuits with two qubit nearest-neighbor gates on a one-dimensional lattice are sufficient to realize an approximate $k$-designs on $\log(D)$ qubits \cite{harrow2018approximate}. Thus for~$L \in \mathcal{O}(\log^2(D))$, in the large-$D$ limit, both conditions in Eq.~\eqref{eq:fluc} are satisfied. Interestingly, in the large-$D$ limit,  $\Delta \mu/\overline{K} \ll 1$ is satisfied even for high values of $\eta$. Thus, our analytic solutions to the dynamics of the training error are valid in the large-$D$ limit, which defines a \textit{wide} QNN. A wide QNN also has a large number of variational parameters as $L\in \mathcal{O}(\log^2(D))$ was assumed in deriving Eq.~\eqref{eq:fluc}. 
It is an interesting question to determine similar conditions for other parameterized quantum circuits, including problem-inspired ansatzes \cite{larocca2021diagnosing}. Furthermore, we remark that  $1/\text{width}$ limit is also useful for analytic understanding of classical neural networks \cite{roberts2021principles}, where $\text{width}$ implies the number of neurons in a single layer.

{}{{\it Supervised learning generalization.---} In supervised learning, the QNTK is a symmetric, positive semi-definite matrix \cite{liu2021representation}. We compute the average behavior of the QNTK~\cite{SM}, in the frozen limit, finding that  
\begin{align}\label{barK}
\overline K_{{\delta _1},{\delta _2}}^{{i_1}{i_2}} \approx \frac{{2L{\mathop{\rm Tr}\nolimits} \left( {{O_{{i_1}}}{O_{{i_2}}}} \right)}}{{{D^2}}} \sigma_{\delta_1 \delta_2}~.
\end{align}}

\begin{figure}[htp]
\centering
\includegraphics[width=0.48\textwidth]{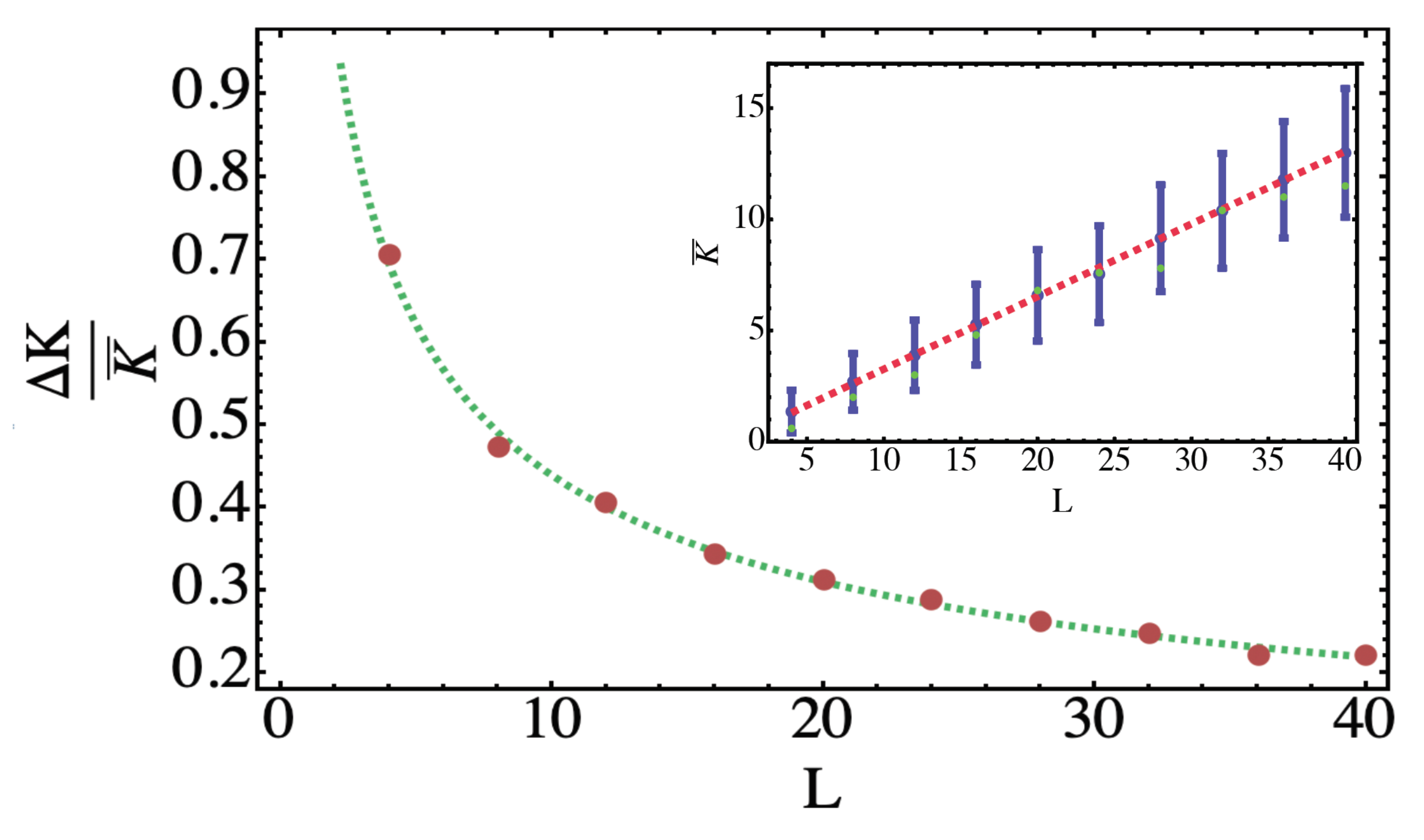}
\caption{Concentration of the QNTK on 4 qubits as a function of the circuit depth $L$. We pick different values of $L$ (up to 40) in the randomized ansatz, defined in Eq.~\eqref{ansatz}. We sample over 1000 variational angles from Eq.~(\ref{ansatz}) independently and uniformly over $[0,2\pi]$. We plot $\Delta K /\bar{K}$ verus $L$, which verifies the analytic scaling of $L^{-1/2}$, as derived in Eq.~\eqref{deltamu}. In the inset, we plot $\overline{K}$ versus $L$, where the dashed line represents the theoretical values of $\overline{K}$ from Eq.~\ref{fullqntke}, green dots represent the numerical values of $\overline{K}$, and the blue error bars represent $\Delta K$.
}
\label{fig:scalingNTK}
\end{figure}

{}{Here, $\delta_{1,2}$ correspond to input variables in the training data.  The feature dependent analytic result is obtained through the definition of the feature $S$-matrix,
\begin{align}
S_{\delta_{1} \delta_{2}}=\left|\phi\left(\mathbf{x}_{\delta_{1}}\right)\right\rangle\left\langle\phi\left(\mathbf{x}_{\delta_{2}}\right)\right|~,
\end{align}
and feature cross-section,
\begin{align}
{\sigma _{{\delta _1}{\delta _2}}} = {\left| {{\rm{Tr}}\left( {{S_{{\delta _1}{\delta _2}}}} \right)} \right|^2}~.
\end{align}
see~\cite{SM} for a full derivation.}
{}{
We use the notation $\mathbf{x}_{{\delta}}$ to represent in general an element in the data space, i.e., both training and test data. In Eq.~\eqref{barK}, we only include the leading order contributions in the large $D$ limit, while the full non-perturbative expressions are given in \cite{SM}. Note that Eq.~\eqref{barK} is dependent on the feature maps used, unlike the case of the optimization task considered in Eq.~\eqref{opt_problem}. Moreover, using 4-design assumptions (see \cite{SM} for more details), we show that $\Delta K /\bar{K} \sim \frac{1}{\sqrt{L}}$ which is similar to what is observed for classical neural networks.}

Importantly, a difference between the supervised learning case and the optimization case is the dependence of the QNTK on the size of the training data. For $\sigma_{\delta_1 \delta_2} \approx \delta_{\delta_1 \delta_2}$ and $O_i=O$ for all $i$, we are able to model the eigenvalues of the QNTK \cite{SM}. We find that for a training set size $\abs{\mathcal{A}}$, there are $\abs{\mathcal{A}}-1$ eigenvalues which do not depend on the size of the data set. On the other hand, there is a one-dimensional eigenspace with the kernel eigenvalue, $\frac{2L(D-|\mathcal{A}|)}{\left(D^2-1\right)^2}\left(D \operatorname{Tr}\left(O^2\right)-\operatorname{Tr}^2(O)\right)$, for all $D \ge \abs{A}$. Thus, for high values of $|A|$, the behavior of the eigenvalues suggests slower decay rates for larger datasets.

\begin{figure}[htp]
\centering
\includegraphics[width=0.48\textwidth]{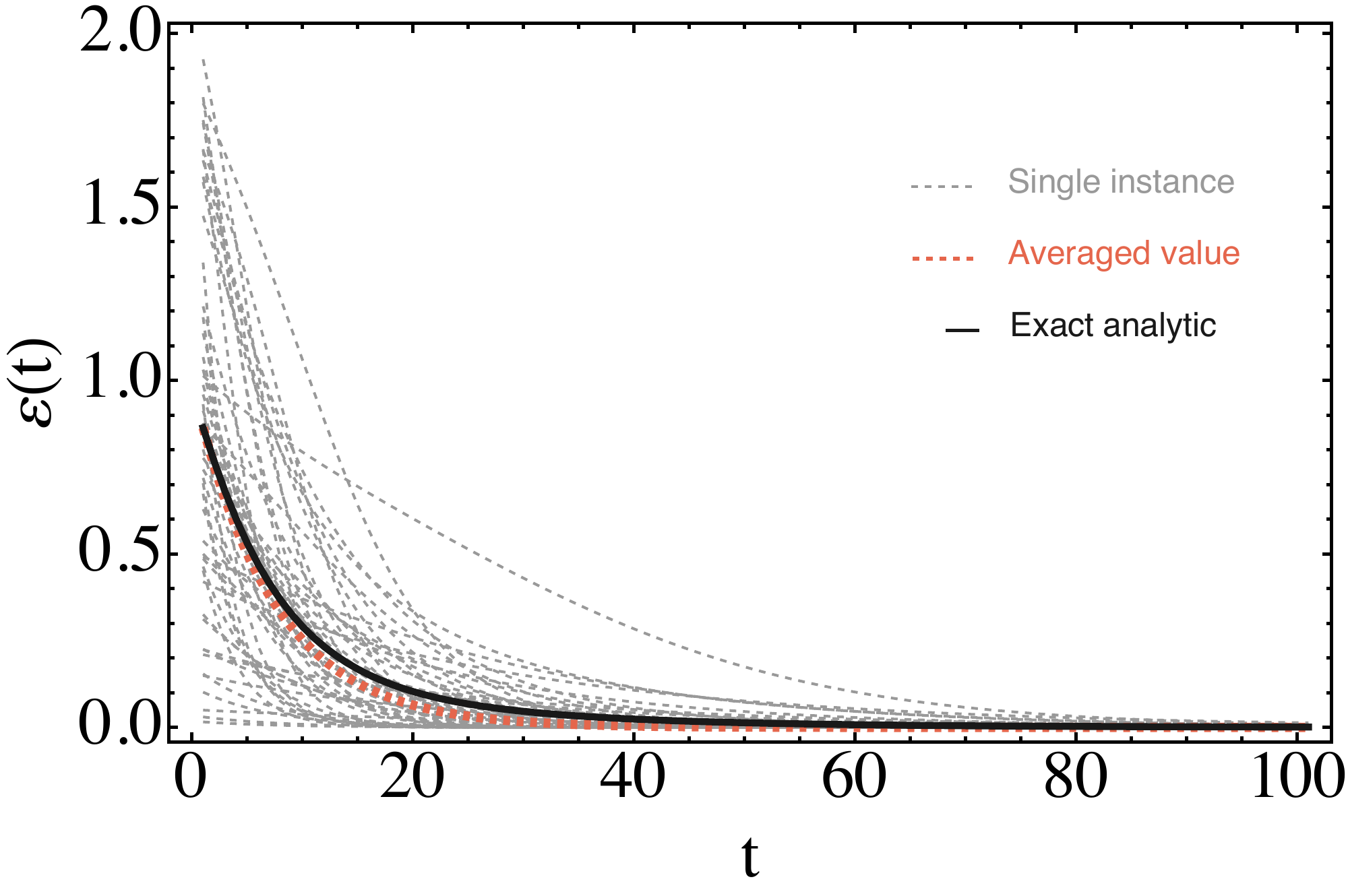}
\caption{Residual training error $\varepsilon$ versus the gradient descent steps $t$ for 2 qubits and $L=64$ for the random ansatz defined in Eq.~\eqref{ansatz}. We use random initial angles in $[0,2\pi]$, and perform the gradient descent experiment with a learning rate $\eta=10^{-4}$ with 1000 steps. For 50 different initializations, we plot the dynamics of $\varepsilon(t)$, the theoretical prediction for the average dynamics of $\varepsilon(t)$, and the numerical values for the averaged $\varepsilon(t)$.}
\label{fig:error}
\end{figure}

{\it Numerical experiments.---}In what follows, we present numerical results verifying our analytic results presented above. 
First, for the optimization problem, we consider a variational ansatz, as defined in Eq.~\eqref{eq:ansatz} with
\begin{align}\label{ansatz}
{U_\ell } = \exp (i{P_\ell }{\theta _\ell })~,~~~~{W_\ell } = {\rm{Haar}} \in {\rm{U}}(D)~,
\end{align}
where we sample $P_{\ell}$ uniformly from the $D$-qubit Pauli group. Moreover, $W_l$ is sampled with respect to a Haar measure on $U(D)$, and then kept fixed during the optimization. Let 
$O = \sum\limits_{j = 1}^{10} {{c_j}{{\tilde P}_j}}$,
where $\tilde{P}_j$ are also sampled from the $D$-qubit Pauli group, and $c_j\in (0,1)$. After sampling once, we keep $P_{\ell}$, $\tilde{P}_j$, and $c_j$ fixed during the optimization.

\begin{figure}[htp]
\centering
\includegraphics[width=0.48\textwidth]{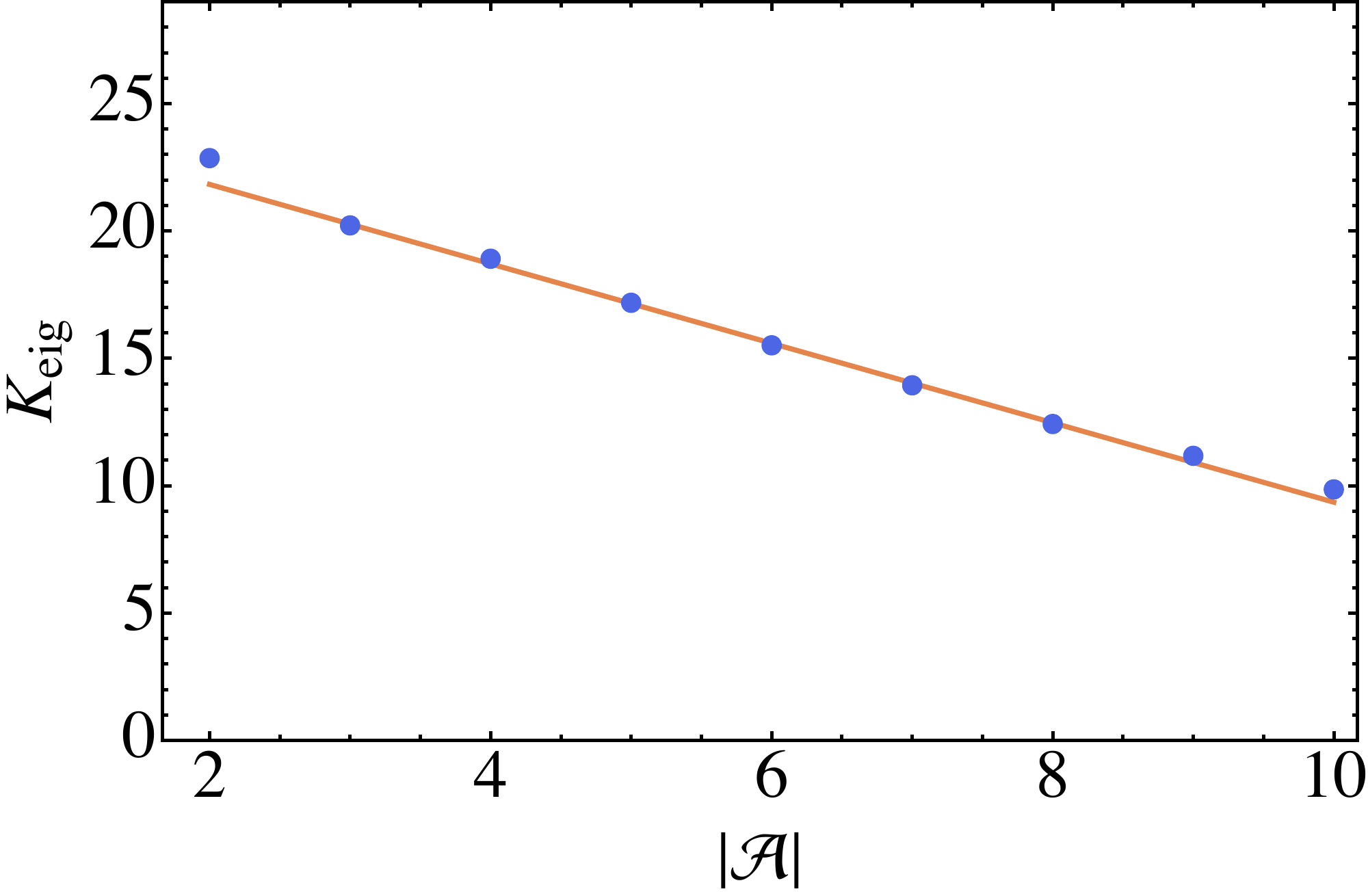}
\caption{Lowest eigenvalue for the kernel of a supervised learning task defined on 4 qubits and $L = 64$, as a function of the training data size $\abs{\mathcal{A}}$. The predicted linear relation $\frac{2L(D-|\mathcal{A}|)}{\left(D^2-1\right)^2}\left(D \operatorname{Tr}\left(O^2\right)-\operatorname{Tr}^2(O)\right)$ (orange solid line) agrees with numerical estimations with 50 independent instances of the ansatz. }

\label{fig:learning}
\end{figure}

In  Figure~\ref{fig:scalingNTK}, we study the scaling of $\overline{K}$ and $\Delta K/\overline{K}$ with respect to $L$ for four qubits. In the inset of  Figure~\ref{fig:scalingNTK} we verify the linear scaling of $\overline{K}$ with $L$, as derived in Eq.~\eqref{fullqntke}. Similarly, Figure~\ref{fig:scalingNTK} also follows the analytic scaling of $\Delta K/\overline{K} \approx 1/\sqrt{L}$, as derived in Eq.~\eqref{deltak}.  In Figure~\ref{fig:error} we plot the residual training error $\varepsilon$ versus the gradient descent optimization steps (time) for two qubits and $L=64$, for 50 independent random initializations. Figure~\ref{fig:error} verifies that in the large-$L$ limit, the residual error $\varepsilon$ decays exponentially, as derived in Eqs.~\eqref{eq:exp-decay}--\eqref{eq:decay-rate}.

{}{In Figure~\ref{fig:learning}, we focus on the data-dependent lowest kernel eigenvalue,
\begin{align}
K_{\text{eigen}}=\frac{2L(D-|\mathcal{A}|)}{\left(D^2-1\right)^2}\left(D \operatorname{Tr}\left(O^2\right)-\operatorname{Tr}^2(O)\right)~.
\end{align}
We consider a 4-qubit example and set $L=64$. Other parameters are the same as in Figure \ref{fig:scalingNTK}. We consider different values of $|\mathcal{A}|$ ranging from 2 to 10, and plot both numerically and analytically the value of the smallest eigenvalue, observing fair agreement. 
We generate the input data vectors such that they are orthogonal to each other. 
We provide further numerical results on supervised learning problems in \cite{SM}.}

{\it Discussion.---}Using the quantum neural tangent kernel (QNTK) theory, we analytically solved the dynamics of the residual training error corresponding to variational quantum cost functions.
Using these analytic solutions we characterized an exponential decay of the residual training error as a function of the parameters of random quantum circuits. We derived conditions for which the second-order effects to the residual error and the fluctuations in the QNTK are negligible for wide QNNs.

{}{One application of the theory developed in our work is to analyze the overparameterization of symmetric QNNs. As discussed previously, for the case when the subspace dimension of symmetric QNNs grows polynomially in the number of qubits, the number of parameters needed to observe a non-vanishing decay of the training error is also polynomial in number of qubits. 
Therefore, extending our results to symmetric QNNs~\cite{larocca2021diagnosing} will be an important direction for future research.} Another open question is to establish connections between the QNTK and the generalization error for quantum machine learning models \cite{simon2021neural,caro2021generalization,PRXQuantum.2.040321},{}{ as well as with the quantum information theory bottleneck, as developed in \cite{PRXQuantum.2.040321}.} 
Finally, note that the symmetric quantum neural networks already lead to desirable features in variational quantum algorithms, like the absence of barren plateaus \cite{pesah2021absence}.

\bigskip

\textit{Acknowledgements.---}We thank Jens Eisert, Keisuke Fujii, Dan A. Roberts and  Xiaodi Wu, for useful discussions.
JL is supported in part by International Business Machines (IBM) Quantum through the Chicago Quantum Exchange, and the Pritzker School of Molecular Engineering at the University of Chicago through AFOSR MURI (FA9550-21-1-0209). LJ acknowledges support from ARO (W911NF-18-1-0020, W911NF-18-1-0212), ARO MURI (W911NF-16-1-0349, W911NF-21-1-0325), AFOSR MURI (FA9550-19-1-0399, FA9550-21-1-0209), AFRL (FA8649-21-P-0781), DoE Q-NEXT, NSF (OMA-1936118, EEC-1941583, OMA-2137642), NTT Research, and the Packard Foundation (2020-71479).

\bibliographystyle{unsrt}
\bibliography{bibliography.bib}

\clearpage

\pagebreak

\onecolumngrid
\appendix

\vspace{0.5in}

\begin{center}
	{\Large \bf Supplemental Material}
\end{center}

Here we provide proofs for the main results of the {\it An analytic theory for the dynamics of wide quantum neural networks}. In Section~\ref{suupp:qntk} we recall the quantum neural tangent kernel (QNTK) theory from \cite{liu:2021wqr}. In Section~\ref{supp:frozenqntk} we estimate the average of the QNTK for random quantum circuits. We analyze fluctuations in QNTK in Section~\ref{supp:fluc}. We derive conditions on second order effects in Section~\ref{supp:dqntk}. Finally, we provide additional numerical experiments in Section~\ref{supp:numerics}.

\section{Quantum neural tangent kernel}\label{suupp:qntk}
In this section, we review the quantum neural tangent kernel (QNTK) theory from  \cite{liu:2021wqr}. Consider the following variational ansatz:
\begin{align}\label{supp:ansatz}
U(\vec{\theta} ) = \left( {\prod\limits_{\ell  = 1}^L {{W_\ell }} \exp \left( {i{\theta _\ell }{X_\ell }} \right)} \right) = \left( {\prod\limits_{\ell  = 1}^L {{W_\ell }{U_\ell }} } \right)~.
\end{align}
Here, $\vec{\theta}=
\{\theta_\ell\}_{\ell =1}^L$ is a set of continuous parameters, $W_\ell$ denote unparameterized gates, and $X_\ell$ are Hermitian operators such that $\tr(X_l^2) = c D$, where $c$ is some constant, and $D$ denotes the dimension of the Hilbert space. We consider the following loss function:
\begin{align}\label{supp:loss}
{\cal L}(\vec{\theta}) = \frac{1}{2}{\left( {\left\langle {{\Psi _0}\left| {{U^\dag }(\vec{\theta})OU(\vec{\theta} )} \right|{\Psi _0}} \right\rangle  - {O_0}} \right)^2} \equiv \frac{1}{2}{\varepsilon ^2}~,
\end{align}
where $\ket{\psi_0}$ is an input state, $O_0$ denotes the target value, and $\varepsilon$ denotes the residual training error \cite{liu:2021wqr}.

The gradient descent algorithm updates the variational parameters as follows:
\begin{align}
\delta {\theta _\ell } \equiv {\theta _\ell }(t + 1) - {\theta _\ell }(t) =  - \eta \frac{{\partial {\cal L}(\theta )}}{{\partial {\theta _\ell }}} =  - \eta \varepsilon \frac{{\partial \varepsilon }}{{\partial {\theta _\ell }}}~,
\end{align}
where $\eta$ is the learning rate, and $t$ refers to the time step of the gradient descent dynamics. 

For small $\eta$, the change in the residual training error can be expressed as
\begin{align}
\delta \varepsilon & \equiv \varepsilon (t + 1) - \varepsilon (t) \nonumber\\
&\approx \sum\limits_\ell  {\frac{{\partial \varepsilon }}{{\partial {\theta _\ell }}}} \delta {\theta _\ell } \nonumber\\
&=  - \eta \sum\limits_\ell  {\frac{{\partial \varepsilon }}{{\partial {\theta _\ell }}}} \frac{{\partial \varepsilon }}{{\partial {\theta _\ell }}}\varepsilon\nonumber\\
& = -\eta K \varepsilon~.
\end{align}
Here, we call $K\equiv \sum\limits_\ell  {\frac{{\partial \varepsilon }}{{\partial {\theta _\ell }}}} \frac{{\partial \varepsilon }}{{\partial {\theta _\ell }}}$ the quantum neural tangent kernel (QNTK). Using Eqs.~\eqref{supp:ansatz} and \eqref{supp:loss}, we get 
\begin{align}
K{\rm{ }} &= \sum\limits_\ell  {\frac{{\partial \varepsilon }}{{\partial {\theta _\ell }}}} \frac{{\partial \varepsilon }}{{\partial {\theta _\ell }}}\nonumber\\
&=  - {\left\langle {{\Psi _0}\left| {U_{ + ,\ell }^\dag \left[ {{X_\ell },U_\ell ^\dag W_\ell ^\dag U_{ - ,\ell }^\dag O{U_{ - ,\ell }}{W_\ell }{U_\ell }} \right]{U_{ + ,\ell }}} \right|{\Psi _0}} \right\rangle ^2}\nonumber\\
&=-\sum_{\ell}\left\langle\Psi_{0}\left|V_{+, \ell}^{\dagger}\left[X_{\ell}, V_{-, \ell}^{\dagger} O V_{-, \ell}\right] V_{+, \ell}\right| \Psi_{0}\right\rangle^{2}~,\label{supp:QNTK}
\end{align}
where
\begin{align}
&{U_{ - ,\ell }} \equiv \prod\limits_{\ell ' = 1}^{\ell  - 1} {{W_{\ell '}}} {U_{\ell '}},\quad {U_{ + ,\ell }} \equiv \prod\limits_{{\ell ^\prime } = \ell  + 1}^L {{W_{\ell '}}} {U_{\ell '}}~,\nonumber\\
&V_{-, \ell} =U_{-, \ell} W_{\ell} U_{\ell} ~,~~~~~V_{+, \ell} =U_{+, \ell}~.
\end{align}

Similarly, we define higher order (in $\eta$) corrections to the residual training error as follows:
\begin{align}
\delta \varepsilon  =  - \eta \sum\limits_\ell  {\frac{{\partial \varepsilon }}{{\partial {\varphi _\ell }}}} \frac{{\partial \varepsilon }}{{\partial {\varphi _\ell }}}\varepsilon  + \frac{1}{2}{\eta ^2}{\varepsilon ^2}\sum\limits_{{\ell _1},{\ell _2}} {\frac{{{\partial ^2}\varepsilon }}{{\partial {\varphi _{{\ell _1}}} \partial {\varphi _{{\ell _2}}}}}} \frac{{\partial \varepsilon }}{{\partial {\varphi _{{\ell _1}}}}}\frac{{\partial\varepsilon }}{{\partial {\varphi _{{\ell _2}}}}}~,
\end{align}
where we define the quantum meta-kernel (dQNTK) as \cite{liu:2021wqr}
\begin{align}
\mu=\sum_{\ell_{1}, \ell_{2}} \frac{\partial ^{2} \varepsilon}{\partial \varphi_{\ell_{1}} \partial \varphi_{\ell_{2}}} \frac{\partial \varepsilon}{\partial \varphi_{\ell_{1}}} \frac{\partial \varepsilon}{\partial \varphi_{\ell_{2}}}~.
\end{align}
From \cite{liu:2021wqr}, we have,
\begin{align}
\frac{{{\partial ^2}\varepsilon }}{{\partial {\varphi _{{\ell _1}}}\partial {\varphi _{{\ell _2}}}}} =  - \left\langle {{\Psi _0}\left| {V_{ + ,{\ell _2}}^\dag \left[ {{X_{{\ell _2}}},V_{{\ell _1},{\ell _2}}^\dag \left[ {{X_{{\ell _1}}},V_{ - ,{\ell _1}}^\dag O{V_{ - ,{\ell _1}}}} \right]{V_{{\ell _1},{\ell _2}}}} \right]{V_{ + ,{\ell _2}}}} \right|{\Psi _0}} \right\rangle ~,
\end{align}
and,
\begin{align}
\mu  &= 2\sum\limits_{{\ell _1} < {\ell _2}} {\left\langle {{\Psi _0}\left| {V_{ + ,{\ell _1}}^\dag \left[ {{X_{{\ell _1}}},V_{ - ,{\ell _1}}^\dag O{V_{ - ,{\ell _1}}}} \right]{V_{ + ,{\ell _1}}}} \right|{\Psi _0}} \right\rangle \left\langle {{\Psi _0}\left| {V_{ + ,{\ell _2}}^\dag \left[ {{X_{{\ell _2}}},V_{ - ,{\ell _2}}^\dag O{V_{ - ,{\ell _2}}}} \right]{V_{ + ,{\ell _2}}}} \right|{\Psi _0}} \right\rangle } \nonumber\\
&\qquad \left( {\left\langle {{\Psi _0}\left| {V_{ + ,{\ell _2}}^\dag \left[ {{X_{{\ell _2}}},V_{{\ell _1},{\ell _2}}^\dag \left[ {{X_{{\ell _1}}},V_{ - ,{\ell _1}}^\dag O{V_{ - ,{\ell _1}}}} \right]{V_{{\ell _1},{\ell _2}}}} \right]{V_{ + ,{\ell _2}}}} \right|{\Psi _0}} \right\rangle } \right)\nonumber\\
&\qquad+ \sum\limits_\ell  {{{\left\langle {{\Psi _0}\left| {V_{ + ,\ell }^\dag \left[ {{X_\ell },V_{ - ,\ell }^\dag O{V_{ - ,\ell }}} \right]{V_{ + ,\ell }}} \right|{\Psi _0}} \right\rangle }^2}} \left( {\left\langle {{\Psi _0}\left| {V_{ + ,\ell }^\dag \left[ {{X_\ell },\left[ {{X_\ell },V_{ - ,\ell }^\dag O{V_\ell }} \right]} \right]{V_{ + ,\ell }}} \right|{\Psi _0}} \right\rangle } \right)~,
\end{align}
where
\begin{align}
&{U_{{\ell _1},{\ell _2}}} = \prod\limits_{\ell  = {\ell _1} + 1}^{{\ell _2} - 1} {{W_\ell }{U_\ell }}~, \nonumber\\
&{V_{{\ell _1},{\ell _2}}} = {U_{{\ell _1},{\ell _2}}}{W_{{\ell _2}}}{U_{{\ell _2}}}\quad \left( {{\text{ for }}{\ell _1} < {\ell _2}} \right)~.
\end{align}

\section{Frozen QNTK}\label{supp:frozenqntk}
We now analyze the QNTK, as defined in Eq.\eqref{supp:QNTK}, for the case when $V_{-}$ and $V_{+}$ are independent and match the Haar distribution up to the second moment. In particular, we perform the following average of $K$:
\begin{align}
\overline{K} &=  - \sum\limits_\ell  {\int {d{V_{ + ,\ell }}d{V_{ - ,\ell }}{{\left\langle {{\Psi _0}\left| {V_{ + ,\ell }^\dag \left[ {{X_\ell },V_{ - ,\ell }^\dag O{V_{ - ,\ell }}} \right]{V_{ + ,\ell }}} \right|{\Psi _0}} \right\rangle }^2}} } \nonumber\\
&=  - \sum\limits_\ell  {\int {d{V_{ + ,\ell }}d{V_{ - ,\ell }}\left\langle {{\Psi _0}\left| {V_ + ^\dag \left[ {{X_\ell },V_{ - ,\ell }^\dag O{V_{ - ,\ell }}} \right]{V_{ + ,\ell }}} \right|{\Psi _0}} \right\rangle \left\langle {{\Psi _0}\left| {V_{ + ,\ell }^\dag \left[ {{X_\ell },V_{ - ,\ell }^\dag O{V_{ - ,\ell }}} \right]{V_{ + ,\ell }}} \right|{\Psi _0}} \right\rangle } } ~.
\end{align}
Let ${\rho _0} = \left| {{\Psi _0}} \right\rangle \left\langle {{\Psi _0}} \right|$. Then we get 
\begin{align}
\overline{K} &=  - \sum\limits_\ell  {\int {d{V_{ + ,\ell }}d{V_{ - ,\ell }}{\rm{Tr}}\left( {{\rho _0}V_{ + ,\ell }^\dag \left[ {{X_\ell },V_{ - ,\ell }^\dag O{V_{ - ,\ell }}} \right]{V_{ + ,\ell }}{\rho _0}V_{ + ,\ell }^\dag \left[ {{X_\ell },V_{ - ,\ell }^\dag O{V_{ - ,\ell }}} \right]{V_{ + ,\ell }}} \right)}} \nonumber\\
&=  - \sum\limits_\ell  {\int {d{V_{ - ,\ell }}} \int {d{V_{ + ,\ell }}{\rm{Tr}}\left( {{\rho _0}V_{ + ,\ell }^\dag {P_{ - ,\ell }}{V_{ + ,\ell }}{\rho _0}V_{ + ,\ell }^\dag {P_{ - ,\ell }}{V_{ + ,\ell }}} \right)} } ~,
\end{align}
where we defined ${P_{ - ,\ell }} = \left[ {{X_\ell },V_{ - ,\ell }^\dag O{V_{ - ,\ell }}} \right]$. Here, we assumed $\rho_0$ is a pure state. Our proof can be generalized to mixed states, similar to the results in \cite{holmes2021connecting}.

Then from formulas that allow for the symbolical integration with respect to the Haar measure on a unitary group \cite{puchala2011symbolic}, or 
from the \texttt{RTNI} package \cite{fukuda2019rtni}, we get
\begin{align}
&\int{d{V_{ + ,\ell }}{\rm{Tr}}\left( {{\rho _0}V_{ + ,\ell }^\dag {P_{ - ,\ell }}{V_{ + ,\ell }}{\rho _0}V_{ + ,\ell }^\dag {P_{ - ,\ell }}{V_{ + ,\ell }}} \right)} \nonumber \\
&= \frac{{{\rm{Tr}}^2\left( {{P_{ - ,\ell }}} \right){\rm{Tr}}\left( {\rho _0^2} \right)}}{{{D^2} - 1}} + \frac{{{\rm{Tr}}\left( {P_{ - ,\ell }^2} \right){\rm{Tr}}\left( {\rho _0^2} \right)}}{{{D^2} - 1}} + \frac{{{\rm{Tr}}^2\left( {{P_{ - ,\ell }}} \right){\rm{Tr}}^2\left( {{\rho _0}} \right)}}{{D - {D^3}}} + \frac{{{\rm{Tr}}\left( {P_{ - ,\ell }^2} \right){\rm{Tr}}\left( {\rho _0^2} \right)}}{{D - {D^3}}}\nonumber\\
&= \frac{{{\rm{Tr}}\left( {P_{ - ,\ell }^2} \right)}}{{{D^2} - 1}} + \frac{{{\rm{Tr}}\left( {P_{ - ,\ell }^2} \right)}}{{D - {D^3}}}\nonumber\\
&= \frac{{{\rm{Tr}}\left( {P_{ - ,\ell }^2} \right)}}{{{D^2} + D}}~.
\end{align}

Similarly, after averaging over $V_{-, \ell}$ we get 
\begin{align}
\overline{K} &=  - \sum\limits_\ell  {\int {d{V_{ - ,\ell }}\frac{{{\rm{Tr}}\left( {P_{ - ,\ell }^2} \right)}}{{{D^2} + D}}} } \nonumber\\
&=  - \frac{1}{{{D^2} + D}}\sum\limits_\ell  {\int {d{V_{ - ,\ell }}{\rm{Tr}}\left( {\left[ {{X_\ell },V_{ - ,\ell }^\dag O{V_{ - ,\ell }}} \right]\left[ {{X_\ell },V_{ - ,\ell }^\dag O{V_{ - ,\ell }}} \right]} \right)} }~.
\end{align}

From the identity ${\rm{Tr}}\left( {{{\left[ {A,B} \right]}^2}} \right)= 2{\rm{Tr}}\left( {ABAB} \right) - 2{\rm{Tr}}\left( {A^2B^2} \right)$, we get the following simplified expression of $\overline{K}$
\begin{align}
\overline{K} &= \frac{2}{{{D^2} + D}}\sum\limits_\ell  {\left(
\frac{{{\rm{Tr}}\left( {{O^2}} \right){\rm{Tr}}\left( {X_\ell ^2} \right)}}{D} - \frac{{{\rm{Tr}}^2\left( O \right){\rm{Tr}}\left( {X_\ell ^2} \right)}}{{{D^2} - 1}} - \frac{{{\rm{Tr}}\left( {{O^2}} \right){\rm{Tr}}^2\left( {{X_\ell }} \right)}}{{{D^2} - 1}}
 + \frac{{{\rm{Tr}}\left( {{O^2}} \right){\rm{Tr}}\left( {X_\ell ^2} \right)}}{{{D^3} - D}} + \frac{{{\rm{Tr}}^2\left( O \right){\rm{Tr}}^2\left( {{X_\ell }} \right)}}{{{D^3} - D}} \right)} \nonumber\\
&=\frac{2}{{{D^2} + D}}\left( {\frac{{D{\mathop{\rm Tr}\nolimits} \left( {{O^2}} \right) - {{{\mathop{\rm Tr}\nolimits} }^2}\left( O \right)}}{{{D^2} - 1}}} \right){\mathop{\rm Tr}\nolimits} \left( {\sum\limits_\ell  {X_\ell ^2} } \right)~.
\end{align}

Since $\tr(X_l^2)= cN$, where $c$ is some constant, in the large-$D$ limit, we get 
\begin{align}
\overline{K} \approx \frac{{L{\mathop{\rm Tr}\nolimits} \left( {{O^2}} \right)}}{{{D^2}}}~.
\end{align}

\section{Fluctuations in the QNTK}\label{supp:fluc}
In this section, we first analyze the fluctuations in $K$ around $\overline{K}$, i.e., $\Delta K^2 = {\mathbb{E}}\left( {{{\left( {K - \bar K} \right)}^2}} \right)$. We get the following expression for $\Delta K^2 = \mathbb{E}(K^2) - \overline{K}^2$:
\begin{align}\label{supp:deltaksquare}
\Delta K^2 =& 2\sum\limits_{{\ell _1} < {\ell _2}} {\int d } {V_{ + ,{\ell _1}}}d{V_{ + ,{\ell _2}}}d{V_{ - ,{\ell _1}}}d{V_{ - ,{\ell _2}}} \times \nonumber\\
&\left( \begin{array}{l}
{\mathop{\rm Tr}\nolimits} \left( {{\rho _0}V_{ + ,{\ell _1}}^\dag \left[ {{X_{{\ell _1}}},V_{ - ,{\ell _1}}^\dag O{V_{ - ,{\ell _1}}}} \right]{V_{ + ,{\ell _1}}}{\rho _0}V_{ + ,{\ell _1}}^\dag \left[ {{X_{{\ell _1}}},V_{ - ,{\ell _1}}^\dag O{V_{ - ,{\ell _1}}}} \right]{V_{ + ,{\ell _1}}}} \right)\\
{\mathop{\rm Tr}\nolimits} \left( {{\rho _0}V_{ + ,{\ell _2}}^\dag \left[ {{X_{{\ell _2}}},V_{ - ,{\ell _2}}^\dag O{V_{ - ,{\ell _2}}}} \right]{V_{ + ,{\ell _2}}}{\rho _0}V_{ + ,{\ell _2}}^\dag \left[ {{X_{{\ell _2}}},V_{ - ,{\ell _2}}^\dag O{V_{ - ,{\ell _2}}}} \right]{V_{ + ,{\ell _2}}}} \right)
\end{array} \right)\nonumber\\
&+ \sum\limits_\ell  {\int d } {V_{ + ,\ell }}d{V_{ - ,\ell }}\left( \begin{array}{l}
{\mathop{\rm Tr}\nolimits} \left( {{\rho _0}V_{ + ,\ell }^\dag \left[ {{X_\ell },V_{ - ,\ell }^\dag O{V_{ - ,\ell }}} \right]{V_{ + ,\ell }}{\rho _0}V_{ + ,\ell }^\dag \left[ {{X_\ell },V_{ - ,\ell }^\dag O{V_{ - ,\ell }}} \right]{V_{ + ,\ell }}} \right)\\
{\mathop{\rm Tr}\nolimits} \left( {{\rho _0}V_{ + ,\ell }^\dag \left[ {{X_\ell },V_{ - ,\ell }^\dag O{V_{ - ,\ell }}} \right]{V_{ + ,\ell }}{\rho _0}V_{ + ,\ell }^\dag \left[ {{X_\ell },V_{ - ,\ell }^\dag O{V_{ - ,\ell }}} \right]{V_{ + ,\ell }}} \right)\end{array} \right) - {{\overline{K}}^2}~.
\end{align}

In order to perform the integration, we make an assumption that $V_{+, \ell_1}, V_{-, \ell_1}, V_{+, \ell_2}$, and $V_{-, \ell_2}$ are independent and they form a 4-design. We focus on the following term in $\Delta K^2$:
\begin{align}
\Delta {K^2} \supset& 2\sum\limits_{{\ell _1} < {\ell _2}} {\int d } {V_{ + ,{\ell _1}}}d{V_{ + ,{\ell _2}}}d{V_{ - ,{\ell _1}}}d{V_{ - ,{\ell _2}}} \times \nonumber\\
&\left( {\begin{array}{*{20}{l}}
{{\rm{Tr}}\left( {{\rho _0}V_{ + ,{\ell _1}}^\dag \left[ {{X_{{\ell _1}}},V_{ - ,{\ell _1}}^\dag O{V_{ - ,{\ell _1}}}} \right]{V_{ + ,{\ell _1}}}{\rho _0}V_{ + ,{\ell _1}}^\dag \left[ {{X_{{\ell _1}}},V_{ - ,{\ell _1}}^\dag O{V_{ - ,{\ell _1}}}} \right]{V_{ + ,{\ell _1}}}} \right)}\\
{{\rm{Tr}}\left( {{\rho _0}V_{ + ,{\ell _2}}^\dag \left[ {{X_{{\ell _2}}},V_{ - ,{\ell _2}}^\dag O{V_{ - ,{\ell _2}}}} \right]{V_{ + ,{\ell _2}}}{\rho _0}V_{ + ,{\ell _2}}^\dag \left[ {{X_{{\ell _2}}},V_{ - ,{\ell _2}}^\dag O{V_{ - ,{\ell _2}}}} \right]{V_{ + ,{\ell _2}}}} \right)}
\end{array}} \right)~.
\end{align}

Because $V_{+, \ell_1}, V_{-, \ell_1}, V_{+, \ell_2}$, and $V_{-, \ell_2}$ are independent, the aforementioned term can be decomposed as
\begin{align}
\Delta {K^2} \supset&
2\sum\limits_{{\ell _1} < {\ell _2}} {\int d } {V_{ + ,{\ell _1}}}d{V_{ - ,{\ell _1}}} \times {\rm{Tr}}\left( {{\rho _0}V_{ + ,{\ell _1}}^\dag \left[ {{X_{{\ell _1}}},V_{ - ,{\ell _1}}^\dag O{V_{ - ,{\ell _1}}}} \right]{V_{ + ,{\ell _1}}}{\rho _0}V_{ + ,{\ell _1}}^\dag \left[ {{X_{{\ell _1}}},V_{ - ,{\ell _1}}^\dag O{V_{ - ,{\ell _1}}}} \right]{V_{ + ,{\ell _1}}}} \right)\nonumber\\
&\qquad \int {d{V_{ + ,{\ell _2}}}d{V_{ - ,{\ell _2}}}}  \times {\rm{Tr}}\left( {{\rho _0}V_{ + ,{\ell _2}}^\dag \left[ {{X_{{\ell _2}}},V_{ - ,{\ell _2}}^\dag O{V_{ - ,{\ell _2}}}} \right]{V_{ + ,{\ell _2}}}{\rho _0}V_{ + ,{\ell _2}}^\dag \left[ {{X_{{\ell _2}}},V_{ - ,{\ell _2}}^\dag O{V_{ - ,{\ell _2}}}} \right]{V_{ + ,{\ell _2}}}} \right)\\
& =  \frac{{2 \times 4}}{{{{({D^2} + D)}^2}{{({D^2} - 1)}^2}}}{\left( {D{\mathop{\rm Tr}\nolimits} \left( {{O^2}} \right) - {{{\mathop{\rm Tr}\nolimits} }^2}(O)} \right)^2}\sum\limits_{{\ell _1} < {\ell _2}} {{\mathop{\rm Tr}\nolimits} } \left( {X_{{\ell _1}}^2} \right){\mathop{\rm Tr}\nolimits} \left( {X_{{\ell _2}}^2} \right)\nonumber\\
&= \frac{{4{D^2}}}{{{{({D^2} + D)}^2}{{({D^2} - 1)}^2}}}{\left( {D{\mathop{\rm Tr}\nolimits} \left( {{O^2}} \right) - {{{\mathop{\rm Tr}\nolimits} }^2}(O)} \right)^2}L(L - 1)\nonumber\\
& = \left( \begin{array}{l}
\frac{{4{{{\mathop{\rm Tr}\nolimits} }^2}\left( {{O^2}} \right)}}{{{D^4}}} - \frac{{8\left( {{{{\mathop{\rm Tr}\nolimits} }^2}(O){\mathop{\rm Tr}\nolimits} \left( {{O^2}} \right) + {{{\mathop{\rm Tr}\nolimits} }^2}\left( {{O^2}} \right)} \right)}}{{{D^5}}}\\
 + \frac{{4{{{\mathop{\rm Tr}\nolimits} }^4}(O) + 16{{{\mathop{\rm Tr}\nolimits} }^2}(O){\mathop{\rm Tr}\nolimits} \left( {{O^2}} \right) + {{{\mathop{\rm Tr}\nolimits} }^2}\left( {{O^2}} \right)}}{{{D^6}}} +  \ldots
\end{array} \right)L(L - 1)~.
\end{align}

We now focus on the second term in Eq.~\eqref{supp:deltaksquare}.

\begin{align}
\Delta {K^2} \supset \sum\limits_\ell  {\int {d{V_{ - ,\ell }}} \int d } {V_{ + ,\ell }}\left( {\begin{array}{*{20}{l}}
{{\rm{Tr}}\left( {{\rho _0}V_{ + ,\ell }^\dag {P_{ - ,\ell }}{V_{ + ,\ell }}{\rho _0}V_{ + ,\ell }^\dag {P_{ - ,\ell }}{V_{ + ,\ell }}} \right)}\\
{{\rm{Tr}}\left( {{\rho _0}V_{ + ,\ell }^\dag {P_{ - ,\ell }}{V_{ + ,\ell }}{\rho _0}V_{ + ,\ell }^\dag {P_{ - ,\ell }}{V_{ + ,\ell }}} \right)}
\end{array}} \right)~,
\end{align}
where we again defined 
\begin{align}\label{seq:pl}
    {P_{ - ,\ell }} = \left[ {{X_\ell },V_{ - ,\ell }^\dag O{V_{ - ,\ell }}} \right]~.
\end{align} 

We fist perform the integration over ${V_{ + ,\ell }}$. Using Section~\ref{rtni}, we represent the aforementioned integral as follows:
\begin{align}
&\int d {V_{ + ,\ell }}\left( {\begin{array}{*{20}{l}}
{{\rm{Tr}}\left( {{\rho _0}V_{ + ,\ell }^\dag {P_{ - ,\ell }}{V_{ + ,\ell }}{\rho _0}V_{ + ,\ell }^\dag {P_{ - ,\ell }}{V_{ + ,\ell }}} \right)}\\
{{\rm{Tr}}\left( {{\rho _0}V_{ + ,\ell }^\dag {P_{ - ,\ell }}{V_{ + ,\ell }}{\rho _0}V_{ + ,\ell }^\dag {P_{ - ,\ell }}{V_{ + ,\ell }}} \right)}
\end{array}} \right)\nonumber\\
&= \sum\limits_{{i_1},{i_2}} {\int d {V_{ + ,\ell }}\left( {{\rm{Tr}}\left( {\bar \rho _0^{{i_2},{i_1}}V_{ + ,\ell }^\dag {P_{ - ,\ell }}{V_{ + ,\ell }}{\rho _0}V_{ + ,\ell }^\dag {P_{ - ,\ell }}{V_{ + ,\ell }}\bar \rho _0^{{i_1},{i_2}}V_{ + ,\ell }^\dag {P_{ - ,\ell }}{V_{ + ,\ell }}{\rho _0}V_{ + ,\ell }^\dag {P_{ - ,\ell }}{V_{ + ,\ell }}} \right)} \right)} ~,
\end{align}
where we follow the following notation, as defined in Section~\ref{rtni}
\begin{align}
{P_{{i_{{j_1}}},{i_{{j_2}}}}} & = \left| {{i_{{j_1}}}} \right\rangle \left\langle {{i_{{j_2}}}} \right|~,\\
{{\bar \rho}_{0}^{i_n,i_1}} & = {P_{i_n,i_1}}{\rho_{0}}
\end{align}

Note that the following identities hold
\begin{align}
\sum\limits_{{i_1},{i_2}} {{\rm{Tr}}(\bar \rho _0^{{i_2},{i_1}}){\rm{Tr}}(\bar \rho _0^{{i_1},{i_2}})}  = \sum\limits_{{i_1},{i_2}} {{\rm{Tr}}(\bar \rho _0^{{i_2},{i_1}}){\rm{Tr}}({\rho _0}\bar \rho _0^{{i_1},{i_2}})}  = \sum\limits_{{i_1},{i_2}} {{\rm{Tr}}(\bar \rho _0^{{i_2},{i_1}}\bar \rho _0^{{i_1},{i_2}})}  = 1~,
\end{align}
which we simplify further using \texttt{RTNI} \cite{fukuda2019rtni} as follows
\begin{align}\label{seq:term1}
\int d {V_{ + ,\ell }}\left( {\begin{array}{*{20}{l}}
{{\rm{Tr}}\left( {{\rho _0}V_{ + ,\ell }^\dag {P_{ - ,\ell }}{V_{ + ,\ell }}{\rho _0}V_{ + ,\ell }^\dag {P_{ - ,\ell }}{V_{ + ,\ell }}} \right)}\\
{{\rm{Tr}}\left( {{\rho _0}V_{ + ,\ell }^\dag {P_{ - ,\ell }}{V_{ + ,\ell }}{\rho _0}V_{ + ,\ell }^\dag {P_{ - ,\ell }}{V_{ + ,\ell }}} \right)}
\end{array}} \right) = \frac{{3\left( {{\rm{Tr}}^2\left( {P_{ - ,\ell }^2} \right) + 2{\rm{Tr}}\left( {P_{ - ,\ell }^4} \right)} \right)}}{{D(D + 1)(D + 2)(D + 3)}}~.
\end{align}

Below we analyze each term of the aforementioned equation separately. 
\begin{itemize}
\item Consider the first term:
\begin{align}
\Delta {K^2} \supset \frac{6}{{D(D + 1)(D + 2)(D + 3)}}\sum\limits_\ell  {\int d } {V_{ - ,\ell }}{\mathop{\rm Tr}\nolimits} \left( {P_{ - ,\ell }^4} \right)~.
\end{align}

Using Eq.~\eqref{seq:pl}, we  expand $\Tr(P^4_{-, \ell})$, and get the following 16 terms
\begin{align}
&{\mathop{\rm Tr}\nolimits} \left( {{{\left[ {A,B} \right]}^4}} \right) = {\mathop{\rm Tr}\nolimits} \left( {(AB - BA)(AB - BA)(AB - BA)(AB - BA)} \right) = \nonumber\\
&+ {\mathop{\rm Tr}\nolimits} \left( {ABABABAB} \right) - {\mathop{\rm Tr}\nolimits} \left( {ABABABBA} \right) - {\mathop{\rm Tr}\nolimits} \left( {ABABBAAB} \right) + {\mathop{\rm Tr}\nolimits} \left( {ABABBABA} \right)\nonumber\\
&- {\mathop{\rm Tr}\nolimits} \left( {ABBAABAB} \right) + {\mathop{\rm Tr}\nolimits} \left( {ABBAABBA} \right) + {\mathop{\rm Tr}\nolimits} \left( {ABBABAAB} \right) - {\mathop{\rm Tr}\nolimits} \left( {ABBABABA} \right)\nonumber\\
&- {\mathop{\rm Tr}\nolimits} \left( {BAABABAB} \right) + {\mathop{\rm Tr}\nolimits} \left( {BAABABBA} \right) + {\mathop{\rm Tr}\nolimits} \left( {BAABBAAB} \right) - {\mathop{\rm Tr}\nolimits} \left( {BAABBABA} \right)\nonumber\\
&+ {\mathop{\rm Tr}\nolimits} \left( {BABAABAB} \right) - {\mathop{\rm Tr}\nolimits} \left( {BABAABBA} \right) - {\mathop{\rm Tr}\nolimits} \left( {BABABAAB} \right) + {\mathop{\rm Tr}\nolimits} \left( {BABABABA} \right)~.
\end{align}
where each term is given by
\begin{align}
 &+ {\rm{Tr}}\left( {ABABABAB} \right) \equiv {\rm{Tr}}\left( {{X_\ell }V_{ - ,\ell }^\dag O{V_{ - ,\ell }}{X_\ell }V_{ - ,\ell }^\dag O{V_{ - ,\ell }}{X_\ell }V_{ - ,\ell }^\dag O{V_{ - ,\ell }}{X_\ell }V_{ - ,\ell }^\dag O{V_{ - ,\ell }}} \right)~,\nonumber\\
&- {\rm{Tr}}\left( {ABABABBA} \right) \equiv  - {\rm{Tr}}\left( {{X_\ell }V_{ - ,\ell }^\dag {O^3}{V_{ - ,\ell }}{X_\ell }V_{ - ,\ell }^\dag O{V_{ - ,\ell }}} \right)~,\nonumber\\
&- {\rm{Tr}}\left( {ABABBAAB} \right) \equiv  - {\rm{Tr}}\left( {{X_\ell }V_{ - ,\ell }^\dag O{V_{ - ,\ell }}{X_\ell }V_{ - ,\ell }^\dag {O^3}{V_{ - ,\ell }}} \right)~,\nonumber\\
&+ {\rm{Tr}}\left( {ABABBABA} \right) \equiv  + {\rm{Tr}}\left( {{X_\ell }V_{ - ,\ell }^\dag {O^2}{V_{ - ,\ell }}{X_\ell }V_{ - ,\ell }^\dag {O^2}{V_{ - ,\ell }}} \right)~,\nonumber\\
&- {\rm{Tr}}\left( {ABBAABAB} \right) \equiv  - {\rm{Tr}}\left( {{X_\ell }V_{ - ,\ell }^\dag {O^3}{V_{ - ,\ell }}{X_\ell }V_{ - ,\ell }^\dag O{V_{ - ,\ell }}} \right)~,\nonumber\\
&+ {\rm{Tr}}\left( {ABBAABBA} \right) \equiv  + {\rm{Tr}}\left( {{O^4}} \right)~,\nonumber\\
&+ {\rm{Tr}}\left( {ABBABAAB} \right) \equiv  + {\rm{Tr}}\left( {{X_\ell }V_{ - ,\ell }^\dag {O^2}{V_{ - ,\ell }}{X_\ell }V_{ - ,\ell }^\dag {O^2}{V_{ - ,\ell }}} \right)~,\nonumber\\
&- {\rm{Tr}}\left( {ABBABABA} \right) \equiv  - {\rm{Tr}}\left( {{X_\ell }V_{ - ,\ell }^\dag {O^3}{V_{ - ,\ell }}{X_\ell }V_{ - ,\ell }^\dag O{V_{ - ,\ell }}} \right)~,\nonumber\\
&- {\rm{Tr}}\left( {BAABABAB} \right) \equiv  - {\rm{Tr}}\left( {{X_\ell }V_{ - ,\ell }^\dag {O^3}{V_{ - ,\ell }}{X_\ell }V_{ - ,\ell }^\dag O{V_{ - ,\ell }}} \right)~,\nonumber\\
&+ {\rm{Tr}}\left( {BAABABBA} \right) \equiv  + {\rm{Tr}}\left( {{X_\ell }V_{ - ,\ell }^\dag {O^2}{V_{ - ,\ell }}{X_\ell }V_{ - ,\ell }^\dag {O^2}{V_{ - ,\ell }}} \right)~,\nonumber\\
&+ {\rm{Tr}}\left( {BAABBAAB} \right) \equiv  + {\rm{Tr}}\left( {{O^4}} \right)~,\nonumber\\
&- {\rm{Tr}}\left( {BAABBABA} \right) \equiv  - {\rm{Tr}}\left( {{X_\ell }V_{ - ,\ell }^\dag {O^3}{V_{ - ,\ell }}{X_\ell }V_{ - ,\ell }^\dag O{V_{ - ,\ell }}} \right)~,\nonumber\\
&+ {\rm{Tr}}\left( {BABAABAB} \right) \equiv  + {\rm{Tr}}\left( {{X_\ell }V_{ - ,\ell }^\dag {O^2}{V_{ - ,\ell }}{X_\ell }V_{ - ,\ell }^\dag {O^2}{V_{ - ,\ell }}} \right)~,\nonumber\\
&- {\rm{Tr}}\left( {BABAABBA} \right) \equiv  - {\rm{Tr}}\left( {{X_\ell }V_{ - ,\ell }^\dag O{V_{ - ,\ell }}{X_\ell }V_{ - ,\ell }^\dag {O^3}{V_{ - ,\ell }}} \right)~,\nonumber\\
&- {\rm{Tr}}\left( {BABABAAB} \right) \equiv  - {\rm{Tr}}\left( {{X_\ell }V_{ - ,\ell }^\dag {O^3}{V_{ - ,\ell }}{X_\ell }V_{ - ,\ell }^\dag O{V_{ - ,\ell }}} \right)~,\nonumber\\
&+ {\rm{Tr}}\left( {BABABABA} \right) \equiv  + {\rm{Tr}}\left( {{X_\ell }V_{ - ,\ell }^\dag O{V_{ - ,\ell }}{X_\ell }V_{ - ,\ell }^\dag O{V_{ - ,\ell }}{X_\ell }V_{ - ,\ell }^\dag O{V_{ - ,\ell }}{X_\ell }V_{ - ,\ell }^\dag O{V_{ - ,\ell }}} \right)~.
\end{align}

We identify similar terms. In particular we have 
\begin{itemize}
\item ${\rm{Tr}}\left( {{X_\ell }V_{ - ,\ell }^\dag O{V_{ - ,\ell }}{X_\ell }V_{ - ,\ell }^\dag O{V_{ - ,\ell }}{X_\ell }V_{ - ,\ell }^\dag O{V_{ - ,\ell }}{X_\ell }V_{ - ,\ell }^\dag O{V_{ - ,\ell }}} \right)$: two terms with a factor $2$.
\item $ {\rm{Tr}}\left( {{X_\ell }V_{ - ,\ell }^\dag {O^3}{V_{ - ,\ell }}{X_\ell }V_{ - ,\ell }^\dag O{V_{ - ,\ell }}} \right)$: eight terms with a factor $-8$.
\item ${\rm{Tr}}\left( {{X_\ell }V_{ - ,\ell }^\dag {O^2}{V_{ - ,\ell }}{X_\ell }V_{ - ,\ell }^\dag {O^2}{V_{ - ,\ell }}} \right)$: four terms with a factor of $4$.
\item $ {\rm{Tr}}\left( {{O^4}} \right)$: two terms with a factor of $2$.
\end{itemize}

Combining everything and using \texttt{RTNI} \cite{fukuda2019rtni}, we get
\begin{align}
\Delta {K^2} \supset & \frac{6}{{D(D + 1)(D + 2)(D + 3)}}\sum\limits_\ell  {\int d } {V_{ - ,\ell }}{\mathop{\rm Tr}\nolimits} \left( {P_{ - ,\ell }^4} \right)\nonumber\\
 &= \frac{L}{{{D^4}}}\left( {12{\rm{Tr}}\left( {{O^4}} \right)} \right) + \frac{L}{{{D^5}}}\left( {24{\rm{Tr}}^2\left( {{O^2}} \right) - 48{\rm{Tr}}\left( O \right){\rm{Tr}}\left( {{O^3}} \right) - 72{\rm{Tr}}\left( {{O^4}} \right)} \right)\nonumber\\
 & \qquad + \frac{L}{{{D^6}}}\left( {336{\rm{Tr}}\left( {{O^4}} \right) + 24{\rm{Tr}}\left( {{O^2}} \right){\rm{Tr}}^2\left( O \right) - 144{\rm{Tr}}^2\left( {{O^2}} \right) + 288{\rm{Tr}}\left( O \right){\rm{Tr}}\left( {{O^3}} \right)} \right) +  \ldots ~.
\end{align}

\item We now analyze the second term in Eq.~\eqref{seq:term1}
\begin{align}
\Delta {K^2} \supset \frac{3}{{D(D + 1)(D + 2)(D + 3)}}\sum\limits_\ell  {\int d } {V_{ - ,\ell }}{{\mathop{\rm Tr}\nolimits} ^2}\left( {P_{ - ,\ell }^2} \right)~,
\end{align}
Using the notation of Section \ref{rtni}, we get
\begin{align}
\int d {V_{ - ,\ell }}{\mathop{\rm Tr}\nolimits} \left( {P_{ - ,\ell }^2} \right){\mathop{\rm Tr}\nolimits} \left( {P_{ - ,\ell }^2} \right) = \sum\limits_{{i_1},{i_2}} {\int d {V_{ - ,\ell }}{\mathop{\rm Tr}\nolimits} \left( {\bar P_{ - ,\ell }^{{i_2},{i_1}}{P_{ - ,\ell }}\bar P_{ - ,\ell }^{{i_1},{i_2}}{P_{ - ,\ell }}} \right)} ~.
\end{align}

We analyze the aforementioned equation without the sum $\sum_{i_1,i_2}$. We denote the index $\bar{A}^{i_2,i_1}$ as $\overleftarrow{P}A $, and $\bar{A}^{i_1,i_2}$ as $\overrightarrow{P}A$. Expanding the terms inside the trace, we get the following 16 terms:
\begin{align}
 &+ {\rm{Tr}}(\overleftarrow P A BAB\overrightarrow P A BAB) =  + {\rm{Tr}}\left( {\overleftarrow P {X_\ell }V_{ - ,\ell }^\dag O{V_{ - ,\ell }}{X_\ell }V_{ - ,\ell }^\dag O{V_{ - ,\ell }}\overrightarrow P {X_\ell }V_{ - ,\ell }^\dag O{V_{ - ,\ell }}{X_\ell }V_{ - ,\ell }^\dag O{V_{ - ,\ell }}} \right)~,\nonumber\\
&- {\rm{Tr}}(\overleftarrow P A BAB\overrightarrow P A BBA) =  - {\rm{Tr}}\left( {{X_\ell }\overleftarrow P {X_\ell }V_{ - ,\ell }^\dag O{V_{ - ,\ell }}{X_\ell }V_{ - ,\ell }^\dag O{V_{ - ,\ell }}\overrightarrow P {X_\ell }V_{ - ,\ell }^\dag {O^2}{V_{ - ,\ell }}} \right)~,\nonumber\\
&- {\rm{Tr}}(\overleftarrow P A BAB\overrightarrow P B AAB) =  - {\rm{Tr}}\left( {\overleftarrow P {X_\ell }V_{ - ,\ell }^\dag O{V_{ - ,\ell }}{X_\ell }V_{ - ,\ell }^\dag O{V_{ - ,\ell }}\overrightarrow P V_{ - ,\ell }^\dag {O^2}{V_{ - ,\ell }}} \right)~,\nonumber\\
&+ {\rm{Tr}}(\overleftarrow P A BAB\overrightarrow P B ABA) =  + {\rm{Tr}}\left( X_\ell {\overleftarrow P {X_\ell }V_{ - ,\ell }^\dag O{V_{ - ,\ell }}{X_\ell }V_{ - ,\ell }^\dag O{V_{ - ,\ell }}\overrightarrow P V_{ - ,\ell }^\dag O{V_{ - ,\ell }}{X_\ell }V_{ - ,\ell }^\dag O{V_{ - ,\ell }}} \right)~,\nonumber\\
&- {\rm{Tr}}(\overleftarrow P A BBA\overrightarrow P A BAB) =  - {\rm{Tr}}\left( {\overleftarrow P {X_\ell }V_{ - ,\ell }^\dag {O^2}{V_{ - ,\ell }}{X_\ell }\overrightarrow P {X_\ell }V_{ - ,\ell }^\dag O{V_{ - ,\ell }}{X_\ell }V_{ - ,\ell }^\dag O{V_{ - ,\ell }}} \right)~,\nonumber\\
&+ {\rm{Tr}}(\overleftarrow P A BBA\overrightarrow P A BBA) =  + {\rm{Tr}}\left( {{X_\ell }\overleftarrow P {X_\ell }V_{ - ,\ell }^\dag {O^2}{V_{ - ,\ell }}{X_\ell }\overrightarrow P {X_\ell }V_{ - ,\ell }^\dag {O^2}{V_{ - ,\ell }}} \right)~,\nonumber\\
&+ {\rm{Tr}}(\overleftarrow P A BBA\overrightarrow P B AAB) =  + {\rm{Tr}}\left( {\overleftarrow P {X_\ell }V_{ - ,\ell }^\dag {O^2}{V_{ - ,\ell }}{X_\ell }\overrightarrow P V_{ - ,\ell }^\dag {O^2}{V_{ - ,\ell }}} \right)~,\nonumber\\
&- {\rm{Tr}}(\overleftarrow P A BBA\overrightarrow P B ABA) =  - {\rm{Tr}}\left( {{X_\ell }\overleftarrow P {X_\ell }V_{ - ,\ell }^\dag {O^2}{V_{ - ,\ell }}{X_\ell }\overrightarrow P V_{ - ,\ell }^\dag O{V_{ - ,\ell }}{X_\ell }V_{ - ,\ell }^\dag O{V_{ - ,\ell }}} \right)~,\nonumber\\
&- {\rm{Tr}}(\overleftarrow P B AAB\overrightarrow P A BAB) =  - {\rm{Tr}}\left( {\overleftarrow P V_{ - ,\ell }^\dag {O^2}{V_{ - ,\ell }}\overrightarrow P {X_\ell }V_{ - ,\ell }^\dag O{V_{ - ,\ell }}{X_\ell }V_{ - ,\ell }^\dag O{V_{ - ,\ell }}} \right)~,\nonumber\\
&+ {\rm{Tr}}(\overleftarrow P B AAB\overrightarrow P A BBA) =  + {\rm{Tr}}\left( {{X_\ell }\overleftarrow P V_{ - ,\ell }^\dag {O^2}{V_{ - ,\ell }}\overrightarrow P {X_\ell }V_{ - ,\ell }^\dag {O^2}{V_{ - ,\ell }}} \right)~,\nonumber\\
&+ {\rm{Tr}}(\overleftarrow P B AAB\overrightarrow P B AAB) =  + {\rm{Tr}}\left( {\overleftarrow P V_{ - ,\ell }^\dag {O^2}{V_{ - ,\ell }}\overrightarrow P V_{ - ,\ell }^\dag {O^2}{V_{ - ,\ell }}} \right)~,\nonumber\\
&- {\rm{Tr}}(\overleftarrow P B AAB\overrightarrow P B ABA) =  - {\rm{Tr}}\left( {{X_\ell }\overleftarrow P V_{ - ,\ell }^\dag {O^2}{V_{ - ,\ell }}\overrightarrow P V_{ - ,\ell }^\dag O{V_{ - ,\ell }}{X_\ell }V_{ - ,\ell }^\dag O{V_{ - ,\ell }}} \right)~,\nonumber\\
&+ {\rm{Tr}}(\overleftarrow P B ABA\overrightarrow P A BAB) =  + {\rm{Tr}}\left( {\overleftarrow P V_{ - ,\ell }^\dag O{V_{ - ,\ell }}{X_\ell }V_{ - ,\ell }^\dag O{V_{ - ,\ell }}{X_\ell }\overrightarrow P {X_\ell }V_{ - ,\ell }^\dag O{V_{ - ,\ell }}{X_\ell }V_{ - ,\ell }^\dag O{V_{ - ,\ell }}} \right)~,\nonumber\\
&- {\rm{Tr}}(\overleftarrow P B ABA\overrightarrow P A BBA) =  - {\rm{Tr}}\left( {{X_\ell }\overleftarrow P V_{ - ,\ell }^\dag O{V_{ - ,\ell }}{X_\ell }V_{ - ,\ell }^\dag O{V_{ - ,\ell }}{X_\ell }\overrightarrow P {X_\ell }V_{ - ,\ell }^\dag {O^2}{V_{ - ,\ell }}} \right)~,\nonumber\\
&- {\rm{Tr}}(\overleftarrow P B ABA\overrightarrow P B AAB) =  - {\rm{Tr}}\left( {\overleftarrow P V_{ - ,\ell }^\dag O{V_{ - ,\ell }}{X_\ell }V_{ - ,\ell }^\dag O{V_{ - ,\ell }}{X_\ell }\overrightarrow P V_{ - ,\ell }^\dag {O^2}{V_{ - ,\ell }}} \right)~,\nonumber\\
&+ {\rm{Tr}}(\overleftarrow P B ABA\overrightarrow P B ABA) =  + {\rm{Tr}}\left( {{X_\ell }\overleftarrow P V_{ - ,\ell }^\dag O{V_{ - ,\ell }}{X_\ell }V_{ - ,\ell }^\dag O{V_{ - ,\ell }}{X_\ell }\overrightarrow P V_{ - ,\ell }^\dag O{V_{ - ,\ell }}{X_\ell }V_{ - ,\ell }^\dag O{V_{ - ,\ell }}} \right)~.
\end{align}

In summary, we get:
\begin{itemize}
\item Four 4-design terms, which contribute equally, and we get a factor of $4$.
\item Eight 3-design terms, which contribute equally, and we get a factor of $-8$.
\item Four 2-design terms, which contribute equally, and we get a factor of $4$.

\end{itemize}

We invoke the following trace identities: 
\begin{align}
{\rm{Tr}}(\overleftarrow P ){\rm{Tr}}(\overrightarrow P ) &= \sum\limits_{{i_1},{i_2}} {{\rm{Tr}}(\left| {{i_2}} \right\rangle \left\langle {{i_1}} \right|){\rm{Tr}}(\left| {{i_1}} \right\rangle \left\langle {{i_2}} \right|)}  = \sum\limits_{{i_1},{i_2}} {{\delta _{{i_1}{i_2}}}{\delta _{{i_2}{i_1}}}}  = D~,\nonumber\\
{\rm{Tr}}(\overleftarrow P A){\rm{Tr}}(\overrightarrow P B) &= \sum\limits_{{i_1},{i_2}} {{\rm{Tr}}(\left| {{i_2}} \right\rangle \left\langle {{i_1}} \right|A){\rm{Tr}}(\left| {{i_1}} \right\rangle \left\langle {{i_2}} \right|B)}  = \sum\limits_{{i_1},{i_2}} {\left\langle {{i_1}} \right|A\left| {{i_2}} \right\rangle \left\langle {{i_2}} \right|B\left| {{i_1}} \right\rangle }  = {\rm{Tr}}(AB)~,\nonumber\\
{\rm{Tr}}(\overleftarrow P A\overrightarrow P B)& = \sum\limits_{{i_1},{i_2}} {{\rm{Tr}}(\left| {{i_2}} \right\rangle \left\langle {{i_1}} \right|A\left| {{i_1}} \right\rangle \left\langle {{i_2}} \right|B)}  = \sum\limits_{{i_1},{i_2}} {\left\langle {{i_1}} \right|A\left| {{i_1}} \right\rangle \left\langle {{i_2}} \right|B\left| {{i_2}} \right\rangle }  = {\rm{Tr}}(A){\rm{Tr}}(B)~,\nonumber\\
\operatorname{Tr} (\overleftarrow{P}  \overrightarrow{P}) &= \sum\limits_{{i_1},{i_2}} {{\mathop{\rm Tr}\nolimits} } \left( {\left| {{i_2}} \right\rangle \left\langle {{i_1}|{i_1}} \right\rangle \left\langle {{i_2}} \right|} \right) = {D^2}~,
\end{align}
where by $\overleftarrow P$ and $\overrightarrow P$ we mean $\overleftarrow P I$ and $\overrightarrow P I$, and $I$ is the identity matrix.

After simplifications with the help of \texttt{RTNI}, we get
\begin{align}
\Delta {K^2} \supset& \frac{3}{{D(D + 1)(D + 2)(D + 3)}}\sum\limits_\ell  {\int d } {V_{ - ,\ell }}{{\mathop{\rm Tr}\nolimits} ^2}\left( {P_{ - ,\ell }^2} \right)\nonumber\\
&= \frac{L}{{{D^4}}}\left( {12{\rm{Tr}}^2\left( {{O^2}} \right)} \right) - \frac{L}{{{D^5}}}\left( {24{\rm{Tr}}\left( {{O^2}} \right){\rm{Tr}}^2\left( O \right) + 72{\rm{Tr}}^2\left( {{O^2}} \right)} \right)\nonumber\\
&\qquad + \frac{L}{{{D^6}}}\left( {12{\rm{Tr}}\left( {{O^4}} \right) + 144{\rm{Tr}}\left( {{O^2}} \right){\rm{Tr}}^2\left( O \right) + 348{\rm{Tr}}^2\left( {{O^2}} \right)} \right) +  \ldots ~.
\end{align}
\end{itemize}

After combining everything, we get 
\begin{align}
\Delta {K^2} &= \frac{{4{D^2}}}{{{{\left( {{D^2} + D} \right)}^2}{{\left( {{D^2} - 1} \right)}^2}}}{\left( {D{\mathop{\rm Tr}\nolimits} \left( {{O^2}} \right) - {{{\mathop{\rm Tr}\nolimits} }^2}(O)} \right)^2}L(L - 1)\nonumber\\
&\qquad + \frac{6}{{D(D + 1)(D + 2)(D + 3)}}\sum\limits_\ell  {\int d } {V_{ - ,\ell }}{\mathop{\rm Tr}\nolimits} \left( {P_{ - ,\ell }^4} \right)\nonumber\\
&\qquad + \frac{3}{{D(D + 1)(D + 2)(D + 3)}}\sum\limits_\ell  {\int d } {V_{ - ,\ell }}{{\mathop{\rm Tr}\nolimits} ^2}\left( {P_{ - ,\ell }^2} \right)\nonumber\\
&\qquad - {\left( {L\left( {D{\mathop{\rm Tr}\nolimits} \left( {{O^2}} \right) - {{{\mathop{\rm Tr}\nolimits} }^2}(O)} \right)\frac{2}{{D + 1}}\left( {\frac{1}{{{D^2} - 1}}} \right)} \right)^2}\nonumber\\
&= \frac{{L}}{{{D^4}}}\left( 8{{{{\mathop{\rm Tr}\nolimits} }^2}\left( {{O^2}} \right) + 12{\rm{Tr}}\left( {{O^4}} \right)} \right) + \nonumber\\
&\qquad \frac{L}{{{D^5}}}  \left( {16{\mathop{\rm Tr}\nolimits} \left( {{O^2}} \right){{{\mathop{\rm Tr}\nolimits} }^2}\left( O \right) + 48{\mathop{\rm Tr}\nolimits} \left( {{O^3}} \right){\mathop{\rm Tr}\nolimits} \left( O \right) + 40{{{\mathop{\rm Tr}\nolimits} }^2}\left( {{O^2}} \right) + 72{\rm{Tr}}\left( {{O^4}} \right)} \right) + \ldots~.
\end{align}

In the leading order we have
\begin{align}
\Delta {K^2} &\approx \frac{L}{{{D^4}}}\left( {8{{{\mathop{\rm Tr}\nolimits} }^2}\left( {{O^2}} \right) + 12L{\rm{Tr}}\left( {{O^4}} \right)} \right)\nonumber\\
\Rightarrow \Delta K &\approx \frac{{\sqrt L }}{{{D^2}}}\sqrt {\left( {8{{{\mathop{\rm Tr}\nolimits} }^2}\left( {{O^2}} \right) + 12{\rm{Tr}}\left( {{O^4}} \right)} \right)} ~.
\end{align}

\section{dQNTK}\label{supp:dqntk}
After including the second-order term, we get
\begin{align}
\delta \varepsilon  =  - \eta \sum\limits_\ell  {\frac{{\partial \varepsilon }}{{\partial {\varphi _\ell }}}} \frac{{\partial \varepsilon }}{{\partial {\varphi _\ell }}}\varepsilon  + \frac{1}{2}{\eta ^2}{\varepsilon ^2}\sum\limits_{{\ell _1},{\ell _2}} {\frac{{{\partial ^2}\varepsilon }}{{\partial {\varphi _{{\ell _1}}} \partial {\varphi _{{\ell _2}}}}}} \frac{{\partial \varepsilon }}{{\partial {\varphi _{{\ell _1}}}}}\frac{{\partial\varepsilon }}{{\partial {\varphi _{{\ell _2}}}}}~,
\end{align}
Our goal is to derive conditions on system parameters such that the following inequality holds:
\begin{align}
\left| {\frac{1}{2}{\eta ^2}{\varepsilon ^2}\sum\limits_{{\ell _1},{\ell _2}} {\frac{{{\partial ^2}\varepsilon }}{{\partial {\varphi _{{\ell _1}}} \partial {\varphi _{{\ell _2}}}}}} \frac{{\partial \varepsilon }}{{\partial {\varphi _{{\ell _1}}}}}\frac{{\partial\varepsilon }}{{\partial {\varphi _{{\ell _2}}}}}} \right| \ll \left| { - \eta \sum\limits_\ell  {\frac{{\partial \varepsilon }}{{\partial {\varphi _\ell }}}} \frac{{\partial \varepsilon }}{{\partial {\varphi _\ell }}}\varepsilon } \right|~.
\end{align}

We call the second term dQNTK \cite{liu:2021wqr}: 
\begin{align}
\mu  = \sum\limits_{{\ell _1},{\ell _2}} {\frac{{{\partial ^2}\varepsilon }}{{\partial {\varphi _{{\ell _1}}} \partial {\varphi _{{\ell _2}}}}}} \frac{{\partial \varepsilon }}{{\partial {\varphi _{{\ell _1}}}}}\frac{{\partial\varepsilon }}{{\partial {\varphi _{{\ell _2}}}}}~,
\end{align}
which can be expressed as 
\begin{align}
\mu  &= \sum\limits_{{\ell _1}\ne {\ell _2}} {\begin{array}{*{20}{l}}
{\left\langle {{\Psi _0}\left| {U_{ + ,{\ell _1}}^\dag \left[ {{X_{{\ell _1}}},U_{{\ell _1}}^\dag W_{{\ell _1}}^\dag U_{ - ,{\ell _1}}^\dag O{U_{ - ,{\ell _1}}}{W_{{\ell _1}}}{U_{{\ell _1}}}} \right]{U_{ + ,{\ell _1}}}} \right|{\Psi _0}} \right\rangle }\\
\qquad {\left\langle {{\Psi _0}\left| {U_{ + ,{\ell _2}}^\dag \left[ {{X_{{\ell _2}}},U_{{\ell _2}}^\dag W_{{\ell _2}}^\dag U_{ - ,{\ell _2}}^\dag O{U_{ - ,{\ell _2}}}{W_{{\ell _2}}}{U_{{\ell _2}}}} \right]{U_{ + ,{\ell _2}}}} \right|{\Psi _0}} \right\rangle }
\end{array}} \nonumber\\
&\qquad \left( {\left\{ {\begin{array}{*{20}{l}}
{\left\langle {{\Psi _0}\left| {U_{ + ,{\ell _1}}^\dag \left[ {{X_{{\ell _1}}},U_{{\ell _1}}^\dag W_{{\ell _1}}^\dag U_{{\ell _2},{\ell _1}}^\dag \left[ {{X_{{\ell _2}}},U_{{\ell _2}}^\dag W_{{\ell _2}}^\dag U_{ - ,{\ell _2}}^\dag O{U_{ - ,{\ell _2}}}{W_{{\ell _2}}}{U_{{\ell _2}}}} \right]{U_{{\ell _2},{\ell _1}}}{W_{{\ell _1}}}{U_{{\ell _1}}}} \right]{U_{ + ,{\ell _1}}}} \right|{\Psi _0}} \right\rangle :{\ell _1} > {\ell _2}}\\
{\left\langle {{\Psi _0}\left| {U_{ + ,{\ell _2}}^\dag \left[ {{X_{{\ell _2}}},U_{{\ell _2}}^\dag W_{{\ell _2}}^\dag U_{{\ell _1},{\ell _2}}^\dag  \left[ {{X_{{\ell _1}}},U_{{\ell _1}}^\dag W_{{\ell _1}}^\dag U_{ - ,{\ell _1}}^\dag O{U_{ - ,{\ell _1}}}{W_{{\ell _1}}}{U_{{\ell _1}}}} \right]{U_{{\ell _1},{\ell _2}}}{W_{{\ell _2}}}{U_{{\ell _2}}}} \right]{U_{ + ,{\ell _2}}}} \right|{\Psi _0}} \right\rangle :{\ell _1} < {\ell _2}}
\end{array}} \right.} \right)\nonumber\\
&+ \sum\limits_\ell  {{{\left\langle {{\Psi _0}\left| {U_{ + ,\ell }^\dag \left[ {{X_\ell },U_\ell ^\dag W_\ell ^\dag U_{ - ,\ell }^\dag O{U_{ - ,\ell }}{W_\ell }{U_\ell }} \right]{U_{ + ,\ell }}} \right|{\Psi _0}} \right\rangle }^2}} \left( {\left\langle {{\Psi _0}\left| {U_{ + ,\ell }^\dag \left[ {{X_\ell },\left[ {{X_\ell },U_\ell ^\dag W_\ell ^\dag U_{ - ,\ell }^\dag O{U_{ - ,\ell }}{W_\ell }{U_\ell }} \right]} \right]{U_{ + ,\ell }}} \right|{\Psi _0}} \right\rangle } \right)~.
\end{align}

Let
\begin{align}
&{V_{ - ,\ell }} = {U_{ - ,\ell }}{W_\ell }{U_\ell }~,\nonumber\\
&{V_{{\ell _1},{\ell _2}}} = {U_{{\ell _1},{\ell _2}}}{W_{{\ell _2}}}{U_{{\ell _2}}}~~~~~(\text{for }{\ell _1} < {\ell _2})~,\nonumber\\
&{V_{ + ,\ell }} = {U_{ + ,\ell }}~.
\end{align}

Then we get
\begin{align}
\mu & = 2\sum\limits_{{\ell _1} < {\ell _2}} {\left\langle {{\Psi _0}\left| {V_{ + ,{\ell _1}}^\dag \left[ {{X_{{\ell _1}}},V_{ - ,{\ell _1}}^\dag O{V_{ - ,{\ell _1}}}} \right]{V_{ + ,{\ell _1}}}} \right|{\Psi _0}} \right\rangle \left\langle {{\Psi _0}\left| {V_{ + ,{\ell _2}}^\dag \left[ {{X_{{\ell _2}}},V_{ - ,{\ell _2}}^\dag O{V_{ - ,{\ell _2}}}} \right]{V_{ + ,{\ell _2}}}} \right|{\Psi _0}} \right\rangle } \nonumber\\
&\qquad \left( {\left\langle {{\Psi _0}\left| {V_{ + ,{\ell _2}}^\dag \left[ {{X_{{\ell _2}}},V_{{\ell _1},{\ell _2}}^\dag \left[ {{X_{{\ell _1}}},V_{ - ,{\ell _1}}^\dag O{V_{ - ,{\ell _1}}}} \right]{V_{{\ell _1},{\ell _2}}}} \right]{V_{ + ,{\ell _2}}}} \right|{\Psi _0}} \right\rangle } \right)\nonumber\\
&\qquad + \sum\limits_\ell  {{{\left\langle {{\Psi _0}\left| {V_{ + ,\ell }^\dag \left[ {{X_\ell },V_{ - ,\ell }^\dag O{V_{ - ,\ell }}} \right]{V_{ + ,\ell }}} \right|{\Psi _0}} \right\rangle }^2}} \left( {\left\langle {{\Psi _0}\left| {U_{ + ,\ell }^\dag \left[ {{X_\ell },\left[ {{X_\ell },V_{ - ,\ell }^\dag O{V_\ell }} \right]} \right]{U_{ + ,\ell }}} \right|{\Psi _0}} \right\rangle } \right)~.
\end{align}

Using
\begin{align}
&{\rho _0} = \left| {{\Psi _0}} \right\rangle \left\langle {{\Psi _0}} \right|~,\nonumber\\
&{P_{ - ,\ell }} = \left[ {{X_\ell },V_{ - ,\ell }^\dag O{V_{ - ,\ell }}} \right]~,
\end{align}
we get
\begin{align}
&\mu  = 2\sum\limits_{{\ell _1}<{\ell _2}} {\mathop{\rm Tr}\nolimits} \left( {{\rho _0}V_{ + ,{\ell _1}}^\dag {P_{ - ,{\ell _1}}}{V_{ + ,{\ell _1}}}{\rho _0}V_{ + ,{\ell _2}}^\dag {P_{ - ,{\ell _2}}}{V_{ + ,{\ell _2}}}} \right){\mathop{\rm Tr}\nolimits} \left( {{\rho _0}V_{ + ,{\ell _2}}^\dag \left[ {{X_{{\ell _2}}},V_{{\ell _1},{\ell _2}}^\dag {P_{ - ,{\ell _1}}}{V_{{\ell _1},{\ell _2}}}} \right]{V_{ + ,{\ell _2}}}} \right)\nonumber\\
&\qquad +\sum\limits_\ell  {{\rm{Tr}}} \left( {{\rho _0}V_{ + ,\ell }^\dag {P_{ - ,\ell }}{V_{ + ,\ell }}{\rho _0}V_{ + ,\ell }^\dag {P_{ - ,\ell }}{V_{ + ,\ell }}} \right){\rm{Tr}}\left( {{\rho _0}V_{ + ,\ell }^\dag \left[ {{X_\ell },{P_{ - ,\ell }}} \right]{V_{ + ,\ell }}} \right)~.
\end{align}

For estimating $\bar{\mu}$, we define the following notation:
\begin{align}
{{\bar \mu }_1} &= \sum\limits_\ell  {\int d } {V_{ + ,\ell }}d{V_{ - ,\ell }}{\mathop{\rm Tr}\nolimits} \left( {{\rho _0}V_{ + ,\ell }^\dag {P_{ - ,\ell }}{V_{ + ,\ell }}{\rho _0}V_{ + ,\ell }^\dag {P_{ - ,\ell }}{V_{ + ,\ell }}} \right){\mathop{\rm Tr}\nolimits} \left( {{\rho _0}V_{ + ,\ell }^\dag \left[ {{X_\ell },{P_{ - ,\ell }}} \right]{V_{ + ,\ell }}} \right)~,\nonumber\\
{{\bar \mu }_2} &= 2\sum\limits_{{\ell _1} < {\ell _2}} {\int {d{V_{ + ,{\ell _1}}}d{V_{ - ,{\ell _1}}}d{V_{ + ,{\ell _2}}}d{V_{ - ,{\ell _2}}}d{V_{{\ell _2},{\ell _1}}}} } \nonumber\\
&\qquad \times {\mathop{\rm Tr}\nolimits} \left( {{\rho _0}V_{ + ,{\ell _1}}^\dag {P_{ - ,{\ell _1}}}{V_{ + ,{\ell _1}}}{\rho _0}V_{ + ,{\ell _2}}^\dag {P_{ - ,{\ell _2}}}{V_{ + ,{\ell _2}}}} \right){\mathop{\rm Tr}\nolimits} \left( {{\rho _0}V_{ + ,{\ell _2}}^\dag \left[ {{X_{{\ell _2}}},V_{{\ell _1},{\ell _2}}^\dag {P_{ - ,{\ell _1}}}{V_{{\ell _1},{\ell _2}}}} \right]{V_{ + ,{\ell _2}}}} \right)~,\nonumber\\
\bar \mu  &= {{\bar \mu }_1} + {{\bar \mu }_2}~.
\end{align}

We compute the following term first:
\begin{align}
{{\bar \mu }_1}  = \sum\limits_\ell  {\int d } {V_{ + ,\ell }}d{V_{ - ,\ell }}{\rm{Tr}}\left( {{\rho _0}V_{ + ,\ell }^\dag {P_{ - ,\ell }}{V_{ + ,\ell }}{\rho _0}V_{ + ,\ell }^\dag {P_{ - ,\ell }}{V_{ + ,\ell }}} \right){\rm{Tr}}\left( {{\rho _0}V_{ + ,\ell }^\dag \left[ {{X_\ell },{P_{ - ,\ell }}} \right]{V_{ + ,\ell }}} \right)~.
\end{align}

After performing Haar integrals we get
\begin{align}
&\sum\limits_\ell  {\int d } {V_{ + ,\ell }}d{V_{ - ,\ell }}{\mathop{\rm Tr}\nolimits} \left( {{\rho _0}V_{ + ,\ell }^\dag {P_{ - ,\ell }}{V_{ + ,\ell }}{\rho _0}V_{ + ,\ell }^\dag {P_{ - ,\ell }}{V_{ + ,\ell }}} \right){\mathop{\rm Tr}\nolimits} \left( {{\rho _0}V_{ + ,\ell }^\dag \left[ {{X_\ell },{P_{ - ,\ell }}} \right]{V_{ + ,\ell }}} \right)\nonumber\\
&= \sum\limits_\ell  {\int d } {V_{ + ,\ell }}d{V_{ - ,\ell }}{\mathop{\rm Tr}\nolimits} \left( {\overleftarrow P {\rho _0}V_{ + ,\ell }^\dag {P_{ - ,\ell }}{V_{ + ,\ell }}{\rho _0}V_{ + ,\ell }^\dag {P_{ - ,\ell }}{V_{ + ,\ell }}\overrightarrow P {\rho _0}V_{ + ,\ell }^\dag \left[ {{X_\ell },{P_{ - ,\ell }}} \right]{V_{ + ,\ell }}} \right)\nonumber\\
&= \frac{2}{{{D^3} + 3{D^2} + 2D}}\sum\limits_\ell  {\int {d{V_{ + ,\ell }}} {\rm{Tr}}\left( {\left[ {{X_\ell },{P_{ - ,\ell }}} \right]P_{ - ,\ell }^2} \right)} \nonumber\\
&= \frac{2}{{{D^3} + 3{D^2} + 2D}}\sum\limits_\ell  {\int {d{V_{ + ,\ell }}} \left( {{\rm{Tr}}\left( {{X_\ell }{P_{ - ,\ell }}P_{ - ,\ell }^2} \right) - {\rm{Tr}}\left( {{P_{ - ,\ell }}{X_\ell }P_{ - ,\ell }^2} \right)} \right)} \nonumber\\
&= \frac{2}{{{D^3} + 3{D^2} + 2D}}\sum\limits_\ell  {\int {d{V_{ + ,\ell }}} \left( {{\rm{Tr}}\left( {{X_\ell }P_{ - ,\ell }^3} \right) - {\rm{Tr}}\left( {{X_\ell }P_{ - ,\ell }^3} \right)} \right)} \nonumber\\
&= 0~.
\end{align}

Similarly, we focus on the second term in $\bar{\mu}$:
\begin{align}
&{\bar \mu}_2  = 2\sum\limits_{{\ell _1} < {\ell _2}} {\int {d{V_{ + ,{\ell _1}}}d{V_{ - ,{\ell _1}}}d{V_{ + ,{\ell _2}}}d{V_{ - ,{\ell _2}}}d{V_{{\ell _2},{\ell _1}}}} } \nonumber\\
&\times {\mathop{\rm Tr}\nolimits} \left( {{\rho _0}V_{ + ,{\ell _1}}^\dag {P_{ - ,{\ell _1}}}{V_{ + ,{\ell _1}}}{\rho _0}V_{ + ,{\ell _2}}^\dag {P_{ - ,{\ell _2}}}{V_{ + ,{\ell _2}}}} \right){\mathop{\rm Tr}\nolimits} \left( {{\rho _0}V_{ + ,{\ell _2}}^\dag \left[ {{X_{{\ell _2}}},V_{{\ell _1},{\ell _2}}^\dag {P_{ - ,{\ell _1}}}{V_{{\ell _1},{\ell _2}}}} \right]{V_{ + ,{\ell _2}}}} \right)\nonumber\\
&= 2\sum\limits_{{\ell _1} < {\ell _2}} {\int {d{V_{ + ,{\ell _1}}}d{V_{ - ,{\ell _1}}}d{V_{ + ,{\ell _2}}}d{V_{ - ,{\ell _2}}}d{V_{{\ell _2},{\ell _1}}}} } \nonumber\\
&\times {\mathop{\rm Tr}\nolimits} \left( {\overleftarrow P {\rho _0}V_{ + ,{\ell _1}}^\dag {P_{ - ,{\ell _1}}}{V_{ + ,{\ell _1}}}{\rho _0}V_{ + ,{\ell _2}}^\dag {P_{ - ,{\ell _2}}}{V_{ + ,{\ell _2}}}\overrightarrow P {\rho _0}V_{ + ,{\ell _2}}^\dag \left[ {{X_{{\ell _2}}},V_{{\ell _1},{\ell _2}}^\dag {P_{ - ,{\ell _1}}}{V_{{\ell _1},{\ell _2}}}} \right]{V_{ + ,{\ell _2}}}} \right)\nonumber\\
&= 2\sum\limits_{{\ell _1} < {\ell _2}} {\int {d{V_{ - ,{\ell _1}}}d{V_{ + ,{\ell _2}}}d{V_{ - ,{\ell _2}}}d{V_{{\ell _2},{\ell _1}}}} } \nonumber\\
&\times \frac{1}{D}{\mathop{\rm Tr}\nolimits} \left( {{\rho _0}V_{ + ,{\ell _2}}^\dag {P_{ - ,{\ell _2}}}{V_{ + ,{\ell _2}}}\overrightarrow P {\rho _0}V_{ + ,{\ell _2}}^\dag \left[ {{X_{{\ell _2}}},V_{{\ell _1},{\ell _2}}^\dag {P_{ - ,{\ell _1}}}{V_{{\ell _1},{\ell _2}}}} \right]{V_{ + ,{\ell _2}}}\overleftarrow P {\rho _0}} \right){\mathop{\rm Tr}\nolimits} \left( {{P_{ - ,{\ell _1}}}} \right)\nonumber\\
&= 0~.
\end{align}
Thus, we get
\begin{align}
\bar \mu  = {{\bar \mu }_1} + {{\bar \mu }_2} = 0~.
\end{align}

Next, we analyze the fluctuations 
\begin{align}
\Delta {\mu ^2} = {\mathbb{E}}({\mu ^2}) - {\bar \mu ^2} = {\mathbb{E}}({\mu ^2})~.
\end{align}

The general form of $\Delta {\mu ^2}$ is given by
\begin{align}
\Delta {\mu ^2} &=  4\int {\mathcal{D}V} \sum\limits_{{\ell _1} < {\ell _2};{\ell _3} < {\ell _4}} \begin{array}{l}
{\mathop{\rm Tr}\nolimits} \left( {{\rho _0}V_{ + ,{\ell _1}}^\dag {P_{ - ,{\ell _1}}}{V_{ + ,{\ell _1}}}{\rho _0}V_{ + ,{\ell _2}}^\dag {P_{ - ,{\ell _2}}}{V_{ + ,{\ell _2}}}} \right){\mathop{\rm Tr}\nolimits} \left( {{\rho _0}V_{ + ,{\ell _2}}^\dag \left[ {{X_{{\ell _2}}},V_{{\ell _1},{\ell _2}}^\dag {P_{ - ,{\ell _1}}}{V_{{\ell _1},{\ell _2}}}} \right]{V_{ + ,{\ell _2}}}} \right)\\
{\mathop{\rm Tr}\nolimits} \left( {{\rho _0}V_{ + ,{\ell _3}}^\dag {P_{ - ,{\ell _3}}}{V_{ + ,{\ell _3}}}{\rho _0}V_{ + ,{\ell _4}}^\dag {P_{ - ,{\ell _4}}}{V_{ + ,{\ell _4}}}} \right){\mathop{\rm Tr}\nolimits} \left( {{\rho _0}V_{ + ,{\ell _4}}^\dag \left[ {{X_{{\ell _4}}},V_{{\ell _3},{\ell _4}}^\dag {P_{ - ,{\ell _3}}}{V_{{\ell _3},{\ell _4}}}} \right]{V_{ + ,{\ell _4}}}} \right)
\end{array} \nonumber\\
&\qquad + \int {\mathcal{D}V} \sum\limits_{{\ell _1},{\ell _2}} \begin{array}{l}
{\mathop{\rm Tr}\nolimits} \left( {{\rho _0}V_{ + ,{\ell _1}}^\dag {P_{ - ,{\ell _1}}}{V_{ + ,{\ell _1}}}{\rho _0}V_{ + ,{\ell _1}}^\dag {P_{ - ,{\ell _1}}}{V_{ + ,{\ell _1}}}} \right){\mathop{\rm Tr}\nolimits} \left( {{\rho _0}V_{ + ,{\ell _1}}^\dag \left[ {{X_{{\ell _1}}},{P_{ - ,{\ell _1}}}} \right]{V_{ + ,{\ell _1}}}} \right)\\
{\mathop{\rm Tr}\nolimits} \left( {{\rho _0}V_{ + ,{\ell _2}}^\dag {P_{ - ,{\ell _2}}}{V_{ + ,{\ell _2}}}{\rho _0}V_{ + ,{\ell _2}}^\dag {P_{ - ,{\ell _2}}}{V_{ + ,{\ell _2}}}} \right){\mathop{\rm Tr}\nolimits} \left( {{\rho _0}V_{ + ,{\ell _2}}^\dag \left[ {{X_{{\ell _2}}},{P_{ - ,{\ell _2}}}} \right]{V_{ + ,{\ell _2}}}} \right)
\end{array} \nonumber\\
&\qquad + 4\int {\mathcal{D}V} \sum\limits_{{\ell _1} < {\ell _2};{\ell _3}} \begin{array}{l}
{\mathop{\rm Tr}\nolimits} \left( {{\rho _0}V_{ + ,{\ell _1}}^\dag {P_{ - ,{\ell _1}}}{V_{ + ,{\ell _1}}}{\rho _0}V_{ + ,{\ell _2}}^\dag {P_{ - ,{\ell _2}}}{V_{ + ,{\ell _2}}}} \right){\mathop{\rm Tr}\nolimits} \left( {{\rho _0}V_{ + ,{\ell _2}}^\dag \left[ {{X_{{\ell _2}}},V_{{\ell _1},{\ell _2}}^\dag {P_{ - ,{\ell _1}}}{V_{{\ell _1},{\ell _2}}}} \right]{V_{ + ,{\ell _2}}}} \right)\\
{\mathop{\rm Tr}\nolimits} \left( {{\rho _0}V_{ + ,{\ell _3}}^\dag {P_{ - ,{\ell _3}}}{V_{ + ,{\ell _3}}}{\rho _0}V_{ + ,{\ell _3}}^\dag {P_{ - ,{\ell _3}}}{V_{ + ,{\ell _3}}}} \right){\mathop{\rm Tr}\nolimits} \left( {{\rho _0}V_{ + ,{\ell _3}}^\dag \left[ {{X_{{\ell _3}}},{P_{ - ,{\ell _3}}}} \right]{V_{ + ,{\ell _3}}}} \right)
\end{array} ~.
\end{align}
Here, $\int \mathcal{D}V$ is the measure over all $V$s that are random variables appearing in the above formulas. We write
\begin{align}
\Delta {\mu_a^2} &= \int {\mathcal{D}V} \sum\limits_{{\ell _1},{\ell _2}} \begin{array}{l}
{\mathop{\rm Tr}\nolimits} \left( {{\rho _0}V_{ + ,{\ell _1}}^\dag {P_{ - ,{\ell _1}}}{V_{ + ,{\ell _1}}}{\rho _0}V_{ + ,{\ell _1}}^\dag {P_{ - ,{\ell _1}}}{V_{ + ,{\ell _1}}}} \right){\mathop{\rm Tr}\nolimits} \left( {{\rho _0}V_{ + ,{\ell _1}}^\dag \left[ {{X_{{\ell _1}}},{P_{ - ,{\ell _1}}}} \right]{V_{ + ,{\ell _1}}}} \right)\\
{\mathop{\rm Tr}\nolimits} \left( {{\rho _0}V_{ + ,{\ell _2}}^\dag {P_{ - ,{\ell _2}}}{V_{ + ,{\ell _2}}}{\rho _0}V_{ + ,{\ell _2}}^\dag {P_{ - ,{\ell _2}}}{V_{ + ,{\ell _2}}}} \right){\mathop{\rm Tr}\nolimits} \left( {{\rho _0}V_{ + ,{\ell _2}}^\dag \left[ {{X_{{\ell _2}}},{P_{ - ,{\ell _2}}}} \right]{V_{ + ,{\ell _2}}}} \right)
\end{array}~,\nonumber\\
\Delta {\mu_b^2} &=   4\int {\mathcal{D}V} \sum\limits_{{\ell _1} < {\ell _2};{\ell _3}} \begin{array}{l}
{\mathop{\rm Tr}\nolimits} \left( {{\rho _0}V_{ + ,{\ell _1}}^\dag {P_{ - ,{\ell _1}}}{V_{ + ,{\ell _1}}}{\rho _0}V_{ + ,{\ell _2}}^\dag {P_{ - ,{\ell _2}}}{V_{ + ,{\ell _2}}}} \right){\mathop{\rm Tr}\nolimits} \left( {{\rho _0}V_{ + ,{\ell _2}}^\dag \left[ {{X_{{\ell _2}}},V_{{\ell _1},{\ell _2}}^\dag {P_{ - ,{\ell _1}}}{V_{{\ell _1},{\ell _2}}}} \right]{V_{ + ,{\ell _2}}}} \right)\\
{\mathop{\rm Tr}\nolimits} \left( {{\rho _0}V_{ + ,{\ell _3}}^\dag {P_{ - ,{\ell _3}}}{V_{ + ,{\ell _3}}}{\rho _0}V_{ + ,{\ell _3}}^\dag {P_{ - ,{\ell _3}}}{V_{ + ,{\ell _3}}}} \right){\mathop{\rm Tr}\nolimits} \left( {{\rho _0}V_{ + ,{\ell _3}}^\dag \left[ {{X_{{\ell _3}}},{P_{ - ,{\ell _3}}}} \right]{V_{ + ,{\ell _3}}}} \right)\end{array} ~,\nonumber\\
\Delta {\mu_c^2} &= 4\int {\mathcal{D}V} \sum\limits_{{\ell _1} < {\ell _2};{\ell _3} < {\ell _4}} \begin{array}{l}
{\mathop{\rm Tr}\nolimits} \left( {{\rho _0}V_{ + ,{\ell _1}}^\dag {P_{ - ,{\ell _1}}}{V_{ + ,{\ell _1}}}{\rho _0}V_{ + ,{\ell _2}}^\dag {P_{ - ,{\ell _2}}}{V_{ + ,{\ell _2}}}} \right){\mathop{\rm Tr}\nolimits} \left( {{\rho _0}V_{ + ,{\ell _2}}^\dag \left[ {{X_{{\ell _2}}},V_{{\ell _1},{\ell _2}}^\dag {P_{ - ,{\ell _1}}}{V_{{\ell _1},{\ell _2}}}} \right]{V_{ + ,{\ell _2}}}} \right)\\
{\mathop{\rm Tr}\nolimits} \left( {{\rho _0}V_{ + ,{\ell _3}}^\dag {P_{ - ,{\ell _3}}}{V_{ + ,{\ell _3}}}{\rho _0}V_{ + ,{\ell _4}}^\dag {P_{ - ,{\ell _4}}}{V_{ + ,{\ell _4}}}} \right){\mathop{\rm Tr}\nolimits} \left( {{\rho _0}V_{ + ,{\ell _4}}^\dag \left[ {{X_{{\ell _4}}},V_{{\ell _3},{\ell _4}}^\dag {P_{ - ,{\ell _3}}}{V_{{\ell _3},{\ell _4}}}} \right]{V_{ + ,{\ell _4}}}} \right)
\end{array} ~,\nonumber\\
\Delta {\mu^2} &= \Delta {\mu_a^2} +\Delta {\mu_b^2} +\Delta {\mu_c^2}~. 
\end{align}
\begin{itemize}
\item Consider the term
\begin{align}
\Delta {\mu_a^2} =  \int {\mathcal{D}V} \sum\limits_{{\ell _1},{\ell _2}} \begin{array}{l}
{\mathop{\rm Tr}\nolimits} \left( {{\rho _0}V_{ + ,{\ell _1}}^\dag {P_{ - ,{\ell _1}}}{V_{ + ,{\ell _1}}}{\rho _0}V_{ + ,{\ell _1}}^\dag {P_{ - ,{\ell _1}}}{V_{ + ,{\ell _1}}}} \right){\mathop{\rm Tr}\nolimits} \left( {{\rho _0}V_{ + ,{\ell _1}}^\dag \left[ {{X_{{\ell _1}}},{P_{ - ,{\ell _1}}}} \right]{V_{ + ,{\ell _1}}}} \right)\\
{\mathop{\rm Tr}\nolimits} \left( {{\rho _0}V_{ + ,{\ell _2}}^\dag {P_{ - ,{\ell _2}}}{V_{ + ,{\ell _2}}}{\rho _0}V_{ + ,{\ell _2}}^\dag {P_{ - ,{\ell _2}}}{V_{ + ,{\ell _2}}}} \right){\mathop{\rm Tr}\nolimits} \left( {{\rho _0}V_{ + ,{\ell _2}}^\dag \left[ {{X_{{\ell _2}}},{P_{ - ,{\ell _2}}}} \right]{V_{ + ,{\ell _2}}}} \right)
\end{array} ~.
\end{align}

From from the calculation of $\bar{\mu}$ we know that terms with $\ell_1\ne \ell_2$, do not contribute. The only non-zero contribution we get is when $\ell_1=\ell_2$ in the sum. Therefore,
\begin{align}\label{nonvanishing1}
\Delta {\mu_a^2} &=  
\sum\limits_{{\ell _1},{\ell _2}} {\int {{\cal D}V\left( \begin{array}{l}
{\rm{Tr}}\left( {{\rho _0}V_{ + ,{\ell _1}}^\dag {P_{ - ,{\ell _1}}}{V_{ + ,{\ell _1}}}{\rho _0}V_{ + ,{\ell _1}}^\dag {P_{ - ,{\ell _1}}}{V_{ + ,{\ell _1}}}} \right){\rm{Tr}}\left( {{\rho _0}V_{ + ,{\ell _1}}^\dag \left[ {{X_{{\ell _1}}},{P_{ - ,{\ell _1}}}} \right]{V_{ + ,{\ell _1}}}} \right)\\
{\rm{Tr}}\left( {{\rho _0}V_{ + ,{\ell _2}}^\dag {P_{ - ,{\ell _2}}}{V_{ + ,{\ell _2}}}{\rho _0}V_{ + ,{\ell _2}}^\dag {P_{ - ,{\ell _2}}}{V_{ + ,{\ell _2}}}} \right){\rm{Tr}}\left( {{\rho _0}V_{ + ,{\ell _2}}^\dag \left[ {{X_{{\ell _2}}},{P_{ - ,{\ell _2}}}} \right]{V_{ + ,{\ell _2}}}} \right)
\end{array} \right)} } \nonumber\\
&= \sum\limits_\ell  {\int {{\cal D}V\left( \begin{array}{l}
{\rm{Tr}}\left( {{\rho _0}V_{ + ,\ell }^\dag {P_{ - ,\ell }}{V_{ + ,\ell }}{\rho _0}V_{ + ,\ell }^\dag {P_{ - ,\ell }}{V_{ + ,\ell }}} \right){\rm{Tr}}\left( {{\rho _0}V_{ + ,\ell }^\dag \left[ {{X_\ell },{P_{ - ,\ell }}} \right]{V_{ + ,\ell }}} \right)\\
{\rm{Tr}}\left( {{\rho _0}V_{ + ,\ell }^\dag {P_{ - ,\ell }}{V_{ + ,\ell }}{\rho _0}V_{ + ,\ell }^\dag {P_{ - ,\ell }}{V_{ + ,\ell }}} \right){\rm{Tr}}\left( {{\rho _0}V_{ + ,\ell }^\dag \left[ {{X_\ell },{P_{ - ,\ell }}} \right]{V_{ + ,\ell }}} \right)
\end{array} \right)} }\nonumber\\
&= \sum\limits_\ell  {\int {{\cal D}V\left( {{\rm{Tr}}\left( \begin{array}{l}
\overleftarrow P {\rho _0}V_{ + ,\ell }^\dag {P_{ - ,\ell }}{V_{ + ,\ell }}{\rho _0}V_{ + ,\ell }^\dag {P_{ - ,\ell }}{V_{ + ,\ell }}\overline P {\rho _0}V_{ + ,\ell }^\dag \left[ {{X_\ell },{P_{ - ,\ell }}} \right]{V_{ + ,\ell }}\\
\widehat P{\rho _0}V_{ + ,\ell }^\dag {P_{ - ,\ell }}{V_{ + ,\ell }}{\rho _0}V_{ + ,\ell }^\dag {P_{ - ,\ell }}{V_{ + ,\ell }}\overrightarrow P {\rho _0}V_{ + ,\ell }^\dag \left[ {{X_\ell },{P_{ - ,\ell }}} \right]{V_{ + ,\ell }}
\end{array} \right)} \right)} } ~.
\end{align}
Here we use the notation:
\begin{align} \label{uselater}
&\overleftarrow P  \equiv {P_{{i_4}{i_1}}}~,~~~~~~~~\overline P  \equiv {P_{{i_1}{i_2}}}~,\nonumber\\
&\widehat P \equiv {P_{{i_2}{i_3}}}~,~~~~~~~~\overrightarrow P  \equiv {P_{{i_3}{i_4}}}~,
\end{align}
and we implicitly hide the sum $\sum_{i_1,i_2,i_3,i_4}$. Recall the following formula
\begin{align}
{\mathbb{E}}\left( {{\mathop{\rm Tr}\nolimits} \left( {U{B_1}{U^\dag }{b_1}U{B_2}{U^\dag }{b_2} \cdots U{B_{k - 1}}{U^\dag }{b_{k - 1}}U{B_k}{U^\dag }{b_k}} \right)} \right)\nonumber\\
 = {\mathbb{E}}\left( {{\mathop{\rm Tr}\nolimits} \left( {{B_1}{U^\dag }{b_1}U{B_2}{U^\dag }{b_2} \cdots U{B_{k - 1}}{U^\dag }{b_{k - 1}}U{B_k}{U^\dag }{b_k}U} \right)} \right)\nonumber\\
 = \sum\limits_{\alpha ,\beta  \in {S_k}} {{\rm{Wg}}} \left( {\beta {\alpha ^{ - 1}}} \right){{\mathop{\rm Tr}\nolimits} _{{\beta ^{ - 1}}}}\left( {{B_1}, \ldots ,{B_k}} \right){{\mathop{\rm Tr}\nolimits} _{\alpha {\gamma _k}}}\left( {{b_1}, \ldots ,{b_k}} \right)~,
\end{align}
where $\operatorname{Wg}$ is the Weingarten function \cite{gu2013moments}, $\alpha$ and $\beta$ are permutations from the permutation group $S_k$, and $\gamma_k$ is the cyclic permutation, and
\begin{align}
{\gamma _k} = (1,2, \ldots ,k) \in {S_k}~,
\end{align}
is the cyclic permutation \cite{Roberts:2016hpo,Cotler:2017jue,Liu:2018hlr,Liu:2020sqb}. Moreover, we use the notation,
\begin{align}
\operatorname{Tr}_\beta \left( {{B_1}, \ldots ,{B_k}} \right) \equiv {\rm{Tr}}\left( {\beta ({B_1}) \ldots \beta ({B_k})} \right)~.
\end{align}

Using Eq.~\ref{uselater} we get,
\begin{align}\label{propertyuse}
\operatorname{Tr}_\beta \left( {\overleftarrow P {\rho _0},{\rho _0},\overline P {\rho _0},\widehat P{\rho _0},{\rho _0},\overrightarrow P {\rho _0}} \right) = 1~,
\end{align}
for all $\beta \in S_6$. In our case, we identify:
\begin{align}
&({B_1}, \ldots ,{B_6}) \leftarrow \left( {{P_{ - ,\ell }},{P_{ - ,\ell }},\left[ {{X_\ell },{P_{ - ,\ell }}} \right]{P_{ - ,\ell }},{P_{ - ,\ell }},\left[ {{X_\ell },{P_{ - ,\ell }}} \right]} \right)~,\nonumber\\
&({b_1}, \ldots ,{b_6}) \leftarrow \left( {\overleftarrow P {\rho _0},{\rho _0},\overline P {\rho _0},\widehat P{\rho _0},{\rho _0},\overrightarrow P {\rho _0}} \right)~,
\end{align}
which implies that
\begin{align}
&\sum\limits_{\alpha ,\beta  \in {S_6}} {{\mathop{\rm Wg}\nolimits} } \left( {\beta {\alpha ^{ - 1}}} \right){{\mathop{\rm Tr}\nolimits} _{{\beta ^{ - 1}}}}\left( {{B_1}, \ldots ,{B_6}} \right){{\mathop{\rm Tr}\nolimits} _{\alpha {\gamma _k}}}\left( {{b_1}, \ldots ,{b_6}} \right)\nonumber\\
&= \sum\limits_{\alpha ,\beta  \in {S_6}} {{\mathop{\rm Wg}\nolimits} } \left( {\beta {\alpha ^{ - 1}}} \right){{\mathop{\rm Tr}\nolimits} _{{\beta ^{ - 1}}}}\left( {{B_1}, \ldots ,{B_6}} \right)\nonumber\\
&= \sum\limits_{\alpha ,\beta  \in {S_6}} {{\mathop{\rm Wg}\nolimits} } \left( {{\beta ^{ - 1}}{\alpha ^{ - 1}}} \right){{\mathop{\rm Tr}\nolimits} _\beta }\left( {{B_1}, \ldots ,{B_6}} \right)~.
\end{align}
The first equality follows from Eq. \ref{propertyuse}, and the second one corresponds to relabelling $\beta$. 

Then we get
\begin{align}
\Delta {\mu_a^2} &= \frac{{144}}{{D(D + 1)(D + 2)(D + 3)(D + 4)\left( {D + 5} \right)}}\nonumber\\
&\qquad\sum\limits_\ell  {\int {d{V_{ - ,\ell }}} \left( \begin{array}{l}
{\rm{Tr}}\left( {P_{ - ,\ell }^2\left[ {{X_\ell },{P_{ - ,\ell }}} \right]P_{ - ,\ell }^2\left[ {{X_\ell },{P_{ - ,\ell }}} \right]} \right)\\
2{\rm{Tr}}\left( {P_{ - ,\ell }^3\left[ {{X_\ell },{P_{ - ,\ell }}} \right]{P_{ - ,\ell }}\left[ {{X_\ell },{P_{ - ,\ell }}} \right]} \right)\\
 + 2{\rm{Tr}}\left( {P_{ - ,\ell }^4{{\left[ {{X_\ell },{P_{ - ,\ell }}} \right]}^2}} \right)
\end{array} \right)} ~.
\end{align}
More precisely, it is
\begin{align}
\Delta {\mu_a^2} &= \frac{{144}}{{D(D + 1)(D + 2)(D + 3)(D + 4)\left( {D + 5} \right)}}\\
&\qquad \sum\limits_\ell  {\int {d{V_{ - ,\ell }}} \left( \begin{array}{l}
{\rm{Tr}}\left( {{{\left[ {{X_\ell },V_{ - ,\ell }^\dag O{V_{ - ,\ell }}} \right]}^2}\left[ {{X_\ell },\left[ {{X_\ell },V_{ - ,\ell }^\dag O{V_{ - ,\ell }}} \right]} \right]{{\left[ {{X_\ell },V_{ - ,\ell }^\dag O{V_{ - ,\ell }}} \right]}^2}\left[ {{X_\ell },\left[ {{X_\ell },V_{ - ,\ell }^\dag O{V_{ - ,\ell }}} \right]} \right]} \right)\\
2{\rm{Tr}}\left( {{{\left[ {{X_\ell },V_{ - ,\ell }^\dag O{V_{ - ,\ell }}} \right]}^3}\left[ {{X_\ell },\left[ {{X_\ell },V_{ - ,\ell }^\dag O{V_{ - ,\ell }}} \right]} \right]\left[ {{X_\ell },V_{ - ,\ell }^\dag O{V_{ - ,\ell }}} \right]\left[ {{X_\ell },\left[ {{X_\ell },V_{ - ,\ell }^\dag O{V_{ - ,\ell }}} \right]} \right]} \right)\\
 + 2{\rm{Tr}}\left( {{{\left[ {{X_\ell },V_{ - ,\ell }^\dag O{V_{ - ,\ell }}} \right]}^4}{{\left[ {{X_\ell },\left[ {{X_\ell },V_{ - ,\ell }^\dag O{V_{ - ,\ell }}} \right]} \right]}^2}} \right)
\end{array} \right)} ~.
\end{align}

In total, each trace has $2^8=256$ independent terms, and therefore, $2^8\times 3= 768$ terms. After algebraic manipulations, we find
\begin{itemize}
\item One 6-design term with a factor of $-8$.
\item One 4-design term with a factor of $-48$.
\item Another type of 4-design term with a factor of $+48$.
\item Another type of 4-design term with a factor of $+24$.
\item A 2-design term  with a factor of $+48$.
\item Another type of 2-design term with a factor of $-24$.
\item Another type of 2-design term with a factor of $-48$.
\item A constant term with the factor $+8$.
\end{itemize}

Using \texttt{RNTK} we get 
\begin{align}
\Delta {\mu ^2} \supset& L\frac{{1152{\rm{Tr}}\left( {{O^6}} \right)}}{{{D^6}}}\nonumber\\
&\qquad - L\frac{{3456\left( {\operatorname{Tr}^2\left( {{O^3}} \right) - 2{\rm{Tr}}\left( {{O^2}} \right){\rm{Tr}}\left( {{O^4}} \right) + 2{\rm{Tr}}\left( O \right){\rm{Tr}}\left( {{O^5}} \right) + 5{\rm{Tr}}\left( {{O^6}} \right)} \right)}}{{{D^7}}} +  \ldots ~.
\end{align}

\item We now analyze the following term:
\begin{align}
\Delta {\mu_b ^2} = 4\int {{\cal D}V} \sum\limits_{{\ell _1} < {\ell _2};{\ell _3}} {\begin{array}{*{20}{l}}
{{\rm{Tr}}\left( {{\rho _0}V_{ + ,{\ell _1}}^\dag {P_{ - ,{\ell _1}}}{V_{ + ,{\ell _1}}}{\rho _0}V_{ + ,{\ell _2}}^\dag {P_{ - ,{\ell _2}}}{V_{ + ,{\ell _2}}}} \right){\rm{Tr}}\left( {{\rho _0}V_{ + ,{\ell _2}}^\dag \left[ {{X_{{\ell _2}}},V_{{\ell _1},{\ell _2}}^\dag {P_{ - ,{\ell _1}}}{V_{{\ell _1},{\ell _2}}}} \right]{V_{ + ,{\ell _2}}}} \right)}\\
{{\rm{Tr}}\left( {{\rho _0}V_{ + ,{\ell _3}}^\dag {P_{ - ,{\ell _3}}}{V_{ + ,{\ell _3}}}{\rho _0}V_{ + ,{\ell _3}}^\dag {P_{ - ,{\ell _3}}}{V_{ + ,{\ell _3}}}} \right){\rm{Tr}}\left( {{\rho _0}V_{ + ,{\ell _3}}^\dag \left[ {{X_{{\ell _3}}},{P_{ - ,{\ell _3}}}} \right]{V_{ + ,{\ell _3}}}} \right)}
\end{array}} ~.
\end{align}

Using the arguments similar to those used in Eq. \ref{nonvanishing1}, we find the following non-vanishing terms:
\begin{align}
&4\int {{\cal D}V} \sum\limits_{{\ell _1} < {\ell _2};{\ell _3}} {\begin{array}{*{20}{l}}
{{\rm{Tr}}\left( {{\rho _0}V_{ + ,{\ell _1}}^\dag {P_{ - ,{\ell _1}}}{V_{ + ,{\ell _1}}}{\rho _0}V_{ + ,{\ell _2}}^\dag {P_{ - ,{\ell _2}}}{V_{ + ,{\ell _2}}}} \right){\rm{Tr}}\left( {{\rho _0}V_{ + ,{\ell _2}}^\dag \left[ {{X_{{\ell _2}}},V_{{\ell _1},{\ell _2}}^\dag {P_{ - ,{\ell _1}}}{V_{{\ell _1},{\ell _2}}}} \right]{V_{ + ,{\ell _2}}}} \right)}\\
{{\rm{Tr}}\left( {{\rho _0}V_{ + ,{\ell _3}}^\dag {P_{ - ,{\ell _3}}}{V_{ + ,{\ell _3}}}{\rho _0}V_{ + ,{\ell _3}}^\dag {P_{ - ,{\ell _3}}}{V_{ + ,{\ell _3}}}} \right){\rm{Tr}}\left( {{\rho _0}V_{ + ,{\ell _3}}^\dag \left[ {{X_{{\ell _3}}},{P_{ - ,{\ell _3}}}} \right]{V_{ + ,{\ell _3}}}} \right)}
\end{array}} \nonumber\\
&= 4\int {{\cal D}V} \sum\limits_{{\ell _1} < {\ell _2}} {\begin{array}{*{20}{l}}
{{\rm{Tr}}\left( {{\rho _0}V_{ + ,{\ell _1}}^\dag {P_{ - ,{\ell _1}}}{V_{ + ,{\ell _1}}}{\rho _0}V_{ + ,{\ell _2}}^\dag {P_{ - ,{\ell _2}}}{V_{ + ,{\ell _2}}}} \right){\rm{Tr}}\left( {{\rho _0}V_{ + ,{\ell _2}}^\dag \left[ {{X_{{\ell _2}}},V_{{\ell _1},{\ell _2}}^\dag {P_{ - ,{\ell _1}}}{V_{{\ell _1},{\ell _2}}}} \right]{V_{ + ,{\ell _2}}}} \right)}\\
{{\rm{Tr}}\left( {{\rho _0}V_{ + ,{\ell _1}}^\dag {P_{ - ,{\ell _1}}}{V_{ + ,{\ell _1}}}{\rho _0}V_{ + ,{\ell _1}}^\dag {P_{ - ,{\ell _1}}}{V_{ + ,{\ell _1}}}} \right){\rm{Tr}}\left( {{\rho _0}V_{ + ,{\ell _1}}^\dag \left[ {{X_{{\ell _1}}},{P_{ - ,{\ell _1}}}} \right]{V_{ + ,{\ell _1}}}} \right)}
\end{array}} \nonumber\\
&\qquad + 4\int {{\cal D}V} \sum\limits_{{\ell _1} < {\ell _2}} {\begin{array}{*{20}{l}}
{{\rm{Tr}}\left( {{\rho _0}V_{ + ,{\ell _1}}^\dag {P_{ - ,{\ell _1}}}{V_{ + ,{\ell _1}}}{\rho _0}V_{ + ,{\ell _2}}^\dag {P_{ - ,{\ell _2}}}{V_{ + ,{\ell _2}}}} \right){\rm{Tr}}\left( {{\rho _0}V_{ + ,{\ell _2}}^\dag \left[ {{X_{{\ell _2}}},V_{{\ell _1},{\ell _2}}^\dag {P_{ - ,{\ell _1}}}{V_{{\ell _1},{\ell _2}}}} \right]{V_{ + ,{\ell _2}}}} \right)}\\
{{\rm{Tr}}\left( {{\rho _0}V_{ + ,{\ell _2}}^\dag {P_{ - ,{\ell _2}}}{V_{ + ,{\ell _2}}}{\rho _0}V_{ + ,{\ell _2}}^\dag {P_{ - ,{\ell _2}}}{V_{ + ,{\ell _2}}}} \right){\rm{Tr}}\left( {{\rho _0}V_{ + ,{\ell _2}}^\dag \left[ {{X_{{\ell _2}}},{P_{ - ,{\ell _2}}}} \right]{V_{ + ,{\ell _2}}}} \right)}
\end{array}} ~.
\end{align}

Namely, we either have $\ell_3=\ell_1$, or $\ell_3=\ell_2$. However, in the first case, we have
\begin{align}
&4\int {{\cal D}V} \sum\limits_{{\ell _1} < {\ell _2}} {\begin{array}{*{20}{l}}
{{\rm{Tr}}\left( {{\rho _0}V_{ + ,{\ell _1}}^\dag {P_{ - ,{\ell _1}}}{V_{ + ,{\ell _1}}}{\rho _0}V_{ + ,{\ell _2}}^\dag {P_{ - ,{\ell _2}}}{V_{ + ,{\ell _2}}}} \right){\rm{Tr}}\left( {{\rho _0}V_{ + ,{\ell _2}}^\dag \left[ {{X_{{\ell _2}}},V_{{\ell _1},{\ell _2}}^\dag {P_{ - ,{\ell _1}}}{V_{{\ell _1},{\ell _2}}}} \right]{V_{ + ,{\ell _2}}}} \right)}\\
{{\rm{Tr}}\left( {{\rho _0}V_{ + ,{\ell _1}}^\dag {P_{ - ,{\ell _1}}}{V_{ + ,{\ell _1}}}{\rho _0}V_{ + ,{\ell _1}}^\dag {P_{ - ,{\ell _1}}}{V_{ + ,{\ell _1}}}} \right){\rm{Tr}}\left( {{\rho _0}V_{ + ,{\ell _1}}^\dag \left[ {{X_{{\ell _1}}},{P_{ - ,{\ell _1}}}} \right]{V_{ + ,{\ell _1}}}} \right)}
\end{array}} \nonumber\\
&= 4\sum\limits_{{\ell _1} < {\ell _2}} {\int {{\cal D}V} \left( {{\rm{Tr}}\left( {{\rho _0}V_{ + ,{\ell _2}}^\dag {X_{{\ell _2}}}V_{{\ell _1},{\ell _2}}^\dag {P_{ - ,{\ell _1}}}{V_{{\ell _1},{\ell _2}}}{V_{ + ,{\ell _2}}}} \right) - {\rm{Tr}}\left( {{\rho _0}V_{ + ,{\ell _2}}^\dag V_{{\ell _1},{\ell _2}}^\dag {P_{ - ,{\ell _1}}}{V_{{\ell _1},{\ell _2}}}{X_{{\ell _2}}}{V_{ + ,{\ell _2}}}} \right)} \right)} \nonumber\\
&\qquad {\rm{Tr}}\left( {{\rho _0}V_{ + ,{\ell _1}}^\dag {P_{ - ,{\ell _1}}}{V_{ + ,{\ell _1}}}{\rho _0}V_{ + ,{\ell _1}}^\dag {P_{ - ,{\ell _1}}}{V_{ + ,{\ell _1}}}} \right){\rm{Tr}}\left( {{\rho _0}V_{ + ,{\ell _1}}^\dag \left[ {{X_{{\ell _1}}},{P_{ - ,{\ell _1}}}} \right]{V_{ + ,{\ell _1}}}} \right)\nonumber\\
&\qquad {\rm{Tr}}\left( {{\rho _0}V_{ + ,{\ell _1}}^\dag {P_{ - ,{\ell _1}}}{V_{ + ,{\ell _1}}}{\rho _0}V_{ + ,{\ell _2}}^\dag {P_{ - ,{\ell _2}}}{V_{ + ,{\ell _2}}}} \right)\nonumber\\
&= 4\frac{1}{D}\sum\limits_{{\ell _1} < {\ell _2}} {\int {{\cal D}V} \left( {{\rm{Tr}}\left( {{V_{ + ,{\ell _2}}}{\rho _0}V_{ + ,{\ell _2}}^\dag {X_{{\ell _2}}}} \right){\rm{Tr}}\left( {{P_{ - ,{\ell _1}}}} \right) - {\rm{Tr}}\left( {{X_{{\ell _2}}}{V_{ + ,{\ell _2}}}{\rho _0}V_{ + ,{\ell _2}}^\dag } \right){\rm{Tr}}\left( {{P_{ - ,{\ell _1}}}} \right)} \right)} \nonumber\\
&\qquad {\rm{Tr}}\left( {{\rho _0}V_{ + ,{\ell _1}}^\dag {P_{ - ,{\ell _1}}}{V_{ + ,{\ell _1}}}{\rho _0}V_{ + ,{\ell _1}}^\dag {P_{ - ,{\ell _1}}}{V_{ + ,{\ell _1}}}} \right){\rm{Tr}}\left( {{\rho _0}V_{ + ,{\ell _1}}^\dag \left[ {{X_{{\ell _1}}},{P_{ - ,{\ell _1}}}} \right]{V_{ + ,{\ell _1}}}} \right)\nonumber\\
&\qquad {\rm{Tr}}\left( {{\rho _0}V_{ + ,{\ell _1}}^\dag {P_{ - ,{\ell _1}}}{V_{ + ,{\ell _1}}}{\rho _0}V_{ + ,{\ell _2}}^\dag {P_{ - ,{\ell _2}}}{V_{ + ,{\ell _2}}}} \right)\nonumber\\
&= 0~,
\end{align}
where we integrated over $dV_{\ell_1,\ell_2}$. 

Similarly, we also have
\begin{align}
4\int {{\cal D}V} \sum\limits_{{\ell _1} < {\ell _2}} {\begin{array}{*{20}{l}}
{{\rm{Tr}}\left( {{\rho _0}V_{ + ,{\ell _1}}^\dag {P_{ - ,{\ell _1}}}{V_{ + ,{\ell _1}}}{\rho _0}V_{ + ,{\ell _2}}^\dag {P_{ - ,{\ell _2}}}{V_{ + ,{\ell _2}}}} \right){\rm{Tr}}\left( {{\rho _0}V_{ + ,{\ell _2}}^\dag \left[ {{X_{{\ell _2}}},V_{{\ell _1},{\ell _2}}^\dag {P_{ - ,{\ell _1}}}{V_{{\ell _1},{\ell _2}}}} \right]{V_{ + ,{\ell _2}}}} \right)}\\
{{\rm{Tr}}\left( {{\rho _0}V_{ + ,{\ell _2}}^\dag {P_{ - ,{\ell _2}}}{V_{ + ,{\ell _2}}}{\rho _0}V_{ + ,{\ell _2}}^\dag {P_{ - ,{\ell _2}}}{V_{ + ,{\ell _2}}}} \right){\rm{Tr}}\left( {{\rho _0}V_{ + ,{\ell _2}}^\dag \left[ {{X_{{\ell _2}}},{P_{ - ,{\ell _2}}}} \right]{V_{ + ,{\ell _2}}}} \right)}
\end{array}} =0~,
\end{align}
which follows from
\begin{align}
\int {d{V_{ + ,{\ell _1}}}} {\mathop{\rm Tr}\nolimits} \left( {{\rho _0}V_{ + ,{\ell _1}}^\dag {P_{ - ,{\ell _1}}}{V_{ + ,{\ell _1}}}{\rho _0}V_{ + ,{\ell _2}}^\dag {P_{ - ,{\ell _2}}}{V_{ + ,{\ell _2}}}} \right)
&= \frac{1}{D}{\mathop{\rm Tr}\nolimits} \left( {{P_{ - ,{\ell _1}}}} \right){\mathop{\rm Tr}\nolimits} \left( {{\rho _0}V_{ + ,{\ell _2}}^\dag {P_{ - ,{\ell _2}}}{V_{ + ,{\ell _2}}}{\rho _0}} \right)\nonumber\\
&= 0~.
\end{align}
Thus
\begin{align}
\Delta {\mu ^2} \supset& 4\int {{\cal D}V} \sum\limits_{{\ell _1} < {\ell _2};{\ell _3}} {\begin{array}{*{20}{l}}
{{\rm{Tr}}\left( {{\rho _0}V_{ + ,{\ell _1}}^\dag {P_{ - ,{\ell _1}}}{V_{ + ,{\ell _1}}}{\rho _0}V_{ + ,{\ell _2}}^\dag {P_{ - ,{\ell _2}}}{V_{ + ,{\ell _2}}}} \right){\rm{Tr}}\left( {{\rho _0}V_{ + ,{\ell _2}}^\dag \left[ {{X_{{\ell _2}}},V_{{\ell _1},{\ell _2}}^\dag {P_{ - ,{\ell _1}}}{V_{{\ell _1},{\ell _2}}}} \right]{V_{ + ,{\ell _2}}}} \right)}\\
{{\rm{Tr}}\left( {{\rho _0}V_{ + ,{\ell _3}}^\dag {P_{ - ,{\ell _3}}}{V_{ + ,{\ell _3}}}{\rho _0}V_{ + ,{\ell _3}}^\dag {P_{ - ,{\ell _3}}}{V_{ + ,{\ell _3}}}} \right){\rm{Tr}}\left( {{\rho _0}V_{ + ,{\ell _3}}^\dag \left[ {{X_{{\ell _3}}},{P_{ - ,{\ell _3}}}} \right]{V_{ + ,{\ell _3}}}} \right)}
\end{array}}\nonumber\\
&=0 ~.
\end{align}
\item Consider the term
\begin{align}
\Delta {\mu _c^2}=  4\int {\mathcal{D}V} \sum\limits_{{\ell _1} < {\ell _2};{\ell _3} < {\ell _4}} \begin{array}{l}
{\mathop{\rm Tr}\nolimits} \left( {{\rho _0}V_{ + ,{\ell _1}}^\dag {P_{ - ,{\ell _1}}}{V_{ + ,{\ell _1}}}{\rho _0}V_{ + ,{\ell _2}}^\dag {P_{ - ,{\ell _2}}}{V_{ + ,{\ell _2}}}} \right){\mathop{\rm Tr}\nolimits} \left( {{\rho _0}V_{ + ,{\ell _2}}^\dag \left[ {{X_{{\ell _2}}},V_{{\ell _1},{\ell _2}}^\dag {P_{ - ,{\ell _1}}}{V_{{\ell _1},{\ell _2}}}} \right]{V_{ + ,{\ell _2}}}} \right)\\
{\mathop{\rm Tr}\nolimits} \left( {{\rho _0}V_{ + ,{\ell _3}}^\dag {P_{ - ,{\ell _3}}}{V_{ + ,{\ell _3}}}{\rho _0}V_{ + ,{\ell _4}}^\dag {P_{ - ,{\ell _4}}}{V_{ + ,{\ell _4}}}} \right){\mathop{\rm Tr}\nolimits} \left( {{\rho _0}V_{ + ,{\ell _4}}^\dag \left[ {{X_{{\ell _4}}},V_{{\ell _3},{\ell _4}}^\dag {P_{ - ,{\ell _3}}}{V_{{\ell _3},{\ell _4}}}} \right]{V_{ + ,{\ell _4}}}} \right)
\end{array}~.
\end{align}
We can split it as the following cases:
\begin{itemize}
\item $\ell_1<\ell_2, \ell_3 < \ell_4$, and none of $\ell_{1,2,3,4}$s are equal. From the calculation of $\bar{\mu}$ we know that it is zero.
\item $\ell_1<\ell_2, \ell_1=\ell_3, \ell_2 \ne \ell_4$. It vanishes because
\begin{align}
\int {d{V_{ - ,{\ell _2}}}} {\rm{Tr}}\left( {{\rho _0}V_{ + ,{\ell _1}}^\dag {P_{ - ,{\ell _1}}}{V_{ + ,{\ell _1}}}{\rho _0}V_{ + ,{\ell _2}}^\dag {P_{ - ,{\ell _2}}}{V_{ + ,{\ell _2}}}} \right) = 0~.
\end{align}
\item $\ell_3<\ell_4, \ell_2=\ell_4, \ell_1 \ne \ell_3$. It similarly also vanishes because
\begin{align}
\int {d{V_{ + ,{\ell _1}}}} {\rm{Tr}}\left( {{\rho _0}V_{ + ,{\ell _1}}^\dag {P_{ - ,{\ell _1}}}{V_{ + ,{\ell _1}}}{\rho _0}V_{ + ,{\ell _2}}^\dag {P_{ - ,{\ell _2}}}{V_{ + ,{\ell _2}}}} \right) = 0~.
\end{align}
\item $\ell_1<\ell_2= \ell_3 < \ell_4$. It similarly also vanishes because
\begin{align}
\int {d{V_{ + ,{\ell _1}}}} {\rm{Tr}}\left( {{\rho _0}V_{ + ,{\ell _1}}^\dag {P_{ - ,{\ell _1}}}{V_{ + ,{\ell _1}}}{\rho _0}V_{ + ,{\ell _2}}^\dag {P_{ - ,{\ell _2}}}{V_{ + ,{\ell _2}}}} \right) = 0~.
\end{align}
\item $\ell_3<\ell_4= \ell_1 < \ell_2$. It similarly also vanishes because
\begin{align}
\int {d{V_{ + ,{\ell _3}}}} {\rm{Tr}}\left( {{\rho _0}V_{ + ,{\ell _3}}^\dag {P_{ - ,{\ell _3}}}{V_{ + ,{\ell _3}}}{\rho _0}V_{ + ,{\ell _4}}^\dag {P_{ - ,{\ell _4}}}{V_{ + ,{\ell _4}}}} \right) = 0~.
\end{align}
\item $\ell_1<\ell_2, \ell_1=\ell_3, \ell_2 = \ell_4$. This is the only probably non-zero part.
\end{itemize}
Thus, we get
\begin{align}\label{c2term}
\Delta {\mu_c^2}&=   4\int {\mathcal{D}V} \sum\limits_{{\ell _1} < {\ell _2};{\ell _3} < {\ell _4}} \begin{array}{l}
{\mathop{\rm Tr}\nolimits} \left( {{\rho _0}V_{ + ,{\ell _1}}^\dag {P_{ - ,{\ell _1}}}{V_{ + ,{\ell _1}}}{\rho _0}V_{ + ,{\ell _2}}^\dag {P_{ - ,{\ell _2}}}{V_{ + ,{\ell _2}}}} \right){\mathop{\rm Tr}\nolimits} \left( {{\rho _0}V_{ + ,{\ell _2}}^\dag \left[ {{X_{{\ell _2}}},V_{{\ell _1},{\ell _2}}^\dag {P_{ - ,{\ell _1}}}{V_{{\ell _1},{\ell _2}}}} \right]{V_{ + ,{\ell _2}}}} \right)\\
{\mathop{\rm Tr}\nolimits} \left( {{\rho _0}V_{ + ,{\ell _3}}^\dag {P_{ - ,{\ell _3}}}{V_{ + ,{\ell _3}}}{\rho _0}V_{ + ,{\ell _4}}^\dag {P_{ - ,{\ell _4}}}{V_{ + ,{\ell _4}}}} \right){\mathop{\rm Tr}\nolimits} \left( {{\rho _0}V_{ + ,{\ell _4}}^\dag \left[ {{X_{{\ell _4}}},V_{{\ell _3},{\ell _4}}^\dag {P_{ - ,{\ell _3}}}{V_{{\ell _3},{\ell _4}}}} \right]{V_{ + ,{\ell _4}}}} \right)
\end{array}\nonumber\\
& = 4\int {{\cal D}V} \sum\limits_{{\ell _1} < {\ell _2}} {\operatorname{Tr}^2\left( {{\rho _0}V_{ + ,{\ell _1}}^\dag {P_{ - ,{\ell _1}}}{V_{ + ,{\ell _1}}}{\rho _0}V_{ + ,{\ell _2}}^\dag {P_{ - ,{\ell _2}}}{V_{ + ,{\ell _2}}}} \right)\operatorname{Tr}^2\left( {{\rho _0}V_{ + ,{\ell _2}}^\dag \left[ {{X_{{\ell _2}}},V_{{\ell _1},{\ell _2}}^\dag {P_{ - ,{\ell _1}}}{V_{{\ell _1},{\ell _2}}}} \right]{V_{ + ,{\ell _2}}}} \right)} ~.
\end{align}
Firstly, we integrate over $dV_{+,\ell_1}$. We use \texttt{RTNI} again to get,
\begin{align}
&\int {d{V_{ + ,{\ell _1}}}} \operatorname{Tr}^2\left( {{\rho _0}V_{ + ,{\ell _1}}^\dag {P_{ - ,{\ell _1}}}{V_{ + ,{\ell _1}}}{\rho _0}V_{ + ,{\ell _2}}^\dag {P_{ - ,{\ell _2}}}{V_{ + ,{\ell _2}}}} \right)\nonumber\\
&= \int {d{V_{ + ,{\ell _1}}}} {\rm{Tr}}\left( {\overleftarrow P {\rho _0}V_{ + ,{\ell _1}}^\dag {P_{ - ,{\ell _1}}}{V_{ + ,{\ell _1}}}{\rho _0}V_{ + ,{\ell _2}}^\dag {P_{ - ,{\ell _2}}}{V_{ + ,{\ell _2}}}\overrightarrow P {\rho _0}V_{ + ,{\ell _1}}^\dag {P_{ - ,{\ell _1}}}{V_{ + ,{\ell _1}}}{\rho _0}V_{ + ,{\ell _2}}^\dag {P_{ - ,{\ell _2}}}{V_{ + ,{\ell _2}}}} \right)\nonumber\\
&= \int {d{V_{ + ,{\ell _1}}}} {\rm{Tr}}\left( {{\rho _0}V_{ + ,{\ell _2}}^\dag {P_{ - ,{\ell _2}}}{V_{ + ,{\ell _2}}}\overleftarrow P {\rho _0}V_{ + ,{\ell _1}}^\dag {P_{ - ,{\ell _1}}}{V_{ + ,{\ell _1}}}{\rho _0}V_{ + ,{\ell _2}}^\dag {P_{ - ,{\ell _2}}}{V_{ + ,{\ell _2}}}\overrightarrow P {\rho _0}V_{ + ,{\ell _1}}^\dag {P_{ - ,{\ell _1}}}{V_{ + ,{\ell _1}}}} \right)\nonumber\\
&= \frac{{{\rm{Tr}}\left( {P_{ - ,{\ell _1}}^2} \right)}}{{D({D^2} - 1)}}\left( \begin{array}{l}
N{\rm{Tr}}\left( {{\rho _0}V_{ + ,{\ell _2}}^\dag {P_{ - ,{\ell _2}}}{V_{ + ,{\ell _2}}}\overrightarrow P } \right){\rm{Tr}}\left( {{\rho _0}V_{ + ,{\ell _2}}^\dag {P_{ - ,{\ell _2}}}{V_{ + ,{\ell _2}}}\overleftarrow P } \right)\\
 - {\rm{Tr}}\left( {\overrightarrow P {\rho _0}V_{ + ,{\ell _2}}^\dag {P_{ - ,{\ell _2}}}{V_{ + ,{\ell _2}}}\overleftarrow P {\rho _0}V_{ + ,{\ell _2}}^\dag {P_{ - ,{\ell _2}}}{V_{ + ,{\ell _2}}}} \right)
\end{array} \right)\nonumber\\
&= \frac{{{\rm{Tr}}\left( {P_{ - ,{\ell _1}}^2} \right)}}{{D({D^2} - 1)}}\left( \begin{array}{l}
N{\rm{Tr}}\left( {{\rho _0}V_{ + ,{\ell _2}}^\dag {P_{ - ,{\ell _2}}}{V_{ + ,{\ell _2}}}{\rho _0}V_{ + ,{\ell _2}}^\dag {P_{ - ,{\ell _2}}}{V_{ + ,{\ell _2}}}} \right)\\
 - {\rm{Tr}}\left( {{\rho _0}V_{ + ,{\ell _2}}^\dag {P_{ - ,{\ell _2}}}{V_{ + ,{\ell _2}}}} \right){\rm{Tr}}\left( {{\rho _0}V_{ + ,{\ell _2}}^\dag {P_{ - ,{\ell _2}}}{V_{ + ,{\ell _2}}}} \right)
\end{array} \right)~.
\end{align}
Our next task is to integrate over $dV_{-,\ell_2}$. We get
\begin{align}
&\int d {V_{ - ,{\ell _2}}}\frac{{{\rm{Tr}}\left( {P_{ - ,{\ell _1}}^2} \right)}}{{D({D^2} - 1)}}\left( {\begin{array}{*{20}{l}}
{N{\rm{Tr}}\left( {{\rho _0}V_{ + ,{\ell _2}}^\dag {P_{ - ,{\ell _2}}}{V_{ + ,{\ell _2}}}{\rho _0}V_{ + ,{\ell _2}}^\dag {P_{ - ,{\ell _2}}}{V_{ + ,{\ell _2}}}} \right)}\\
{ - {\rm{Tr}}\left( {{\rho _0}V_{ + ,{\ell _2}}^\dag {P_{ - ,{\ell _2}}}{V_{ + ,{\ell _2}}}} \right){\rm{Tr}}\left( {{\rho _0}V_{ + ,{\ell _2}}^\dag {P_{ - ,{\ell _2}}}{V_{ + ,{\ell _2}}}} \right)}
\end{array}} \right)\nonumber\\
&=  - \frac{{2\left( {N{\rm{Tr}}\left( {{O^2}} \right) - \operatorname{Tr}^2\left( O \right)} \right)}}{{(D - 1){D^2}{{(D + 1)}^2}}}{\rm{Tr}}\left( {P_{ - ,{\ell _1}}^2} \right)\nonumber\\
&\qquad + \frac{{\left( {(2D + 1){\rm{Tr}}\left( {{O^2}} \right) - \operatorname{Tr}^2\left( O \right)} \right)}}{{(D - 1){D^2}{{(D + 1)}^2}}}{\rm{Tr}}\left( {P_{ - ,{\ell _1}}^2} \right)\operatorname{Tr}^2\left( {{\rho _0}V_{ + ,{\ell _2}}^\dag {X_{{\ell _2}}}{V_{ + ,{\ell _2}}}} \right)\nonumber\\
&\qquad - \frac{{\left( {{\rm{Tr}}\left( {{O^2}} \right) + \operatorname{Tr}^2\left( O \right)} \right)}}{{(D - 1){D^2}{{(D + 1)}^2}}}{\rm{Tr}}\left( {P_{ - ,{\ell _1}}^2} \right){\rm{Tr}}\left( {{\rho _0}V_{ + ,{\ell _2}}^\dag {X_{{\ell _2}}}{V_{ + ,{\ell _2}}}{\rho _0}V_{ + ,{\ell _2}}^\dag {X_{{\ell _2}}}{V_{ + ,{\ell _2}}}} \right)~.
\end{align}
Moreover, similarly we could integrate over $dV_{\ell_1,\ell_2}$, and we have
\begin{align}\label{indiv}
&\int {d{V_{{\ell _1},{\ell _2}}}} {{\mathop{\rm Tr}\nolimits} ^2}\left( {{\rho _0}V_{ + ,{\ell _2}}^\dag \left[ {{X_{{\ell _2}}},V_{{\ell _1},{\ell _2}}^\dag {P_{ - ,{\ell _1}}}{V_{{\ell _1},{\ell _2}}}} \right]{V_{ + ,{\ell _2}}}} \right)\nonumber\\
&= \int {d{V_{{\ell _1},{\ell _2}}}} {\left( {{\mathop{\rm Tr}\nolimits} \left( {{\rho _0}V_{ + ,{\ell _2}}^\dag {X_{{\ell _2}}}V_{{\ell _1},{\ell _2}}^\dag {P_{ - ,{\ell _1}}}{V_{{\ell _1},{\ell _2}}}{V_{ + ,{\ell _2}}}} \right) - {\mathop{\rm Tr}\nolimits} \left( {{\rho _0}V_{ + ,{\ell _2}}^\dag V_{{\ell _1},{\ell _2}}^\dag {P_{ - ,{\ell _1}}}{V_{{\ell _1},{\ell _2}}}{X_{{\ell _2}}}{V_{ + ,{\ell _2}}}} \right)} \right)^2}\nonumber\\
&= \int {d{V_{{\ell _1},{\ell _2}}}} {\left( {{\mathop{\rm Tr}\nolimits} \left( {{V_{ + ,{\ell _2}}}{\rho _0}V_{ + ,{\ell _2}}^\dag {X_{{\ell _2}}}V_{{\ell _1},{\ell _2}}^\dag {P_{ - ,{\ell _1}}}{V_{{\ell _1},{\ell _2}}}} \right) - {\mathop{\rm Tr}\nolimits} \left( {{X_{{\ell _2}}}{V_{ + ,{\ell _2}}}{\rho _0}V_{ + ,{\ell _2}}^\dag V_{{\ell _1},{\ell _2}}^\dag {P_{ - ,{\ell _1}}}{V_{{\ell _1},{\ell _2}}}} \right)} \right)^2}\nonumber\\
&= \int {d{V_{{\ell _1},{\ell _2}}}} {\mathop{\rm Tr}\nolimits} \left( {\overleftarrow P {V_{ + ,{\ell _2}}}{\rho _0}V_{ + ,{\ell _2}}^\dag {X_{{\ell _2}}}V_{{\ell _1},{\ell _2}}^\dag {P_{ - ,{\ell _1}}}{V_{{\ell _1},{\ell _2}}}\overrightarrow P {V_{ + ,{\ell _2}}}{\rho _0}V_{ + ,{\ell _2}}^\dag {X_{{\ell _2}}}V_{{\ell _1},{\ell _2}}^\dag {P_{ - ,{\ell _1}}}{V_{{\ell _1},{\ell _2}}}} \right)\nonumber\\
&\qquad+ \int {d{V_{{\ell _1},{\ell _2}}}} {\mathop{\rm Tr}\nolimits} \left( {\overleftarrow P {X_{{\ell _2}}}{V_{ + ,{\ell _2}}}{\rho _0}V_{ + ,{\ell _2}}^\dag V_{{\ell _1},{\ell _2}}^\dag {P_{ - ,{\ell _1}}}{V_{{\ell _1},{\ell _2}}}\overrightarrow P {X_{{\ell _2}}}{V_{ + ,{\ell _2}}}{\rho _0}V_{ + ,{\ell _2}}^\dag V_{{\ell _1},{\ell _2}}^\dag {P_{ - ,{\ell _1}}}{V_{{\ell _1},{\ell _2}}}} \right)\nonumber\\
&\qquad - 2\int {d{V_{{\ell _1},{\ell _2}}}} {\mathop{\rm Tr}\nolimits} \left( {\overleftarrow P {V_{ + ,{\ell _2}}}{\rho _0}V_{ + ,{\ell _2}}^\dag {X_{{\ell _2}}}V_{{\ell _1},{\ell _2}}^\dag {P_{ - ,{\ell _1}}}{V_{{\ell _1},{\ell _2}}}\overrightarrow P {X_{{\ell _2}}}{V_{ + ,{\ell _2}}}{\rho _0}V_{ + ,{\ell _2}}^\dag V_{{\ell _1},{\ell _2}}^\dag {P_{ - ,{\ell _1}}}{V_{{\ell _1},{\ell _2}}}} \right)~.
\end{align}
Again, we use the formula
\begin{align}
\int dU {{\rm{Tr}}\left( {A{U^\dag }BUC{U^\dag }DU} \right)}  = \frac{{{\rm{Tr}}\left( {BD} \right)}}{{D\left( {{D^2} - 1} \right)}}\left( {N{\rm{Tr}}\left( A \right){\rm{Tr}}\left( C \right) - {\rm{Tr}}\left( {AC} \right)} \right)\text{,       when }\operatorname{Tr}(B)=0~,
\end{align}
to calculate individual terms in Eq. \ref{indiv},
\begin{align}
&\int {d{V_{{\ell _1},{\ell _2}}}} {\mathop{\rm Tr}\nolimits} \left( {\overleftarrow P {V_{ + ,{\ell _2}}}{\rho _0}V_{ + ,{\ell _2}}^\dag {X_{{\ell _2}}}V_{{\ell _1},{\ell _2}}^\dag {P_{ - ,{\ell _1}}}{V_{{\ell _1},{\ell _2}}}\overrightarrow P {V_{ + ,{\ell _2}}}{\rho _0}V_{ + ,{\ell _2}}^\dag {X_{{\ell _2}}}V_{{\ell _1},{\ell _2}}^\dag {P_{ - ,{\ell _1}}}{V_{{\ell _1},{\ell _2}}}} \right)\nonumber\\
&= \int {d{V_{{\ell _1},{\ell _2}}}} {\mathop{\rm Tr}\nolimits} \left( {\overleftarrow P {X_{{\ell _2}}}{V_{ + ,{\ell _2}}}{\rho _0}V_{ + ,{\ell _2}}^\dag V_{{\ell _1},{\ell _2}}^\dag {P_{ - ,{\ell _1}}}{V_{{\ell _1},{\ell _2}}}\overrightarrow P {X_{{\ell _2}}}{V_{ + ,{\ell _2}}}{\rho _0}V_{ + ,{\ell _2}}^\dag V_{{\ell _1},{\ell _2}}^\dag {P_{ - ,{\ell _1}}}{V_{{\ell _1},{\ell _2}}}} \right)\nonumber\\
&= \frac{{{\mathop{\rm Tr}\nolimits} (P_{ - ,{\ell _1}}^2)}}{{D\left( {{D^2} - 1} \right)}}\left( \begin{array}{l}
D{\mathop{\rm Tr}\nolimits} (\overleftarrow P {V_{ + ,{\ell _2}}}{\rho _0}V_{ + ,{\ell _2}}^\dag {X_{{\ell _2}}}){\mathop{\rm Tr}\nolimits} (\overrightarrow P {V_{ + ,{\ell _2}}}{\rho _0}V_{ + ,{\ell _2}}^\dag {X_{{\ell _2}}})\\
 - {\mathop{\rm Tr}\nolimits} (\overleftarrow P {V_{ + ,{\ell _2}}}{\rho _0}V_{ + ,{\ell _2}}^\dag {X_{{\ell _2}}}\overrightarrow P {V_{ + ,{\ell _2}}}{\rho _0}V_{ + ,{\ell _2}}^\dag {X_{{\ell _2}}})
\end{array} \right)\nonumber\\
&= \frac{{{\mathop{\rm Tr}\nolimits} (P_{ - ,{\ell _1}}^2)}}{{D\left( {{D^2} - 1} \right)}}\left( \begin{array}{l}
D{\mathop{\rm Tr}\nolimits} ({V_{ + ,{\ell _2}}}{\rho _0}V_{ + ,{\ell _2}}^\dag {X_{{\ell _2}}}{V_{ + ,{\ell _2}}}{\rho _0}V_{ + ,{\ell _2}}^\dag {X_{{\ell _2}}})\\
 - {\mathop{\rm Tr}\nolimits} ({V_{ + ,{\ell _2}}}{\rho _0}V_{ + ,{\ell _2}}^\dag {X_{{\ell _2}}}){\mathop{\rm Tr}\nolimits} ({V_{ + ,{\ell _2}}}{\rho _0}V_{ + ,{\ell _2}}^\dag {X_{{\ell _2}}})
\end{array} \right)~,
\end{align}
and
\begin{align}
&- 2\int {d{V_{{\ell _1},{\ell _2}}}} {\mathop{\rm Tr}\nolimits} \left( {\overleftarrow P {V_{ + ,{\ell _2}}}{\rho _0}V_{ + ,{\ell _2}}^\dag {X_{{\ell _2}}}V_{{\ell _1},{\ell _2}}^\dag {P_{ - ,{\ell _1}}}{V_{{\ell _1},{\ell _2}}}\overrightarrow P {X_{{\ell _2}}}{V_{ + ,{\ell _2}}}{\rho _0}V_{ + ,{\ell _2}}^\dag V_{{\ell _1},{\ell _2}}^\dag {P_{ - ,{\ell _1}}}{V_{{\ell _1},{\ell _2}}}} \right)\nonumber\\
&= \frac{{ - 2{\mathop{\rm Tr}\nolimits} (P_{ - ,{\ell _1}}^2)}}{{D\left( {{D^2} - 1} \right)}}\left( \begin{array}{l}
D{\mathop{\rm Tr}\nolimits} ({V_{ + ,{\ell _2}}}{\rho _0}V_{ + ,{\ell _2}}^\dag {X_{{\ell _2}}}{X_{{\ell _2}}}{V_{ + ,{\ell _2}}}{\rho _0}V_{ + ,{\ell _2}}^\dag )\\
 - {\mathop{\rm Tr}\nolimits} ({V_{ + ,{\ell _2}}}{\rho _0}V_{ + ,{\ell _2}}^\dag {X_{{\ell _2}}}){\mathop{\rm Tr}\nolimits} ({X_{{\ell _2}}}{V_{ + ,{\ell _2}}}{\rho _0}V_{ + ,{\ell _2}}^\dag )
\end{array} \right)\nonumber\\
&= \frac{{ - 2{\mathop{\rm Tr}\nolimits} (P_{ - ,{\ell _1}}^2)}}{{D\left( {{D^2} - 1} \right)}}\left( {N - {\mathop{\rm Tr}\nolimits} ({V_{ + ,{\ell _2}}}{\rho _0}V_{ + ,{\ell _2}}^\dag {X_{{\ell _2}}}){\mathop{\rm Tr}\nolimits} ({V_{ + ,{\ell _2}}}{\rho _0}V_{ + ,{\ell _2}}^\dag {X_{{\ell _2}}})} \right)~.
\end{align}

Then we get
\begin{align}
&\int d {V_{{\ell _1},{\ell _2}}}{{\mathop{\rm Tr}\nolimits} ^2}\left( {{\rho _0}V_{ + ,{\ell _2}}^\dag \left[ {{X_{{\ell _2}}},V_{{\ell _1},{\ell _2}}^\dag {P_{ - ,{\ell _1}}}{V_{{\ell _1},{\ell _2}}}} \right]{V_{ + ,{\ell _2}}}} \right)\nonumber\\
&= \frac{{2{\mathop{\rm Tr}\nolimits} \left( {P_{ - ,{\ell _1}}^2} \right)}}{{\left( {{D^2} - 1} \right)}}\left( {{\mathop{\rm Tr}\nolimits} \left( {{V_{ + ,{\ell _2}}}{\rho _0}V_{ + ,{\ell _2}}^\dag {X_{{\ell _2}}}{V_{ + ,{\ell _2}}}{\rho _0}V_{ + ,{\ell _2}}^\dag {X_{{\ell _2}}}} \right) - 1} \right)\nonumber\\
&=  - \frac{{2{\mathop{\rm Tr}\nolimits} \left( {P_{ - ,{\ell _1}}^2} \right)}}{{\left( {{D^2} - 1} \right)}} + \frac{{2{\mathop{\rm Tr}\nolimits} \left( {P_{ - ,{\ell _1}}^2} \right)}}{{\left( {{D^2} - 1} \right)}}{\mathop{\rm Tr}\nolimits} \left( {{V_{ + ,{\ell _2}}}{\rho _0}V_{ + ,{\ell _2}}^\dag {X_{{\ell _2}}}{V_{ + ,{\ell _2}}}{\rho _0}V_{ + ,{\ell _2}}^\dag {X_{{\ell _2}}}} \right)~.
\end{align}

Back to Eq. \ref{c2term}, we get
\begin{align}\label{c2term2}
&4\int {\cal D} V\sum\limits_{{\ell _1} < {\ell _2}} {{{{\mathop{\rm Tr}\nolimits} }^2}} \left( {{\rho _0}V_{ + ,{\ell _1}}^\dag {P_{ - ,{\ell _1}}}{V_{ + ,{\ell _1}}}{\rho _0}V_{ + ,{\ell _2}}^\dag {P_{ - ,{\ell _2}}}{V_{ + ,{\ell _2}}}} \right){{\mathop{\rm Tr}\nolimits} ^2}\left( {{\rho _0}V_{ + ,{\ell _2}}^\dag \left[ {{X_{{\ell _2}}},V_{{\ell _1},{\ell _2}}^\dag {P_{ - ,{\ell _1}}}{V_{{\ell _1},{\ell _2}}}} \right]{V_{ + ,{\ell _2}}}} \right)\nonumber\\
&= 4\int {\cal D} V\sum\limits_{{\ell _1} < {\ell _2}} {\left( \begin{array}{l}
 - \frac{{2\left( {D{\mathop{\rm Tr}\nolimits} \left( {{O^2}} \right) - {{{\mathop{\rm Tr}\nolimits} }^2}(O)} \right)}}{{(D - 1){D^2}{{(D + 1)}^2}}}{\mathop{\rm Tr}\nolimits} \left( {P_{ - ,{\ell _1}}^2} \right)\\
 + \frac{{\left( {(2D + 1){\mathop{\rm Tr}\nolimits} \left( {{O^2}} \right) - {{{\mathop{\rm Tr}\nolimits} }^2}(O)} \right)}}{{(D - 1){D^2}{{(D + 1)}^2}}}{\mathop{\rm Tr}\nolimits} \left( {P_{ - ,{\ell _1}}^2} \right){{\mathop{\rm Tr}\nolimits} ^2}\left( {{\rho _0}V_{ + ,{\ell _2}}^\dag {X_{{\ell _2}}}{V_{ + ,{\ell _2}}}} \right)\\
 - \frac{{\left( {{\mathop{\rm Tr}\nolimits} \left( {{O^2}} \right) + {{{\mathop{\rm Tr}\nolimits} }^2}(O)} \right)}}{{(D - 1){D^2}{{(D + 1)}^2}}}{\mathop{\rm Tr}\nolimits} \left( {P_{ - ,{\ell _1}}^2} \right){\mathop{\rm Tr}\nolimits} \left( {{\rho _0}V_{ + ,{\ell _2}}^\dag {X_{{\ell _2}}}{V_{ + ,{\ell _2}}}{\rho _0}V_{ + ,{\ell _2}}^\dag {X_{{\ell _2}}}{V_{ + ,{\ell _2}}}} \right)
\end{array} \right)}  \times \nonumber\\
&\qquad \left( { - \frac{{2{\mathop{\rm Tr}\nolimits} \left( {P_{ - ,{\ell _1}}^2} \right)}}{{\left( {{D^2} - 1} \right)}} + \frac{{2{\mathop{\rm Tr}\nolimits} \left( {P_{ - ,{\ell _1}}^2} \right)}}{{\left( {{D^2} - 1} \right)}}{\mathop{\rm Tr}\nolimits} \left( {{V_{ + ,{\ell _2}}}{\rho _0}V_{ + ,{\ell _2}}^\dag {X_{{\ell _2}}}{V_{ + ,{\ell _2}}}{\rho _0}V_{ + ,{\ell _2}}^\dag {X_{{\ell _2}}}} \right)} \right)\nonumber\\
&= \frac{8}{{{{(D - 1)}^2}{D^2}{{(D + 1)}^3}}}\sum\limits_{{\ell _1} < {\ell _2}} {\int {d{V_{ - ,{\ell _1}}}} {{{\mathop{\rm Tr}\nolimits} }^2}\left( {P_{ - ,{\ell _1}}^2} \right)}  \times \nonumber\\
&\qquad \int {d{V_{ + ,{\ell _2}}}} \left( \begin{array}{l}
\left( \begin{array}{l}
2\left( {D{\mathop{\rm Tr}\nolimits} \left( {{O^2}} \right) - {{{\mathop{\rm Tr}\nolimits} }^2}(O)} \right)\\
 - \left( {(2D + 1){\mathop{\rm Tr}\nolimits} \left( {{O^2}} \right) - {{{\mathop{\rm Tr}\nolimits} }^2}(O)} \right){{\mathop{\rm Tr}\nolimits} ^2}\left( {{\rho _0}V_{ + ,{\ell _2}}^\dag {X_{{\ell _2}}}{V_{ + ,{\ell _2}}}} \right)\\
 + \left( {{\mathop{\rm Tr}\nolimits} \left( {{O^2}} \right) + {{{\mathop{\rm Tr}\nolimits} }^2}(O)} \right){\mathop{\rm Tr}\nolimits} \left( {{\rho _0}V_{ + ,{\ell _2}}^\dag {X_{{\ell _2}}}{V_{ + ,{\ell _2}}}{\rho _0}V_{ + ,{\ell _2}}^\dag {X_{{\ell _2}}}{V_{ + ,{\ell _2}}}} \right)
\end{array} \right)\\
\qquad \left( {1 - {\mathop{\rm Tr}\nolimits} \left( {{V_{ + ,{\ell _2}}}{\rho _0}V_{ + ,{\ell _2}}^\dag {X_{{\ell _2}}}{V_{ + ,{\ell _2}}}{\rho _0}V_{ + ,{\ell _2}}^\dag {X_{{\ell _2}}}} \right)} \right)
\end{array} \right)~.
\end{align}

Recall that
\begin{align}
&\int {d{V_{ - ,{\ell _1}}}} {{\mathop{\rm Tr}\nolimits} ^2}\left( {P_{ - ,{\ell _1}}^2} \right)\nonumber\\
&= 4{{\mathop{\rm Tr}\nolimits} ^2}\left( {{O^2}} \right) - \frac{{8{\mathop{\rm Tr}\nolimits} \left( {{O^2}} \right){{{\mathop{\rm Tr}\nolimits} }^2}(O)}}{D} + \frac{{4{{{\mathop{\rm Tr}\nolimits} }^4}(O) + 16{{{\mathop{\rm Tr}\nolimits} }^2}\left( {{O^2}} \right)}}{{{D^2}}} +  \ldots ~.
\end{align}
So we only need to compute the following integral from Eq. \ref{c2term2}:
\begin{align}
\int {d{V_{ + ,{\ell _2}}}} \left( \begin{array}{l}
\left( \begin{array}{l}
2\left( {D{\mathop{\rm Tr}\nolimits} \left( {{O^2}} \right) - {{{\mathop{\rm Tr}\nolimits} }^2}(O)} \right)\\
 - \left( {(2D + 1){\mathop{\rm Tr}\nolimits} \left( {{O^2}} \right) - {{{\mathop{\rm Tr}\nolimits} }^2}(O)} \right){{\mathop{\rm Tr}\nolimits} ^2}\left( {{\rho _0}V_{ + ,{\ell _2}}^\dag {X_{{\ell _2}}}{V_{ + ,{\ell _2}}}} \right)\\
 + \left( {{\mathop{\rm Tr}\nolimits} \left( {{O^2}} \right) + {{{\mathop{\rm Tr}\nolimits} }^2}(O)} \right){\mathop{\rm Tr}\nolimits} \left( {{\rho _0}V_{ + ,{\ell _2}}^\dag {X_{{\ell _2}}}{V_{ + ,{\ell _2}}}{\rho _0}V_{ + ,{\ell _2}}^\dag {X_{{\ell _2}}}{V_{ + ,{\ell _2}}}} \right)
\end{array} \right)\\
\left( {1 - {\mathop{\rm Tr}\nolimits} \left( {{V_{ + ,{\ell _2}}}{\rho _0}V_{ + ,{\ell _2}}^\dag {X_{{\ell _2}}}{V_{ + ,{\ell _2}}}{\rho _0}V_{ + ,{\ell _2}}^\dag {X_{{\ell _2}}}} \right)} \right)
\end{array} \right)~.
\end{align}
After expansion, we have $3\times 2 = 6$ terms, mostly up to 4-design. We have
\begin{itemize}
\item Constant term:
\begin{align}
\int {d{V_{ + ,{\ell _2}}}} \left( {2\left( {D{\mathop{\rm Tr}\nolimits} \left( {{O^2}} \right) - {{{\mathop{\rm Tr}\nolimits} }^2}(O)} \right)} \right) = 2D{\mathop{\rm Tr}\nolimits} \left( {{O^2}} \right) - 2{{\mathop{\rm Tr}\nolimits} ^2}(O)~.
\end{align}
\item Three 2-design terms:
\begin{align}
&\int {d{V_{ + ,{\ell _2}}}} 2\left( {D{\mathop{\rm Tr}\nolimits} \left( {{O^2}} \right) - {{{\mathop{\rm Tr}\nolimits} }^2}(O)} \right)( - 1){\mathop{\rm Tr}\nolimits} \left( {{V_{ + ,{\ell _2}}}{\rho _0}V_{ + ,{\ell _2}}^\dag {X_{{\ell _2}}}{V_{ + ,{\ell _2}}}{\rho _0}V_{ + ,{\ell _2}}^\dag {X_{{\ell _2}}}} \right)\nonumber\\
&=  - 2\left( {D{\mathop{\rm Tr}\nolimits} \left( {{O^2}} \right) - {{{\mathop{\rm Tr}\nolimits} }^2}(O)} \right)\int {d{V_{ + ,{\ell _2}}}} {\mathop{\rm Tr}\nolimits} \left( {{\rho _0}V_{ + ,{\ell _2}}^\dag {X_{{\ell _2}}}{V_{ + ,{\ell _2}}}{\rho _0}V_{ + ,{\ell _2}}^\dag {X_{{\ell _2}}}{V_{ + ,{\ell _2}}}} \right)\nonumber\\
&= \frac{{ - 2\left( {D{\mathop{\rm Tr}\nolimits} \left( {{O^2}} \right) - {{{\mathop{\rm Tr}\nolimits} }^2}(O)} \right)}}{{D + 1}}~,
\end{align}
\begin{align}
&- \int {d{V_{ + ,{\ell _2}}}} \left( {(2D + 1){\mathop{\rm Tr}\nolimits} \left( {{O^2}} \right) - {{{\mathop{\rm Tr}\nolimits} }^2}(O)} \right){{\mathop{\rm Tr}\nolimits} ^2}\left( {{\rho _0}V_{ + ,{\ell _2}}^\dag {X_{{\ell _2}}}{V_{ + ,{\ell _2}}}} \right)\nonumber\\
&=  - \left( {(2D + 1){\mathop{\rm Tr}\nolimits} \left( {{O^2}} \right) - {{{\mathop{\rm Tr}\nolimits} }^2}(O)} \right)\int {d{V_{ + ,{\ell _2}}}} {\mathop{\rm Tr}\nolimits} \left( {\overleftarrow P {\rho _0}V_{ + ,{\ell _2}}^\dag {X_{{\ell _2}}}{V_{ + ,{\ell _2}}}\overrightarrow P {\rho _0}V_{ + ,{\ell _2}}^\dag {X_{{\ell _2}}}{V_{ + ,{\ell _2}}}} \right)\nonumber\\
&= \frac{{ - \left( {(2D + 1){\mathop{\rm Tr}\nolimits} \left( {{O^2}} \right) - {{{\mathop{\rm Tr}\nolimits} }^2}(O)} \right)}}{{D + 1}}~,
\end{align}
and
\begin{align}
&\int {d{V_{ + ,{\ell _2}}}} \left( {{\mathop{\rm Tr}\nolimits} \left( {{O^2}} \right) + {{{\mathop{\rm Tr}\nolimits} }^2}(O)} \right){\mathop{\rm Tr}\nolimits} \left( {{\rho _0}V_{ + ,{\ell _2}}^\dag {X_{{\ell _2}}}{V_{ + ,{\ell _2}}}{\rho _0}V_{ + ,{\ell _2}}^\dag {X_{{\ell _2}}}{V_{ + ,{\ell _2}}}} \right)\nonumber\\
&= \left( {{\mathop{\rm Tr}\nolimits} \left( {{O^2}} \right) + {{{\mathop{\rm Tr}\nolimits} }^2}(O)} \right)\int {d{V_{ + ,{\ell _2}}}} {\mathop{\rm Tr}\nolimits} \left( {{\rho _0}V_{ + ,{\ell _2}}^\dag {X_{{\ell _2}}}{V_{ + ,{\ell _2}}}{\rho _0}V_{ + ,{\ell _2}}^\dag {X_{{\ell _2}}}{V_{ + ,{\ell _2}}}} \right)\nonumber\\
&= \frac{{\left( {{\mathop{\rm Tr}\nolimits} \left( {{O^2}} \right) + {{{\mathop{\rm Tr}\nolimits} }^2}(O)} \right)}}{{D + 1}}~.
\end{align}

Therefore, we get
\begin{align}
&\frac{{ - 2\left( {D{\mathop{\rm Tr}\nolimits} \left( {{O^2}} \right) - {{{\mathop{\rm Tr}\nolimits} }^2}(O)} \right)}}{{D + 1}} + \frac{{ - \left( {(2D + 1){\mathop{\rm Tr}\nolimits} \left( {{O^2}} \right) - {{{\mathop{\rm Tr}\nolimits} }^2}(O)} \right)}}{{D + 1}} + \frac{{\left( {{\mathop{\rm Tr}\nolimits} \left( {{O^2}} \right) + {{{\mathop{\rm Tr}\nolimits} }^2}(O)} \right)}}{{D + 1}}\nonumber\\
&=  - \frac{{4\left( {D{\mathop{\rm Tr}\nolimits} \left( {{O^2}} \right) - {{{\mathop{\rm Tr}\nolimits} }^2}(O)} \right)}}{{D + 1}}~.
\end{align}

\item Two 4-design terms:
\begin{align}
&\int {d{V_{ + ,{\ell _2}}}} \left( \begin{array}{l}
\left( {(2D + 1){\mathop{\rm Tr}\nolimits} \left( {{O^2}} \right) - {{{\mathop{\rm Tr}\nolimits} }^2}(O)} \right){{\mathop{\rm Tr}\nolimits} ^2}\left( {{\rho _0}V_{ + ,{\ell _2}}^\dag {X_{{\ell _2}}}{V_{ + ,{\ell _2}}}} \right)\\
{\mathop{\rm Tr}\nolimits} \left( {{V_{ + ,{\ell _2}}}{\rho _0}V_{ + ,{\ell _2}}^\dag {X_{{\ell _2}}}{V_{ + ,{\ell _2}}}{\rho _0}V_{ + ,{\ell _2}}^\dag {X_{{\ell _2}}}} \right)
\end{array} \right)\nonumber\\
&= \left( {(2D + 1){\mathop{\rm Tr}\nolimits} \left( {{O^2}} \right) - {{{\mathop{\rm Tr}\nolimits} }^2}(O)} \right)\times\nonumber\\
&\qquad \int d {V_{ + ,{\ell _2}}}\left( {{\mathop{\rm Tr}\nolimits} \left( {\overleftarrow P {\rho _0}V_{ + ,{\ell _2}}^\dag {X_{{\ell _2}}}{V_{ + ,{\ell _2}}}\overline P {\rho _0}V_{ + ,{\ell _2}}^\dag {X_{{\ell _2}}}{V_{ + ,{\ell _2}}}\overrightarrow P {\rho _0}V_{ + ,{\ell _2}}^\dag {X_{{\ell _2}}}{V_{ + ,{\ell _2}}}{\rho _0}V_{ + ,{\ell _2}}^\dag {X_{{\ell _2}}}{V_{ + ,{\ell _2}}}} \right)} \right)\nonumber\\
&= \frac{{3\left( {(2D + 1){\mathop{\rm Tr}\nolimits} \left( {{O^2}} \right) - {{{\mathop{\rm Tr}\nolimits} }^2}(O)} \right)}}{{(D + 1)(D + 3)}}~,
\end{align}
and
\begin{align}
&- \int {d{V_{ + ,{\ell _2}}}} \left( \begin{array}{l}
\left( {{\mathop{\rm Tr}\nolimits} \left( {{O^2}} \right) + {{{\mathop{\rm Tr}\nolimits} }^2}(O)} \right){\mathop{\rm Tr}\nolimits} \left( {{V_{ + ,{\ell _2}}}{\rho _0}V_{ + ,{\ell _2}}^\dag {X_{{\ell _2}}}{V_{ + ,{\ell _2}}}{\rho _0}V_{ + ,{\ell _2}}^\dag {X_{{\ell _2}}}} \right)\\
{\mathop{\rm Tr}\nolimits} \left( {{\rho _0}V_{ + ,{\ell _2}}^\dag {X_{{\ell _2}}}{V_{ + ,{\ell _2}}}{\rho _0}V_{ + ,{\ell _2}}^\dag {X_{{\ell _2}}}{V_{ + ,{\ell _2}}}} \right)
\end{array} \right)\nonumber\\
&=  - \left( {{\mathop{\rm Tr}\nolimits} \left( {{O^2}} \right) + {{{\mathop{\rm Tr}\nolimits} }^2}(O)} \right)\int {d{V_{ + ,{\ell _2}}}} \left( \begin{array}{l}
{\mathop{\rm Tr}\nolimits} \left( {{\rho _0}V_{ + ,{\ell _2}}^\dag {X_{{\ell _2}}}{V_{ + ,{\ell _2}}}{\rho _0}V_{ + ,{\ell _2}}^\dag {X_{{\ell _2}}}{V_{ + ,{\ell _2}}}} \right)\\
{\mathop{\rm Tr}\nolimits} \left( {{\rho _0}V_{ + ,{\ell _2}}^\dag {X_{{\ell _2}}}{V_{ + ,{\ell _2}}}{\rho _0}V_{ + ,{\ell _2}}^\dag {X_{{\ell _2}}}{V_{ + ,{\ell _2}}}} \right)
\end{array} \right)\nonumber\\
&=  - \left( {{\mathop{\rm Tr}\nolimits} \left( {{O^2}} \right) + {{{\mathop{\rm Tr}\nolimits} }^2}(O)} \right)\int {d{V_{ + ,{\ell _2}}}} {\mathop{\rm Tr}\nolimits} \left( \begin{array}{l}
\overleftarrow P {\rho _0}V_{ + ,{\ell _2}}^\dag {X_{{\ell _2}}}{V_{ + ,{\ell _2}}}{\rho _0}V_{ + ,{\ell _2}}^\dag {X_{{\ell _2}}}{V_{ + ,{\ell _2}}}\\
\overrightarrow P {\rho _0}V_{ + ,{\ell _2}}^\dag {X_{{\ell _2}}}{V_{ + ,{\ell _2}}}{\rho _0}V_{ + ,{\ell _2}}^\dag {X_{{\ell _2}}}{V_{ + ,{\ell _2}}}
\end{array} \right)\nonumber\\
&= \frac{{ - 3\left( {{\mathop{\rm Tr}\nolimits} \left( {{O^2}} \right) + {{{\mathop{\rm Tr}\nolimits} }^2}(O)} \right)}}{{(D + 1)(D + 3)}}~.
\end{align}
Again, we use $\overleftarrow{P}$, $\overline{P}$ and $\overrightarrow{P}$ if we have three traces in total. Thus we get
\begin{align}
&\frac{{3\left( {(2D + 1){\rm{Tr}}\left( {{O^2}} \right) - \operatorname{Tr}^2(O)} \right)}}{{(D + 1)(D + 3)}} - \frac{{3\left( {{\mathop{\rm Tr}\nolimits} \left( {{O^2}} \right) + {{{\mathop{\rm Tr}\nolimits} }^2}(O)} \right)}}{{(D + 1)(D + 3)}}\nonumber\\
&= \frac{{6\left( {D{\mathop{\rm Tr}\nolimits} \left( {{O^2}} \right) - {{{\mathop{\rm Tr}\nolimits} }^2}(O)} \right)}}{{(D + 1)(D + 3)}}~.
\end{align}
\end{itemize}

Combining several terms, we get 
\begin{align}
&\left. {\int d {V_{ + ,{\ell _2}}}\left( {\begin{array}{*{20}{l}}
{2\left( {D{\mathop{\rm Tr}\nolimits} \left( {{O^2}} \right) - {{{\mathop{\rm Tr}\nolimits} }^2}(O)} \right)}\\
{ - \left( {(2D + 1){\mathop{\rm Tr}\nolimits} \left( {{O^2}} \right) - {{{\mathop{\rm Tr}\nolimits} }^2}(O)} \right){{{\mathop{\rm Tr}\nolimits} }^2}\left( {{\rho _0}V_{ + ,{\ell _2}}^\dag {X_{{\ell _2}}}{V_{ + ,{\ell _2}}}} \right)}\\
{ + \left( {{\mathop{\rm Tr}\nolimits} \left( {{O^2}} \right) + {{{\mathop{\rm Tr}\nolimits} }^2}(O)} \right){\mathop{\rm Tr}\nolimits} \left( {{\rho _0}V_{ + ,{\ell _2}}^\dag {X_{{\ell _2}}}{V_{ + ,{\ell _2}}}{\rho _0}V_{ + ,{\ell _2}}^\dag {X_{{\ell _2}}}{V_{ + ,{\ell _2}}}} \right)}
\end{array}} \right)} \right)\nonumber\\
&= 2D{\mathop{\rm Tr}\nolimits} \left( {{O^2}} \right) - 2{{\mathop{\rm Tr}\nolimits} ^2}(O) + \frac{{6\left( {D{\mathop{\rm Tr}\nolimits} \left( {{O^2}} \right) - {{{\mathop{\rm Tr}\nolimits} }^2}(O)} \right)}}{{(D + 1)(D + 3)}} - \frac{{4\left( {D{\mathop{\rm Tr}\nolimits} \left( {{O^2}} \right) - {{{\mathop{\rm Tr}\nolimits} }^2}(O)} \right)}}{{D + 1}}\nonumber\\
&= \frac{{2D(D + 2)\left( {D{\mathop{\rm Tr}\nolimits} \left( {{O^2}} \right) - {{{\mathop{\rm Tr}\nolimits} }^2}(O)} \right)}}{{(D + 1)(D + 3)}}~.
\end{align}

Therefore,
\begin{align}
&4\int {\cal D} V\sum\limits_{{\ell _1} < {\ell _2}} {{{{\mathop{\rm Tr}\nolimits} }^2}} \left( {{\rho _0}V_{ + ,{\ell _1}}^\dag {P_{ - ,{\ell _1}}}{V_{ + ,{\ell _1}}}{\rho _0}V_{ + ,{\ell _2}}^\dag {P_{ - ,{\ell _2}}}{V_{ + ,{\ell _2}}}} \right){{\mathop{\rm Tr}\nolimits} ^2}\left( {{\rho _0}V_{ + ,{\ell _2}}^\dag \left[ {{X_{{\ell _2}}},V_{{\ell _1},{\ell _2}}^\dag {P_{ - ,{\ell _1}}}{V_{{\ell _1},{\ell _2}}}} \right]{V_{ + ,{\ell _2}}}} \right)\nonumber\\
&= \frac{8}{{{{(D - 1)}^2}{D^2}{{(D + 1)}^3}}}\frac{{2D(D + 2)\left( {D{\mathop{\rm Tr}\nolimits} \left( {{O^2}} \right) - {{{\mathop{\rm Tr}\nolimits} }^2}(O)} \right)}}{{(D + 1)(D + 3)}}\sum\limits_{{\ell _1} < {\ell _2}} {\int d } {V_{ - ,{\ell _1}}}{{\mathop{\rm Tr}\nolimits} ^2}\left( {P_{ - ,{\ell _1}}^2} \right)\nonumber\\
&= \frac{{16(D + 2)\left( {D{\mathop{\rm Tr}\nolimits} \left( {{O^2}} \right) - {{{\mathop{\rm Tr}\nolimits} }^2}(O)} \right)}}{{{{(D - 1)}^2}N{{(D + 1)}^4}(D + 3)}}\sum\limits_{{\ell _1} < {\ell _2}} {\int d } {V_{ - ,{\ell _1}}}{{\mathop{\rm Tr}\nolimits} ^2}\left( {P_{ - ,{\ell _1}}^2} \right)\nonumber\\
&= \left( {\frac{{64{{{\mathop{\rm Tr}\nolimits} }^3}({O^2})}}{{{D^6}}} - \frac{{192\left( {{{{\mathop{\rm Tr}\nolimits} }^3}({O^2}) + {{{\mathop{\rm Tr}\nolimits} }^2}\left( {{O^2}} \right){{{\mathop{\rm Tr}\nolimits} }^2}(O)} \right)}}{{{D^7}}} +  \ldots } \right)\frac{{L(L - 1)}}{2}\nonumber\\
&= L(L - 1)\left( {\frac{{32{{{\mathop{\rm Tr}\nolimits} }^3}({O^2})}}{{{D^6}}} - \frac{{96\left( {{{{\mathop{\rm Tr}\nolimits} }^3}({O^2}) + {{{\mathop{\rm Tr}\nolimits} }^2}\left( {{O^2}} \right){{{\mathop{\rm Tr}\nolimits} }^2}(O)} \right)}}{{{D^7}}} +  \ldots } \right)~.
\end{align}
\end{itemize}

Finally, we conclude that
\begin{align}
&\Delta {\mu ^2} = L(L - 1)\left( {\frac{{32{{{\mathop{\rm Tr}\nolimits} }^3}({O^2})}}{{{D^6}}} - \frac{{96\left( {{{{\mathop{\rm Tr}\nolimits} }^3}({O^2}) + {{{\mathop{\rm Tr}\nolimits} }^2}\left( {{O^2}} \right){{{\mathop{\rm Tr}\nolimits} }^2}(O)} \right)}}{{{D^7}}} +  \ldots } \right)\nonumber\\
&L\left( {\frac{{1152{\mathop{\rm Tr}\nolimits} \left( {{O^6}} \right)}}{{{D^6}}} - \frac{{3456\left( {{{{\mathop{\rm Tr}\nolimits} }^2}\left( {{O^3}} \right) - 2{\mathop{\rm Tr}\nolimits} \left( {{O^2}} \right){\mathop{\rm Tr}\nolimits} \left( {{O^4}} \right) + 2{\mathop{\rm Tr}\nolimits} (O){\mathop{\rm Tr}\nolimits} \left( {{O^5}} \right) + 5{\mathop{\rm Tr}\nolimits} \left( {{O^6}} \right)} \right)}}{{{D^7}}} \ldots } \right) ~.
\end{align}
Thus, at the leading order, we have
\begin{align}
\Delta \mu  \approx \frac{{\sqrt {32} L}}{{{D^3}}}{{\mathop{\rm Tr}\nolimits} ^{3/2}}({O^2})~.
\end{align}

\section{Concentration conditions for QNTK}
We analyze the condition that $\Delta K/\overline{K} \ll 1$.
\begin{align}
\Delta K \approx \frac{{\sqrt L }}{{{D^2}}}\sqrt {\left( {8{{{\mathop{\rm Tr}\nolimits} }^2}\left( {{O^2}} \right) + 12{\mathop{\rm Tr}\nolimits} \left( {{O^4}} \right)} \right)}  \ll \bar K \approx \frac{{2L{\mathop{\rm Tr}\nolimits} \left( {{O^2}} \right)}}{{{D^2}}}~,
\end{align}
and the condition for $\Delta \mu/\overline{K}\ll 1$ implies 
\begin{align}
\frac{1}{2}{\eta ^2}{\varepsilon ^2}\Delta \mu  \ll \eta \bar K\varepsilon  \Rightarrow \eta \varepsilon \Delta \mu  \ll \bar K \Rightarrow \eta \varepsilon \frac{L}{{{D^3}}}{{\mathop{\rm Tr}\nolimits} ^{3/2}}({O^2}) \ll \frac{{L{\mathop{\rm Tr}\nolimits} \left( {{O^2}} \right)}}{{{D^2}}}~.
\end{align}
Thus we get 
\begin{align}
    \frac{\Delta K}{\overline{K}} \approx \frac{1}{\sqrt{L}} \ll 1, \quad \frac{\Delta \mu}{\overline{K}} \approx \frac{\left(\eta\sqrt{\Tr(O^2)}\right)}{D} \ll 1
\end{align}

\section{Some results about the Haar integral with the help of \texttt{RTNI}}\label{rtni}
In this section, we provide a technique to compute Haar integrals of the following form (see also \cite{Roberts:2016hpo,Cotler:2017jue,Liu:2018hlr,Liu:2020sqb}):
\begin{align}
F=\int {dU\left( \begin{array}{l}
{\rm{Tr}}\left( {{A_{1,1}}U{B_{1,1}}{U^\dag } \ldots {A_{1,{m_1}}}U{B_{1,{m_1}}}{U^\dag }} \right){\rm{Tr}}\left( {{A_{2,1}}U{B_{2,1}}{U^\dag } \ldots {A_{2,{m_2}}}U{B_{2,{m_2}}}{U^\dag }} \right) \ldots \\
{\rm{Tr}}\left( {{A_{n,1}}U{B_{n,1}}{U^\dag } \ldots {A_{n,{m_n}}}U{B_{n,{m_n}}}{U^\dag }} \right)
\end{array} \right)}~.
\end{align}

Consider the following trace:
\begin{align}
{\rm{Tr}}\left( {{A_{j,1}}U{B_{j,1}}{U^\dag } \ldots {A_{j,{m_j}}}U{B_{j,{m_j}}}{U^\dag }} \right) = \sum\limits_{{i_j}} {\left\langle {{i_j}} \right|{A_{j,1}}U{B_{j,1}}{U^\dag } \ldots {A_{j,{m_j}}}U{B_{j,{m_j}}}{U^\dag }\left| {{i_j}} \right\rangle } ~,
\end{align}
using which we get
\begin{align}
F = \sum\limits_{{i_1},{i_2} \ldots {i_n}} {\int {dU\left( \begin{array}{l}
\left\langle {{i_1}} \right|{A_{1,1}}U{B_{1,1}}{U^\dag } \ldots {A_{1,{m_1}}}U{B_{1,{m_1}}}{U^\dag }\left| {{i_1}} \right\rangle \\
\left\langle {{i_2}} \right|{A_{2,1}}U{B_{2,1}}{U^\dag } \ldots {A_{2,{m_2}}}U{B_{2,{m_2}}}{U^\dag }\left| {{i_2}} \right\rangle  \ldots \\
\left\langle {{i_n}} \right|{A_{n,1}}U{B_{n,1}}{U^\dag } \ldots {A_{n,{m_n}}}U{B_{n,{m_n}}}{U^\dag }\left| {{i_n}} \right\rangle
\end{array} \right)} }~.
\end{align}

Let
\begin{align}
{P_{{i_{j_1}},{i_{j_2}}}} = \left| {{i_{{j_1}}}} \right\rangle \left\langle {{i_{{j_2}}}} \right|~.
\end{align}
Then we get
\begin{align}
F = \sum\limits_{{i_1},{i_2} \ldots {i_n}} {\int {dU{\rm{Tr}}\left( \begin{array}{l}
{A_{1,1}}U{B_{1,1}}{U^\dag } \ldots {A_{1,{m_1}}}U{B_{1,{m_1}}}{U^\dag }{P_{i_1,i_2}}\\
{A_{2,1}}U{B_{2,1}}{U^\dag } \ldots {A_{2,{m_2}}}U{B_{2,{m_2}}}{U^\dag }{P_{i_2,i_3}} \ldots \\
{A_{n,1}}U{B_{n,1}}{U^\dag } \ldots {A_{n,{m_n}}}U{B_{n,{m_n}}}{U^\dag }{P_{i_n,i_1}}
\end{array} \right)} } ~.
\end{align}

To simplify further, we introduce the following notation:
\begin{align}
&{{\bar A}_{1,1}^{i_n,i_1}} = {P_{i_n,i_1}}{A_{1,1}}~,\nonumber\\
&{{\bar A}_{j,1}^{i_{j-1},i_j}} = {P_{i_{j - 1},i_j}}{A_{j,1}}~~~~~~~~(j \ge 2)~,
\end{align}
using which we get
\begin{align}
F = \sum\limits_{{i_1},{i_2} \ldots {i_n}} {\int {dU{\rm{Tr}}\left( \begin{array}{l}
{{\bar A}_{1,1}^{i_n,i_1}}U{B_{1,1}}{U^\dag } \ldots {A_{1,{m_1}}}U{B_{1,{m_1}}}{U^\dag }\\
{{\bar A}_{2,1}^{i_1,i_2}}U{B_{2,1}}{U^\dag } \ldots {A_{2,{m_2}}}U{B_{2,{m_2}}}{U^\dag } \ldots \\
{{\bar A}_{n,1}^{i_{n-1},i_n}}U{B_{n,1}}{U^\dag } \ldots {A_{n,{m_n}}}U{B_{n,{m_n}}}{U^\dag }
\end{array} \right)} } ~.
\end{align}

Thus, we have reduced $F$ to a standard form which can be estimated using the  \texttt{RTNI} \cite{fukuda2019rtni} package.

\section{More numerical experiments without supervised data}\label{supp:numerics}
In this section, we discuss some further numerical results besides our main text.

\begin{figure*}[ht!]
\centering
\includegraphics[width=0.48\textwidth]{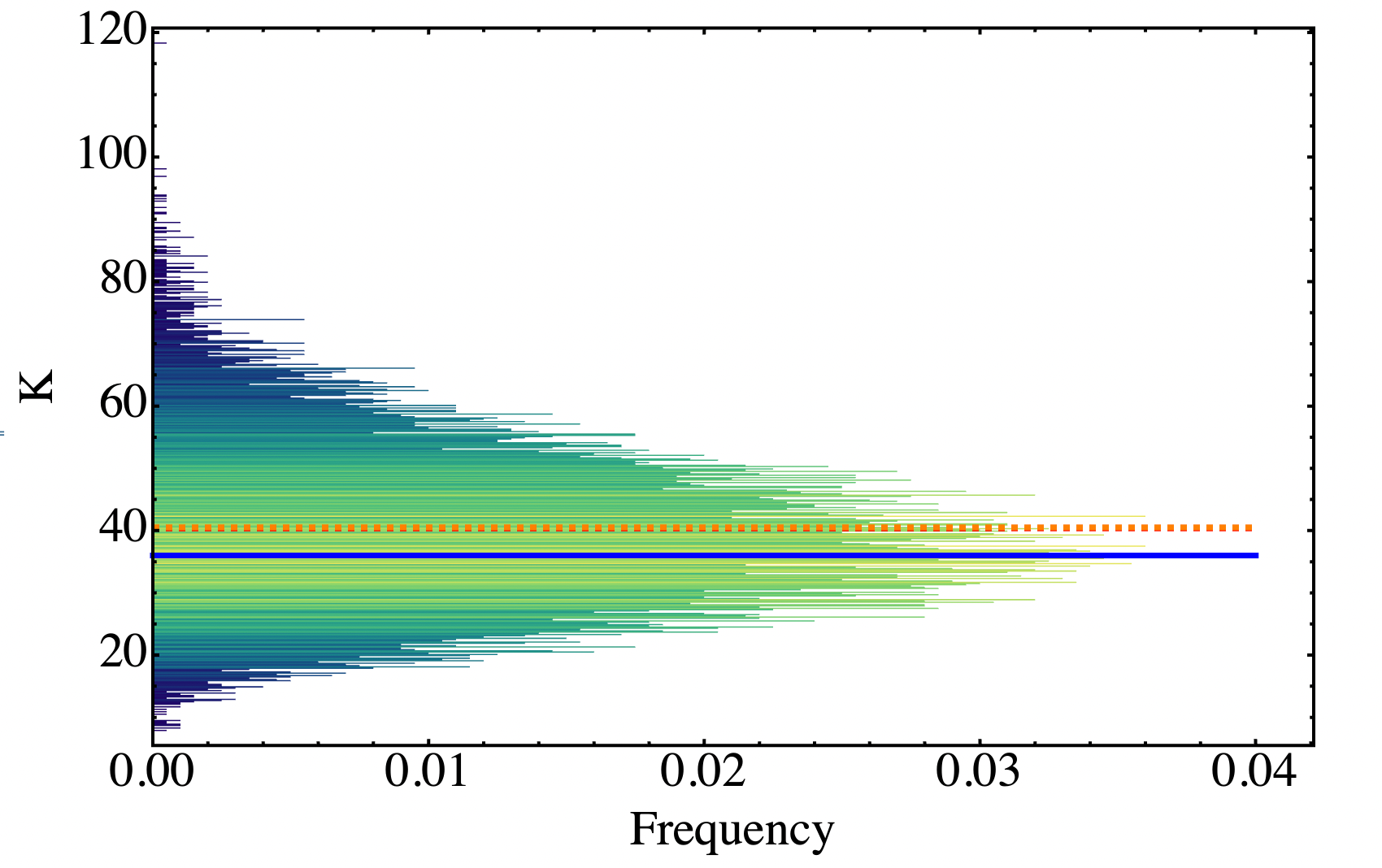}
\includegraphics[width=0.47\textwidth]{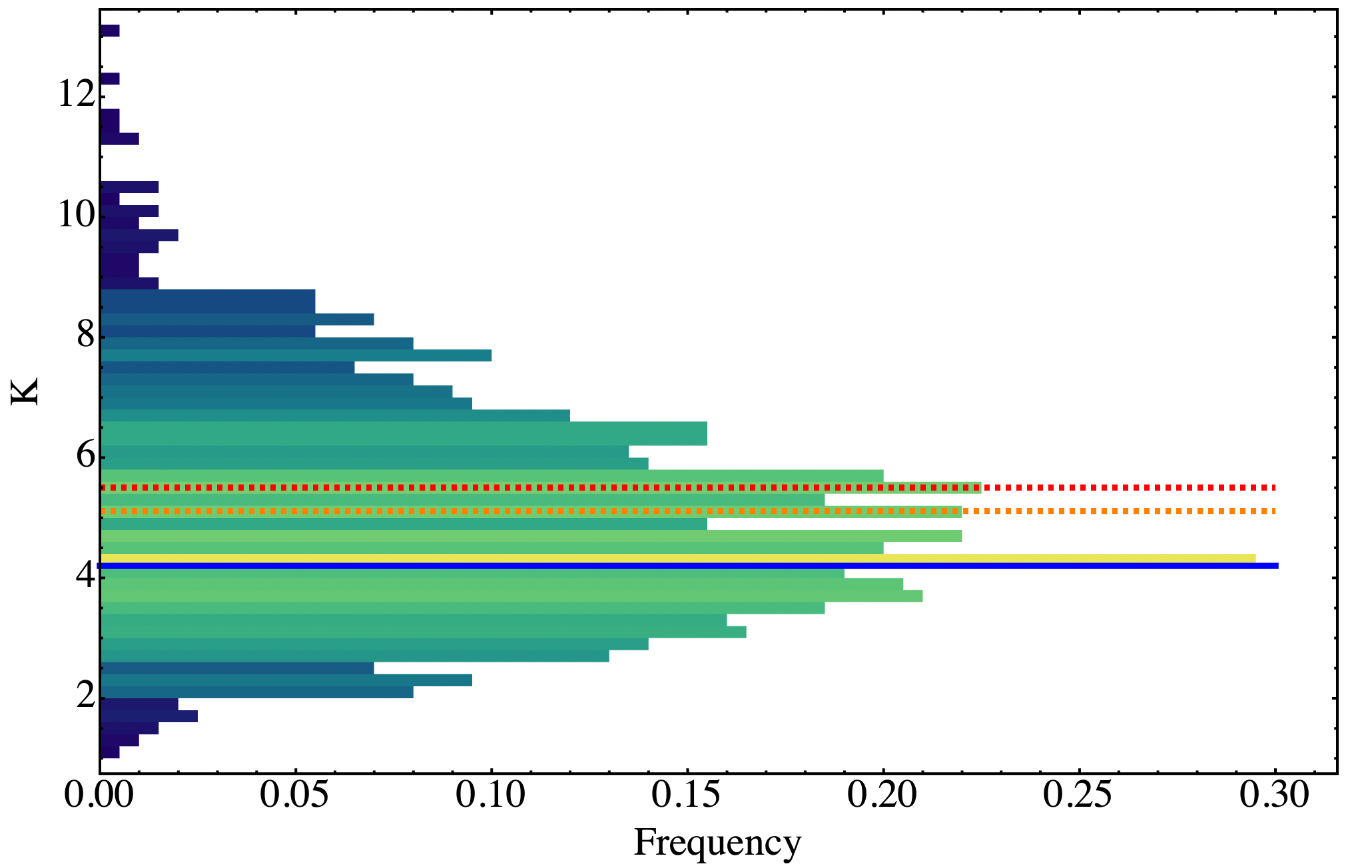}

\caption{QNTK statistics with Haar random sampling. Here we sample over 10000 and 1000 instances with 2 qubits (left) and 4 (right) respectively, and we plot the probability distribution functions with $L=16$. Red dashed line: Theoretical prediction of $\bar{K}$. Orange dashed line: averaged value of $K$ over all samples. Blue line: the maximal likelihood value of $K$ over all samples.}
\label{fig:statisticshaar}
\end{figure*}
\subsection{Haar randomness}
The average QNTK for random quantum circuits is given by
\begin{align}
\overline{K}=L\left(D \operatorname{Tr}\left(O^{2}\right)-\operatorname{Tr}^{2}(O)\right) \frac{2}{D+1}\left(\frac{1}{D^{2}-1}\right)~,
\end{align}

\begin{figure*}[ht!]
\centering
\includegraphics[width=0.45\textwidth]{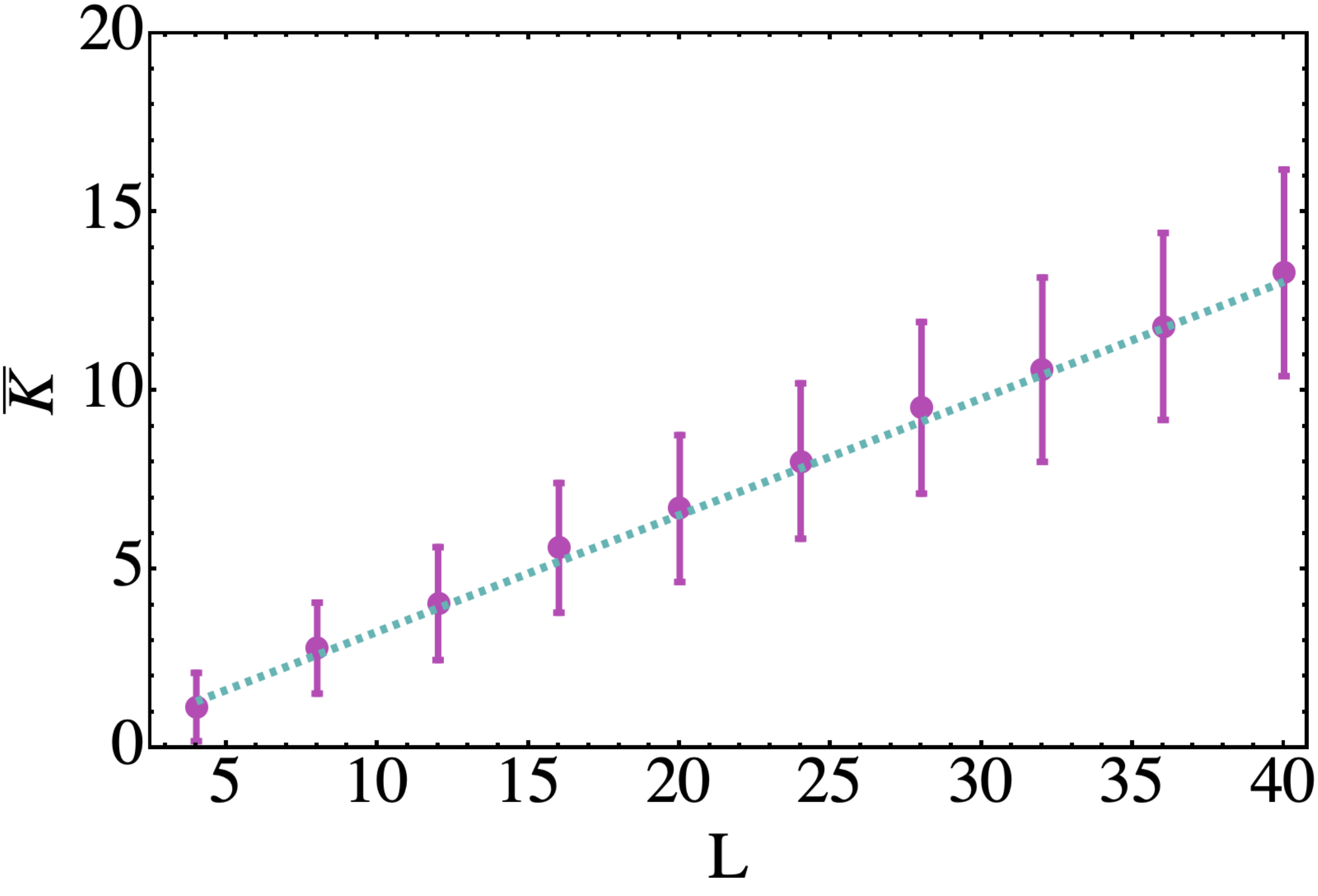}
\includegraphics[width=0.48\textwidth]{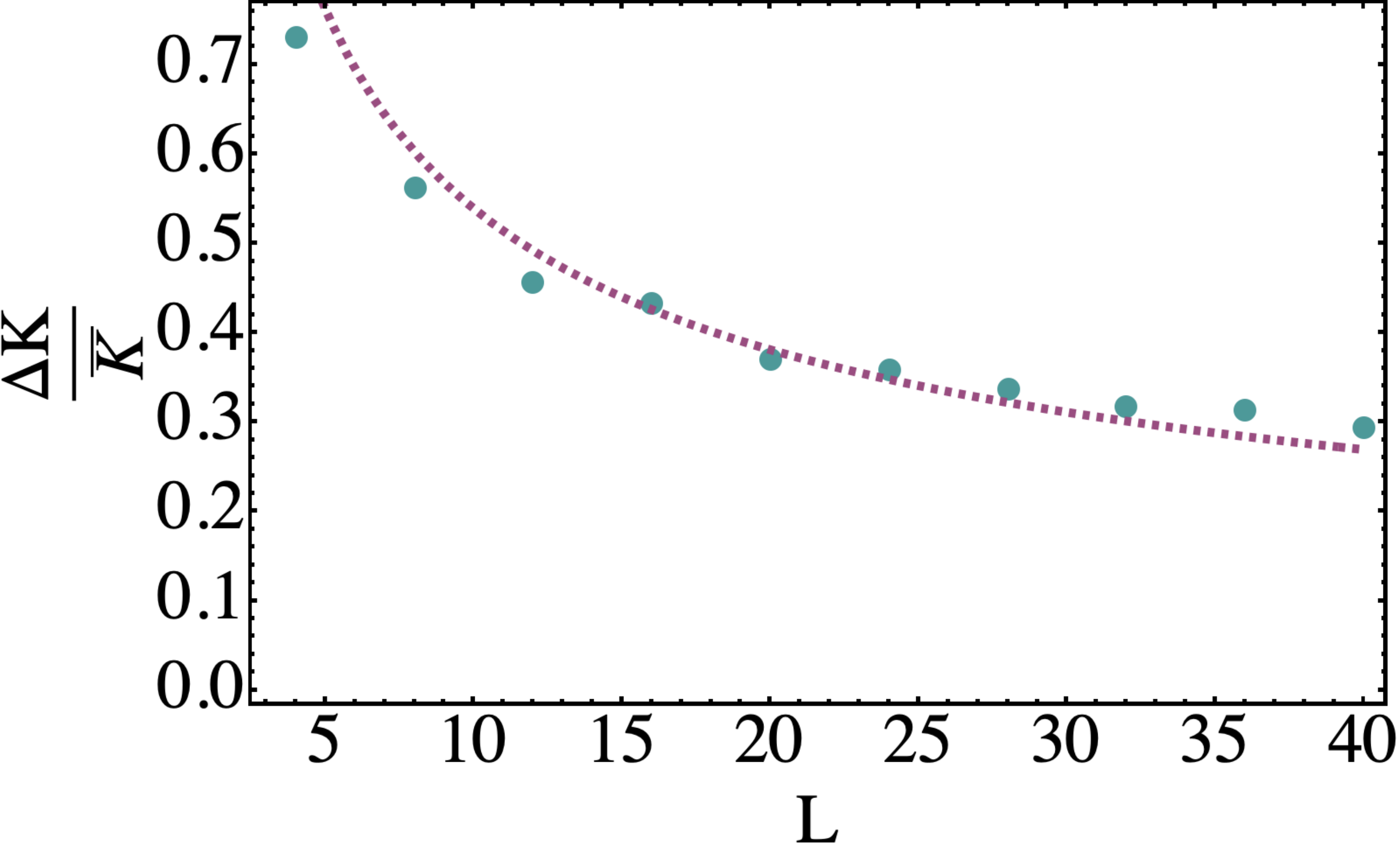}
\caption{Scaling of the QNTK statistics with 2 qubits. Here we set $L$ up to 40 in our randomized ansatz, and we sample over 1000 variational angles independently with uniform distributions in $[0,2\pi]$. Left: Scaling of $\bar{K}$ with $L$; Blue dots: numerical mean values of 10000 samples; Blue error bar: numerical standard deviations of 1000 samples; Red line: theoretical results. Right: Scaling of $\Delta K /\bar{K}$ with $L$; Black dots: numerical standard deviations of 1000 samples;  Red line: fitting using $L^{-1/2}$. }
\label{fig:scalinghaar}
\end{figure*}

Figure \ref{fig:statisticshaar} presents the concentration statistics for 2 and 4 qubits by sampling over 10000 and 1000 elements respectively for $L=16$, while Figure \ref{fig:scalinghaar} presents the scaling property for 4 qubits by the same setup with different values of $L$.

\subsection{Hardware efficient ansatz}

\begin{figure*}[ht!]
\centering
\includegraphics[width=1\textwidth]{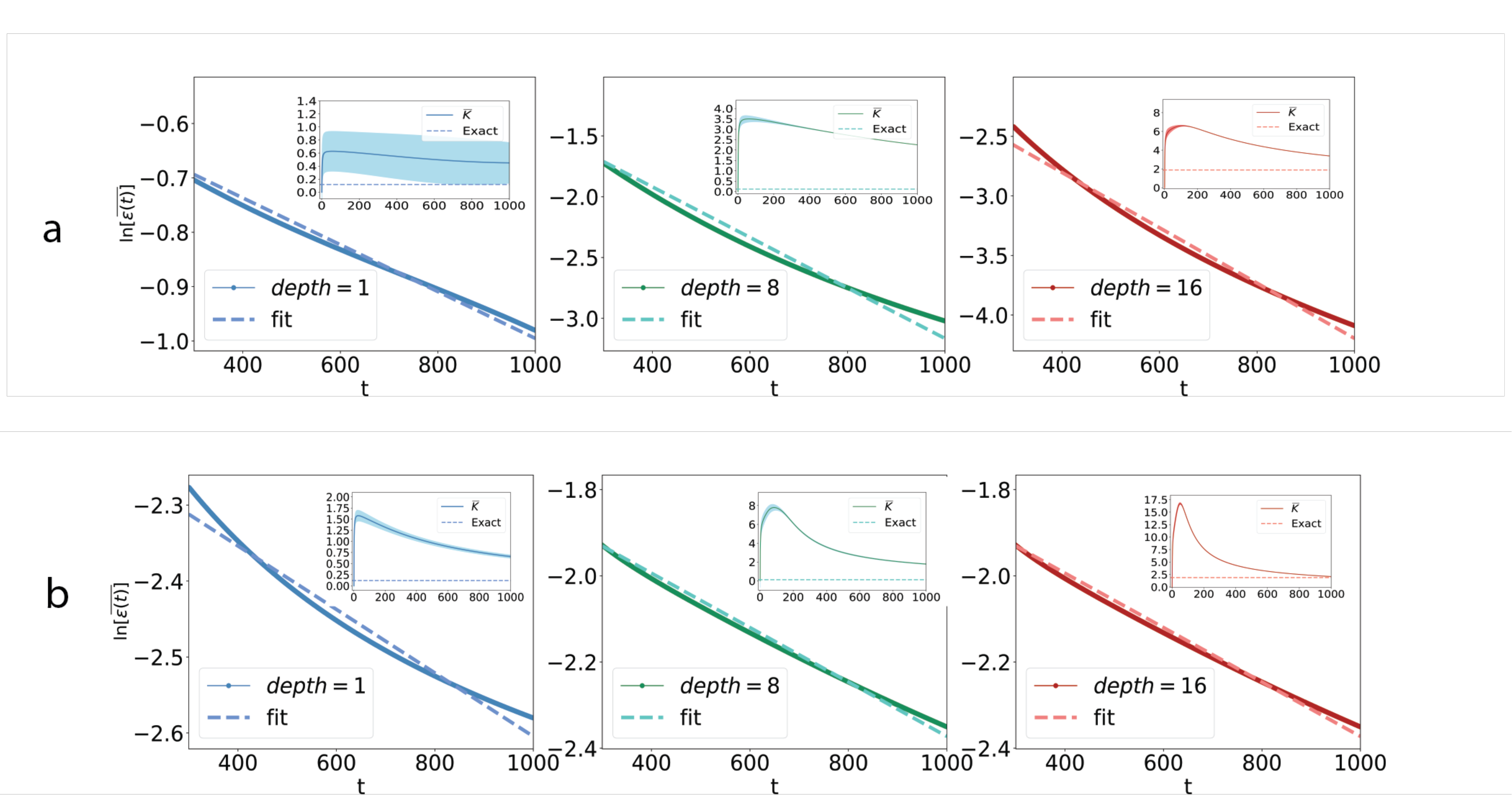}
\caption{The value of residual error $\varepsilon$ and $K$ obtained from simulation for two different ansatzes on IBM Qiskit backend for $N=4$, $\eta=0.001$, and depths of $1,8$ and $16$, depicted respectively from left to right. The top and bottom rows are for ansatz~(a) and ansatz~(b), respectively. $\varepsilon$ and $K$ are averaged over 40 and 20 different realizations for ansatz (a) and ansatz (b), respectively. In both cases, $\varepsilon$ indicates exponential decay over time as predicted from our analytic results. Insets: In the case of ansatz (a), as the depth of the circuit increases, $\overline{K}$ approaches the analytic value. For ansatz (b), $K$ values do not vary over different realizations and they show a good agreement with the theoretical results at depth=16.}
\label{fig:K_Hardware}
\end{figure*}

Here, we investigate the residual training dynamics for two different ansatzes, and we call them ansatz (a) and ansatz (b). In Figure~\ref{fig:K_Hardware}, we study the behavior of $K$ for the gradient descent optimization performed on 4 qubits. For  Figure~\ref{fig:K_Hardware}(a), a single layer of the ansatz is composed of parameterized single-qubit Pauli rotations followed by a ladder of controlled-phase rotations. Here, Pauli $X, Y$, and $Z$ rotation directions for each qubit on each layer are chosen independently and with equal probability. Moreover, the initial state of each qubit is given by $\ket{\psi} = \exp(-i (\pi/8)Y)\ket{0}$. For Figure~\ref{fig:K_Hardware}(b), a single layer of the ansatz is composed of parameterized general single-qubit rotations followed by CNOTs connecting neighboring qubits. For both Figures~\ref{fig:K_Hardware}(a) and (b), the observable is composed of Pauli $Z$ on the first and second qubits. i.e., $O = Z_1 \otimes Z_2$, and we show results for different layers of the aforementioned ansatzes. 

In Figures~\ref{fig:K_Hardware}(a) and (b), we plot the average value of the residual error $\log\varepsilon(t)$ where the average is over 40 and 20 different initializations for ansatzes (a) and (b), respectively. Results for both ansatzes indicate exponential decay of the $\varepsilon(t)$ over time. In the inset of each figure, we plot the average value of $K$ versus the gradient descent steps. For ansatz (a) and depth=1, $K(t)$ demonstrates large fluctuations over different realizations. For high values of depth, the variation becomes smaller, and $K$ values obtained via gradient descent converge to fixed values over different realizations. As we increase the depth, the value of $K(t)$ approaches the value predicted by our analytic results. On the other hand, ansatz (b), at each depth, $\overline{K}$ is close to the analytic value.

\section{Supervised learning}

\subsection{Quantum representation learning}
In this section we consider the supervised learning problem. We set our model as 
\begin{align}
z_{i ; \delta} \equiv z_{i}\left(\theta, \mathbf{x}_{\delta}\right)=\left\langle\phi\left(\mathbf{x}_{\delta}\right)\left|U^{\dagger} O_{i} U\right| \phi\left(\mathbf{x}_{\delta}\right)\right\rangle~.
\end{align}
Here, $z_{i;\delta}$ denotes the prediction of the model, with the output vector index $i$ and the sample index $\delta$. We build our model by measuring the expectation value of $O_i$, with a trainable unitary $U(\theta)$ parameterized by $\theta$, and the feature map $\ket{\phi (\mathbf{x}_\delta)}$ with the sample $\mathbf{x}_\delta$. We consider the supervised learning task where $z_{i ; \tilde{\alpha}} $ is compared with $y_{i;\tilde{\alpha}}$, and $\tilde{\alpha}$ is the index from the training set $\mathcal{A}$. We denote the whole data space as $\mathcal{D}$. Every sample from the training set $\mathcal{A}$ is denoted with a tilde notation. The trainable unitary is given by
\begin{align}
U(\theta ) = \prod\limits_{\ell  = 1}^L {{W_\ell }} {U_\ell }({\theta _\ell }) = \prod\limits_{\ell  = 1}^L {{W_\ell }} \exp \left( {i{\theta _\ell }{X_\ell }} \right)~.
\end{align}
Here $W_\ell$ denote unparameterized gates. We assume that $X_\ell$ are Pauli operators. We define the loss function
\begin{align}
{\mathcal{L}_{\cal A}}(\theta ) = \frac{1}{2}\sum\limits_{\tilde \alpha ,i} {{{\left( {{y_{i;\tilde \alpha }} - {z_{i;\tilde \alpha }}} \right)}^2}}  = \frac{1}{2}\sum\limits_{\tilde \alpha ,i} {\varepsilon _{i;\tilde \alpha }^2} ~,
\end{align}
with the residual training error,
\begin{align}
{\varepsilon _{i;\tilde \alpha }} \equiv {z_{i;\tilde \alpha }} - {y_{i;\tilde \alpha }}~.
\end{align}
We consider the gradient descent algorithm
\begin{align}
\delta {\theta _\ell } \equiv {\theta _\ell }(t + 1) - {\theta _\ell }(t) =  - \eta \frac{{d{{\cal L}_A}}}{{d{\theta _\ell }}}~.
\end{align}

Here, $\delta\theta$ means the difference between step $t+1$ and step $t$. For small values of the learning rate $\eta$, using the Taylor expansion we get
\begin{align}
\delta {z_{i;\delta }} = {z_{i;\delta }}(t + 1) - {z_{i;\delta }}(t) = \sum\limits_\ell  {\frac{{d{z_{i;\delta }}}}{{d{\theta _\ell }}}\delta {\theta _\ell }}  =  - \eta \sum\limits_{\ell ,i',\tilde \alpha } {{\varepsilon _{i';\tilde \alpha }}} \frac{{d{z_{i;\delta }}}}{{d{\theta _\ell }}}\frac{{d{z_{i';\tilde \alpha }}}}{{d{\theta _\ell }}}~. 
\end{align}

We define the neural tangent kernel as follows:
\begin{align}
K_{\delta ,\tilde \alpha }^{ii'} = \sum\limits_\ell  {\frac{{d{z_{i;\delta }}}}{{d{\theta _\ell }}}} \frac{{d{z_{i';\tilde \alpha }}}}{{d{\theta _\ell }}}~.
\end{align}

We use a compact notation by defining the joint index 
\begin{align}
(\delta ,i) = \bar a,\quad \left( {\tilde \alpha ,i'} \right) = \hat b~,
\end{align}
which belong to the space $\mathcal{D} \times \mathcal{O}(\mathcal{H})$ and $\mathcal{A} \times \mathcal{O}(\mathcal{H})$, respectively. We use $\hat{a}$ to indicate that the corresponding data component is in the sample set $\mathcal{A}$, and for a general data point we denote it as $\bar{a}$. Moreover, we use the notation $\mathcal{O}(\mathcal{H})$ to denote the space of Hermitian operators in the Hilbert space $\mathcal{H}$. Therefore, 
\begin{align}
\delta {z_{\bar a}} =  - \eta \sum\limits_{\hat b} {{K_{\bar a\hat b}}} {\varepsilon _{\hat b}}~.
\end{align}

Then the QNTK is given by
\begin{align}
K_{\delta ,\tilde \alpha }^{ii'} = \sum\limits_\ell  {\frac{{d{z_{i;\delta }}}}{{d{\theta _\ell }}}} \frac{{d{z_{i';\tilde \alpha }}}}{{d{\theta _\ell }}} =  - \sum\limits_\ell  {\left( {\begin{array}{*{20}{l}}
{\left\langle {\phi \left( {{{\bf{x}}_\delta }} \right)\left| {U_{ + ,\ell }^\dag \left[ {{X_\ell },U_\ell ^\dag W_\ell ^\dag U_{ - ,\ell }^\dag {O_i}{U_{ - ,\ell }}{W_\ell }{U_\ell }} \right]{U_{ + ,\ell }}} \right|\phi \left( {{{\bf{x}}_\delta }} \right)} \right\rangle  \times }\\
{\left\langle {\phi \left( {{{\bf{x}}_{\tilde \alpha }}} \right)\left| {U_{ + ,\ell }^\dag \left[ {{X_\ell },U_\ell ^\dag W_\ell ^\dag U_{ - ,\ell }^\dag {O_{i'}}{U_{ - ,\ell }}{W_\ell }{U_\ell }} \right]{U_{ + ,\ell }}} \right|\phi \left( {{{\bf{x}}_{\tilde \alpha }}} \right)} \right\rangle }
\end{array}} \right)} ~,
\end{align}
where
\begin{align}
&{U_{ - ,\ell }} \equiv \prod\limits_{\ell ' = 1}^{\ell  - 1} {{W_{\ell '}}} {U_{\ell '}},~~~~~~~{U_{ + ,\ell }} \equiv \prod\limits_{{\ell ^\prime } = \ell  + 1}^L {{W_{\ell '}}} {U_{\ell '}}~,\\
&{V_{ - ,\ell }} = {U_{ - ,\ell }}{W_\ell }{U_\ell },~~~~~~~{V_{ + ,\ell }} = {U_{ + ,\ell }}~.
\end{align}

Then we get
\begin{align}
K_{\delta_1 ,\delta_2 }^{i_1 i_2} =  - \sum\limits_\ell  {\left( {\begin{array}{*{20}{l}}
{\left\langle {\phi \left( {{{\bf{x}}_{{\delta _1}}}} \right)\left| {V_{ + ,\ell }^\dag \left[ {{X_\ell },V_{ - ,\ell }^\dag {O_{{i_1}}}{V_{-,\ell} }} \right]{V_{ + ,\ell }}} \right|\phi \left( {{{\bf{x}}_{{\delta _1}}}} \right)} \right\rangle  \times }\\
{\left\langle {\phi \left( {{{\bf{x}}_{{\delta _2}}}} \right)\left| {V_{ + ,\ell }^\dag \left[ {{X_\ell },V_{ - ,\ell }^\dag {O_{{i_2}}}{V_{-,\ell} }} \right]{V_{ + ,\ell }}} \right|\phi \left( {{{\bf{x}}_{{\delta _2}}}} \right)} \right\rangle }
\end{array}} \right)} ~.
\end{align}

\subsection{Frozen QNTK}
The average of the QNTK is given by
\begin{align}
&\bar K_{{\delta _1},{\delta _2}}^{{i_1}{i_2}} =  - \sum\limits_\ell  {\int {d{V_{ + ,\ell }}d{V_{ - ,\ell }}} \left( {\begin{array}{*{20}{l}}
{\left\langle {\phi \left( {{{\bf{x}}_{{\delta _1}}}} \right)\left| {V_{ + ,\ell }^\dag \left[ {{X_\ell },V_{ - ,\ell }^\dag {O_{{i_1}}}{V_{ - ,\ell }}} \right]{V_{ + ,\ell }}} \right|\phi \left( {{{\bf{x}}_{{\delta _1}}}} \right)} \right\rangle }\\
{\left\langle {\phi \left( {{{\bf{x}}_{{\delta _2}}}} \right)\left| {V_{ + ,\ell }^\dag \left[ {{X_\ell },V_{ - ,\ell }^\dag {O_{{i_2}}}{V_{ - ,\ell }}} \right]{V_{ + ,\ell }}} \right|\phi \left( {{{\bf{x}}_{{\delta _2}}}} \right)} \right\rangle }
\end{array}} \right)} \nonumber\\
&=  - \sum\limits_\ell  {\int d } {V_{ + ,\ell }}d{V_{ - ,\ell }}{\mathop{\rm Tr}\nolimits} \left( {S_{{\delta _1}{\delta _2}}^\dag V_{ + ,\ell }^\dag \left[ {{X_\ell },V_{ - ,\ell }^\dag {O_{{i_1}}}{V_{ - ,\ell }}} \right]{V_{ + ,\ell }}{S_{{\delta _1}{\delta _2}}}V_{ + ,\ell }^\dag \left[ {{X_\ell },V_{ - ,\ell }^\dag {O_{{i_2}}}{V_{ - ,\ell }}} \right]{V_{ + ,\ell }}} \right)~.
\end{align}
Here, we define the \emph{feature $S$-matrix element} as
\begin{align}
{S_{{\delta _1}{\delta _2}}} = \left| {\phi \left( {{{\bf{x}}_{{\delta _1}}}} \right)} \right\rangle \left\langle {\phi \left( {{{\bf{x}}_{{\delta _2}}}} \right)} \right|~.
\end{align}

First, we integrate over $dV_{+,\ell}$. Let
\begin{align}
P_{ - ,\ell }^i = \left[ {{X_\ell },V_{ - ,\ell }^\dag {O_i}{V_{ - ,\ell }}} \right]~,
\end{align}
which implies that
\begin{align}
\bar K_{{\delta _1},{\delta _2}}^{{i_1}{i_2}} =  - \sum\limits_\ell  {\int d } {V_{ + ,\ell }}d{V_{ - ,\ell }}{\mathop{\rm Tr}\nolimits} \left( {S_{{\delta _1}{\delta _2}}^\dag V_{ + ,\ell }^\dag P_{ - ,\ell }^{{i_1}}{V_{ + ,\ell }}{S_{{\delta _1}{\delta _2}}}V_{ + ,\ell }^\dag P_{ - ,\ell }^{{i_2}}{V_{ + ,\ell }}} \right)~.
\end{align}
After performing the integration, we get
\begin{align}
\int{d}{{V}_{+,\ell }}\text{Tr}\left( S_{{{\delta }_{1}}{{\delta }_{2}}}^{\dagger }V_{+,\ell }^{\dagger }P_{-,\ell }^{{{i}_{1}}}{{V}_{+,\ell }}{{S}_{{{\delta }_{1}}{{\delta }_{2}}}}V_{+,\ell }^{\dagger }P_{-,\ell }^{{{i}_{2}}}{{V}_{+,\ell }} \right)=\frac{D{{\left| {{\chi }_{{{\delta }_{1}}{{\delta }_{2}}}} \right|}^{2}}-1}{({{D}^{2}}+D)(D-1)}\text{Tr}\left( P_{-,\ell }^{{{i}_{1}}}P_{-,\ell }^{{{i}_{2}}} \right)~,
\end{align}
where we used
\begin{align}
&{\mathop{\rm Tr}\nolimits} \left( {S_{{\delta _1}{\delta _2}}^\dag {S_{{\delta _1}{\delta _2}}}} \right) = 1~,\nonumber\\
&\operatorname{Tr}\left(S_{\delta_1 \delta_2}\right)=\left\langle\phi\left(\mathbf{x}_{\delta_2}\right) \mid \phi\left(\mathbf{x}_{\delta_1}\right)\right\rangle \equiv \chi_{\delta_1 \delta_2}~.
\end{align}

Then the $\bar{K}$ simplies to
\begin{align}
  \bar{K}_{{{\delta }_{1}},{{\delta }_{2}}}^{{{i}_{1}}{{i}_{2}}}&=-\frac{D{{\left| {{\chi }_{{{\delta }_{1}}{{\delta }_{2}}}} \right|}^{2}}-1}{({{D}^{2}}+D)(D-1)}\sum\limits_{\ell }{\int{d}}{{V}_{-,\ell }}\operatorname{Tr}\left( P_{-,\ell }^{{{i}_{1}}}P_{-,\ell }^{{{i}_{2}}} \right) \nonumber\\
 & =-\frac{D{{\left| {{\chi }_{{{\delta }_{1}}{{\delta }_{2}}}} \right|}^{2}}-1}{({{D}^{2}}+D)(D-1)}\sum\limits_{\ell }{\int{d}}{{V}_{-,\ell }}\operatorname{Tr}\left( \left[ {{X}_{\ell }},V_{-,\ell }^{\dagger }{{O}_{{{i}_{1}}}}{{V}_{-,\ell }} \right]\left[ {{X}_{\ell }},V_{-,\ell }^{\dagger }{{O}_{{{i}_{2}}}}{{V}_{-,\ell }} \right] \right) \nonumber\\
 & =\frac{2L\left( D{{\left| {{\chi }_{{{\delta }_{1}}{{\delta }_{2}}}} \right|}^{2}}-1 \right)}{{{\left( {{D}^{2}}-1 \right)}^{2}}}\left( D\operatorname{Tr}\left( {{O}_{{{i}_{1}}}}{{O}_{{{i}_{2}}}} \right)-\operatorname{Tr}\left( {{O}_{{{i}_{1}}}} \right)\operatorname{Tr}\left( {{O}_{{{i}_{2}}}} \right) \right) ~.
\end{align}

In the large-$D$ limit, we obtain
\begin{align}
\bar{K}_{{{\delta }_{1}},{{\delta }_{2}}}^{{{i}_{1}}{{i}_{2}}}\approx \frac{2L\left( D{{\left| {{\chi }_{{{\delta }_{1}}{{\delta }_{2}}}} \right|}^{2}}-1 \right)\left( \operatorname{Tr}\left( {{O}_{{{i}_{1}}}}{{O}_{{{i}_{2}}}} \right) \right)}{{{D}^{3}}}~.
\end{align}

Let 
\begin{align}
\sigma_{\delta_1 \delta_2} = \left| {{\chi }_{{{\delta }_{1}}{{\delta }_{2}}}} \right|=\mathcal{O}(1)~. 
\end{align}

Then we get
\begin{align}
\bar{K}_{{{\delta }_{1}},{{\delta }_{2}}}^{{{i}_{1}}{{i}_{2}}}\approx \frac{2L}{{{D}^{2}}}{{\sigma }_{{{\delta }_{1}}{{\delta }_{2}}}}\operatorname{Tr}\left( {{O}_{{{i}_{1}}}}{{O}_{{{i}_{2}}}} \right)~.
\end{align}

We also assume that feature maps are orthogonal. 
\begin{align}
{{\chi }_{{{\delta }_{1}}{{\delta }_{2}}}} =  {{\delta }_{{{\delta }_{1}}{{\delta }_{2}}}}~.
\end{align}

Then without taking the large-$D$ limit, we get
\begin{align}
  \bar{K}_{{{\delta }_{1}},{{\delta }_{2}}}^{{{i}_{1}}{{i}_{2}}}&\approx \frac{2L\left( D{{\delta }_{{{\delta }_{1}}{{\delta }_{2}}}}-1 \right)}{{{\left( {{D}^{2}}-1 \right)}^{2}}}\left( D\operatorname{Tr}\left( {{O}_{{{i}_{1}}}}{{O}_{{{i}_{2}}}} \right)-\operatorname{Tr}\left( {{O}_{{{i}_{1}}}} \right)\operatorname{Tr}\left( {{O}_{{{i}_{2}}}} \right) \right) \nonumber\\
 & =\left\{ \begin{matrix}
   \frac{2L}{\left( {{D}^{2}}-1 \right)(D+1)}\left( D\operatorname{Tr}\left( {{O}_{{{i}_{1}}}}{{O}_{{{i}_{2}}}} \right)-\operatorname{Tr}\left( {{O}_{{{i}_{1}}}} \right)\operatorname{Tr}\left( {{O}_{{{i}_{2}}}} \right) \right) & {{\delta }_{1}}={{\delta }_{2}}  \\
   -\frac{2L}{{{\left( {{D}^{2}}-1 \right)}^{2}}}\left( D\operatorname{Tr}\left( {{O}_{{{i}_{1}}}}{{O}_{{{i}_{2}}}} \right)-\operatorname{Tr}\left( {{O}_{{{i}_{1}}}} \right)\operatorname{Tr}\left( {{O}_{{{i}_{2}}}} \right) \right) & {{\delta }_{1}}\ne {{\delta }_{2}}  \\
\end{matrix} \right. ~.
\end{align}

For simplicity, we assume that $O_i = O$ for all $i$. Therefore,
\begin{align}
{{\bar{K}}_{{{\delta }_{1}},{{\delta }_{2}}}}\approx \left\{ \begin{matrix}
   \frac{2L}{\left( {{D}^{2}}-1 \right)(D+1)}\left( D\operatorname{Tr}\left( {{O}^{2}} \right)-{{\operatorname{Tr}}^{2}}\left( O \right) \right) & {{\delta }_{1}}={{\delta }_{2}}  \\
   -\frac{2L}{{{\left( {{D}^{2}}-1 \right)}^{2}}}\left( D\operatorname{Tr}\left( {{O}^{2}} \right)-{{\operatorname{Tr}}^{2}}\left( O \right) \right) & {{\delta }_{1}}\ne {{\delta }_{2}}  \\
\end{matrix} \right. ~.
\end{align}
After diagonalizing ${{\bar{K}}_{{{\delta }_{1}},{{\delta }_{2}}}}$, we find that there is an ($\mathcal{A}-1$)-dimensional eigenspace with the eigenvalue,
\begin{align}
\frac{2DL}{{{\left( {{D}^{2}}-1 \right)}^{2}}}\left( D\operatorname{Tr}\left( {{O}^{2}} \right)-{{\operatorname{Tr}}^{2}}\left( O \right) \right)~.
\end{align}
Moreover, there is a remaining one-dimensional eigenspace with the eigenvalue
\begin{align}
\frac{2L(D-\left| \mathcal{A} \right|)}{{{\left( {{D}^{2}}-1 \right)}^{2}}}\left( D\operatorname{Tr}\left( {{O}^{2}} \right)-{{\operatorname{Tr}}^{2}}\left( O \right) \right).
\end{align}
Note that high values of  $\abs{\mathcal{A}}$ indicate slow convergence. The condition \begin{align}
    D\ge \abs{\mathcal{A}}~,
\end{align} 
ensures the positivity of the QNTK eigenvalues. 
If $D< \abs{\mathcal{A}}$, the relation $\kappa_{\delta_1 \delta_2} \approx \delta_{\delta_1 \delta_2}$ does not hold. 

\subsection{Fluctuating QNTK}

In this section, we compute
\begin{align}
{\left( {\Delta K_{{\delta _1},{\delta _2}}^{{i_1}{i_2}}} \right)^2} = {\mathbb{E}}\left( {{{(K_{{\delta _1},{\delta _2}}^{{i_1}{i_2}} - \bar K_{{\delta _1},{\delta _2}}^{{i_1}{i_2}})}^2}} \right) = {\mathbb{E}}\left( {{{\left( {K_{{\delta _1},{\delta _2}}^{{i_1}{i_2}}} \right)}^2}} \right) - {\left( {\bar K_{{\delta _1},{\delta _2}}^{{i_1}{i_2}}} \right)^2}~.
\end{align}
We have
\begin{align}\label{eq:deltaKi1i2delta1delta2}
{\left( {\Delta K_{{\delta _1},{\delta _2}}^{{i_1}{i_2}}} \right)^2}&= 2\sum\limits_{{\ell _1} < {\ell _2}} {\int d } {V_{ + ,{\ell _1}}}d{V_{ + ,{\ell _2}}}d{V_{ - ,{\ell _1}}}d{V_{ - ,{\ell _2}}}\left( {\begin{array}{*{20}{l}}
{{\rm{Tr}}\left( {S_{{\delta _1}{\delta _2}}^\dag V_{ + ,{\ell _1}}^\dag P_{ - ,{\ell _1}}^{{i_1}}{V_{ + ,{\ell _1}}}{S_{{\delta _1}{\delta _2}}}V_{ + ,{\ell _1}}^\dag P_{ - ,{\ell _1}}^{{i_2}}{V_{ + ,{\ell _1}}}} \right)}\\
{{\rm{Tr}}\left( {S_{{\delta _1}{\delta _2}}^\dag V_{ + ,{\ell _2}}^\dag P_{ - ,{\ell _2}}^{{i_1}}{V_{ + ,{\ell _2}}}{S_{{\delta _1}{\delta _2}}}V_{ + ,{\ell _2}}^\dag P_{ - ,{\ell _2}}^{{i_2}}{V_{ + ,{\ell _2}}}} \right)}
\end{array}} \right)\nonumber\\
&\qquad + \sum\limits_\ell  {\int d } {V_{ + ,\ell }}d{V_{ - ,\ell }}\left( {\begin{array}{*{20}{l}}
{{\rm{Tr}}\left( {S_{{\delta _1}{\delta _2}}^\dag V_{ + ,\ell }^\dag P_{ - ,\ell }^{{i_1}}{V_{ + ,\ell }}{S_{{\delta _1}{\delta _2}}}V_{ + ,\ell }^\dag P_{ - ,\ell }^{{i_2}}{V_{ + ,\ell }}} \right)}\\
{{\rm{Tr}}\left( {S_{{\delta _1}{\delta _2}}^\dag V_{ + ,\ell }^\dag P_{ - ,\ell }^{{i_1}}{V_{ + ,\ell }}{S_{{\delta _1}{\delta _2}}}V_{ + ,\ell }^\dag P_{ - ,\ell }^{{i_2}}{V_{ + ,\ell }}} \right)}
\end{array}} \right) - {{\bar K}^2}~.
\end{align}
We start from the first term,
\begin{align}
{\left( {\Delta K_{{\delta _1},{\delta _2}}^{{i_1}{i_2}}} \right)^2} \supset 2\sum\limits_{{\ell _1} < {\ell _2}} {\int d } {V_{ + ,{\ell _1}}}d{V_{ + ,{\ell _2}}}d{V_{ - ,{\ell _1}}}d{V_{ - ,{\ell _2}}}\left( {\begin{array}{*{20}{l}}
{{\rm{Tr}}\left( {S_{{\delta _1}{\delta _2}}^\dag V_{ + ,{\ell _1}}^\dag P_{ - ,{\ell _1}}^{{i_1}}{V_{ + ,{\ell _1}}}{S_{{\delta _1}{\delta _2}}}V_{ + ,{\ell _1}}^\dag P_{ - ,{\ell _1}}^{{i_2}}{V_{ + ,{\ell _1}}}} \right)}\\
{{\rm{Tr}}\left( {S_{{\delta _1}{\delta _2}}^\dag V_{ + ,{\ell _2}}^\dag P_{ - ,{\ell _2}}^{{i_1}}{V_{ + ,{\ell _2}}}{S_{{\delta _1}{\delta _2}}}V_{ + ,{\ell _2}}^\dag P_{ - ,{\ell _2}}^{{i_2}}{V_{ + ,{\ell _2}}}} \right)}
\end{array}} \right)~.
\end{align}

Similar to our proof for the optimization case, we get 
\begin{align}
{{\left( \Delta K_{{{\delta }_{1}},{{\delta }_{2}}}^{{{i}_{1}}{{i}_{2}}} \right)}^{2}}\supset L(L-1)\frac{4{{\left( D{{\left| {{\chi }_{{{\delta }_{1}}{{\delta }_{2}}}} \right|}^{2}}-1 \right)}^{2}}}{{{\left( {{D}^{2}}-1 \right)}^{4}}}{{\left( D\operatorname{Tr}\left( {{O}_{{{i}_{1}}}}{{O}_{{{i}_{2}}}} \right)-\operatorname{Tr}\left( {{O}_{{{i}_{1}}}} \right)\operatorname{Tr}\left( {{O}_{{{i}_{2}}}} \right) \right)}^{2}}~.
\end{align}

Furthermore, we compute the second term in Eq.~\eqref{eq:deltaKi1i2delta1delta2}
\begin{align}
{\left( {\Delta K_{{\delta _1},{\delta _2}}^{{i_1}{i_2}}} \right)^2} \supset \sum\limits_\ell  {\int d } {V_{ + ,\ell }}d{V_{ - ,\ell }}\left( {\begin{array}{*{20}{c}}
{{\mathop{\rm Tr}\nolimits} \left( {S_{{\delta _1}{\delta _2}}^\dag V_{ + ,\ell }^\dag P_{ - ,\ell }^{{i_1}}{V_{ + ,\ell }}{S_{{\delta _1}{\delta _2}}}V_{ + ,\ell }^\dag P_{ - ,\ell }^{{i_2}}{V_{ + ,\ell }}} \right)}\\
{{\mathop{\rm Tr}\nolimits} \left( {S_{{\delta _1}{\delta _2}}^\dag V_{ + ,\ell }^\dag P_{ - ,\ell }^{{i_1}}{V_{ + ,\ell }}{S_{{\delta _1}{\delta _2}}}V_{ + ,\ell }^\dag P_{ - ,\ell }^{{i_2}}{V_{ + ,\ell }}} \right)}
\end{array}} \right)~.
\end{align}

Note that
\begin{align}
&\int d {V_{ + ,\ell }}\left( {\begin{array}{*{20}{c}}
{{\mathop{\rm Tr}\nolimits} \left( {S_{{\delta _1}{\delta _2}}^\dag V_{ + ,\ell }^\dag P_{ - ,\ell }^{{i_1}}{V_{ + ,\ell }}{S_{{\delta _1}{\delta _2}}}V_{ + ,\ell }^\dag P_{ - ,\ell }^{{i_2}}{V_{ + ,\ell }}} \right)}\\
{{\mathop{\rm Tr}\nolimits} \left( {S_{{\delta _1}{\delta _2}}^\dag V_{ + ,\ell }^\dag P_{ - ,\ell }^{{i_1}}{V_{ + ,\ell }}{S_{{\delta _1}{\delta _2}}}V_{ + ,\ell }^\dag P_{ - ,\ell }^{{i_2}}{V_{ + ,\ell }}} \right)}
\end{array}} \right)\nonumber\\
&= \int {d{V_{ + ,\ell }}{\mathop{\rm Tr}\nolimits} \left( {\overleftarrow P S_{{\delta _1}{\delta _2}}^\dag V_{ + ,\ell }^\dag P_{ - ,\ell }^{{i_1}}{V_{ + ,\ell }}{S_{{\delta _1}{\delta _2}}}V_{ + ,\ell }^\dag P_{ - ,\ell }^{{i_2}}{V_{ + ,\ell }}\overrightarrow P S_{{\delta _1}{\delta _2}}^\dag V_{ + ,\ell }^\dag P_{ - ,\ell }^{{i_1}}{V_{ + ,\ell }}{S_{{\delta _1}{\delta _2}}}V_{ + ,\ell }^\dag P_{ - ,\ell }^{{i_2}}{V_{ + ,\ell }}} \right)} ~.
\end{align}

We use the identities
\begin{align}
&{\rm{Tr}}({S_{{\delta _1}{\delta _2}}}) = \left\langle {\phi \left( {{{\bf{x}}_{{\delta _2}}}} \right)|\phi \left( {{{\bf{x}}_{{\delta _1}}}} \right)} \right\rangle  \equiv {\chi _{{\delta _1}{\delta _2}}}~,\nonumber\\
&{\rm{Tr}}(S_{{\delta _1}{\delta _2}}^\dag ) = \chi _{{\delta _1}{\delta _2}}^*~,~~~~~~{\rm{Tr}}({S_{{\delta _1}{\delta _2}}}{S_{{\delta _1}{\delta _2}}}) = \chi _{{\delta _1}{\delta _2}}^2~,\nonumber\\
&{\rm{Tr}}(S_{{\delta _1}{\delta _2}}^\dag S_{{\delta _1}{\delta _2}}^\dag ) = \chi _{{\delta _1}{\delta _2}}^{*2}~,~~~~~~{\rm{Tr}}(S_{{\delta _1}{\delta _2}}^\dag {S_{{\delta _1}{\delta _2}}}) = 1~.
\end{align}

Moreover, we use the following notation:
\begin{align}
&{\mathop{\rm Tr}\nolimits} (\overleftarrow P ){\mathop{\rm Tr}\nolimits} (\overrightarrow P ) = D~,~~~~~~{\mathop{\rm Tr}\nolimits} (\overleftarrow P A){\mathop{\rm Tr}\nolimits} (\overrightarrow P B) = {\mathop{\rm Tr}\nolimits} (AB)~,\nonumber\\
&{\mathop{\rm Tr}\nolimits} (\overleftarrow P A\overrightarrow P B) = {\mathop{\rm Tr}\nolimits} (A){\mathop{\rm Tr}\nolimits} (B)~,~~~~~~{\mathop{\rm Tr}\nolimits} (\overleftarrow P \overrightarrow P ) = {D^2}~,
\end{align}
and
\begin{align}
&S_{{\delta _1}{\delta _2}}^\dag {S_{{\delta _1}{\delta _2}}} = \left| {\phi \left( {{{\bf{x}}_{{\delta _2}}}} \right)} \right\rangle \left\langle {\phi \left( {{{\bf{x}}_{{\delta _1}}}} \right)} \right|\left| {\phi \left( {{{\bf{x}}_{{\delta _1}}}} \right)} \right\rangle \left\langle {\phi \left( {{{\bf{x}}_{{\delta _2}}}} \right)} \right| = {S_{{\delta _2}{\delta _2}}}~,\nonumber\\
&{S_{{\delta _1}{\delta _2}}}S_{{\delta _1}{\delta _2}}^\dag  = \left| {\phi \left( {{{\bf{x}}_{{\delta _1}}}} \right)} \right\rangle \left\langle {\phi \left( {{{\bf{x}}_{{\delta _2}}}} \right)} \right|\left| {\phi \left( {{{\bf{x}}_{{\delta _2}}}} \right)} \right\rangle \left\langle {\phi \left( {{{\bf{x}}_{{\delta _1}}}} \right)} \right| = {S_{{\delta _1}{\delta _1}}}~,\nonumber\\
&S_{{\delta _1}{\delta _2}}^\dag {S_{{\delta _1}{\delta _1}}} = \left| {\phi \left( {{{\bf{x}}_{{\delta _2}}}} \right)} \right\rangle \left\langle {\phi \left( {{{\bf{x}}_{{\delta _1}}}} \right)} \right|\left| {\phi \left( {{{\bf{x}}_{{\delta _1}}}} \right)} \right\rangle \left\langle {\phi \left( {{{\bf{x}}_{{\delta _1}}}} \right)} \right| = S_{{\delta _1}{\delta _2}}^\dag~,\nonumber\\
&S_{{\delta _1}{\delta _2}}^\dag {S_{{\delta _2}{\delta _2}}} = \left| {\phi \left( {{{\bf{x}}_{{\delta _2}}}} \right)} \right\rangle \left\langle {\phi \left( {{{\bf{x}}_{{\delta _1}}}} \right)} \right|\left| {\phi \left( {{{\bf{x}}_{{\delta _2}}}} \right)} \right\rangle \left\langle {\phi \left( {{{\bf{x}}_{{\delta _2}}}} \right)} \right| = \chi _{{\delta _1}{\delta _2}}^*{S_{{\delta _2}{\delta _2}}}~,\nonumber\\
&{S_{{\delta _1}{\delta _1}}}{S_{{\delta _2}{\delta _2}}} = \left| {\phi \left( {{{\bf{x}}_{{\delta _1}}}} \right)} \right\rangle \left\langle {\phi \left( {{{\bf{x}}_{{\delta _1}}}} \right)} \right|\left| {\phi \left( {{{\bf{x}}_{{\delta _2}}}} \right)} \right\rangle \left\langle {\phi \left( {{{\bf{x}}_{{\delta _2}}}} \right)} \right| = \chi _{{\delta _1}{\delta _2}}^*{S_{{\delta _1}{\delta _2}}}~,\nonumber\\
&{S_{{\delta _1}{\delta _2}}}{S_{{\delta _1}{\delta _2}}} = \left| {\phi \left( {{{\bf{x}}_{{\delta _1}}}} \right)} \right\rangle \left\langle {\phi \left( {{{\bf{x}}_{{\delta _2}}}} \right)} \right|\left| {\phi \left( {{{\bf{x}}_{{\delta _1}}}} \right)} \right\rangle \left\langle {\phi \left( {{{\bf{x}}_{{\delta _2}}}} \right)} \right| = {\chi _{{\delta _1}{\delta _2}}}{S_{{\delta _1}{\delta _2}}}~.
\end{align}

Using the aforementioned notation, we get
\begin{align}
&\int d {V_{ + ,\ell }}\left( {\begin{array}{*{20}{c}}
{{\mathop{\rm Tr}\nolimits} \left( {S_{{\delta _1}{\delta _2}}^\dag V_{ + ,\ell }^\dag P_{ - ,\ell }^{{i_1}}{V_{ + ,\ell }}{S_{{\delta _1}{\delta _2}}}V_{ + ,\ell }^\dag P_{ - ,\ell }^{{i_2}}{V_{ + ,\ell }}} \right)}\\
{{\mathop{\rm Tr}\nolimits} \left( {S_{{\delta _1}{\delta _2}}^\dag V_{ + ,\ell }^\dag P_{ - ,\ell }^{{i_1}}{V_{ + ,\ell }}{S_{{\delta _1}{\delta _2}}}V_{ + ,\ell }^\dag P_{ - ,\ell }^{{i_2}}{V_{ + ,\ell }}} \right)}
\end{array}} \right)\nonumber\\
&= \frac{{{\rm{Tr}}\left( {{{\left( {P_{ - ,\ell }^{{i_1}}} \right)}^2}} \right){\rm{Tr}}\left( {{{\left( {P_{ - ,\ell }^{{i_2}}} \right)}^2}} \right)\left( {2{{\left| {{\chi _{{\delta _1}{\delta _2}}}} \right|}^4} - 4{{\left| {{\chi _{{\delta _1}{\delta _2}}}} \right|}^2}(D + 1) + (D + 1)(D + 2)} \right)}}{{(D - 1){D^2}(D + 1)(D + 2)(D + 3)}}\\
&+ \frac{{2{\rm{Tr}}^2\left( {P_{ - ,\ell }^{{i_1}}P_{ - ,\ell }^{{i_2}}} \right)({{\left| {{\chi _{{\delta _1}{\delta _2}}}} \right|}^2}(D + 1)({{\left| {{\chi _{{\delta _1}{\delta _2}}}} \right|}^2}(D + 2) - 4) + 2)}}{{(D - 1){D^2}(D + 1)(D + 2)(D + 3)}}\nonumber\\
&- \frac{{4{\rm{Tr}}\left( {P_{ - ,\ell }^{{i_1}}P_{ - ,\ell }^{{i_1}}P_{ - ,\ell }^{{i_2}}P_{ - ,\ell }^{{i_2}}} \right)\left( {{{\left| {{\chi _{{\delta _1}{\delta _2}}}} \right|}^4}(D + 1) - {{\left| {{\chi _{{\delta _1}{\delta _2}}}} \right|}^2}\left( {{D^2} + D + 2} \right) + D + 1} \right)}}{{(D - 1){D^2}(D + 1)(D + 2)(D + 3)}}\nonumber\\
&+ \frac{{2{\rm{Tr}}\left( {P_{ - ,\ell }^{{i_1}}P_{ - ,\ell }^{{i_2}}P_{ - ,\ell }^{{i_1}}P_{ - ,\ell }^{{i_2}}} \right)({{\left| {{\chi _{{\delta _1}{\delta _2}}}} \right|}^2}(D + 1)({{\left| {{\chi _{{\delta _1}{\delta _2}}}} \right|}^2}(D + 2) - 4) + 2)}}{{(D - 1){D^2}(D + 1)(D + 2)(D + 3)}}~.
\end{align}

We call $\sigma_{\delta_1 \delta_2} \equiv \abs{\chi_{\delta_1 \delta_2}}^2$ the \emph{feature cross-section}. Therefore, we get
\begin{align}\label{eq:intergal10}
&\int d {V_{ + ,\ell }}\left( {\begin{array}{*{20}{c}}
{{\mathop{\rm Tr}\nolimits} \left( {S_{{\delta _1}{\delta _2}}^\dag V_{ + ,\ell }^\dag P_{ - ,\ell }^{{i_1}}{V_{ + ,\ell }}{S_{{\delta _1}{\delta _2}}}V_{ + ,\ell }^\dag P_{ - ,\ell }^{{i_2}}{V_{ + ,\ell }}} \right)}\\
{{\mathop{\rm Tr}\nolimits} \left( {S_{{\delta _1}{\delta _2}}^\dag V_{ + ,\ell }^\dag P_{ - ,\ell }^{{i_1}}{V_{ + ,\ell }}{S_{{\delta _1}{\delta _2}}}V_{ + ,\ell }^\dag P_{ - ,\ell }^{{i_2}}{V_{ + ,\ell }}} \right)}
\end{array}} \right)\nonumber\\
&= \frac{{{\rm{Tr}}\left( {{{\left( {P_{ - ,\ell }^{{i_1}}} \right)}^2}} \right){\rm{Tr}}\left( {{{\left( {P_{ - ,\ell }^{{i_2}}} \right)}^2}} \right)\left( {2\sigma _{{\delta _1}{\delta _2}}^2 - 4{\sigma _{{\delta _1}{\delta _2}}}(D + 1) + (D + 1)(D + 2)} \right)}}{{(D - 1){D^2}(D + 1)(D + 2)(D + 3)}}\nonumber\\
&+ \frac{{2{\rm{Tr}}^2\left( {P_{ - ,\ell }^{{i_1}}P_{ - ,\ell }^{{i_2}}} \right)(\sigma _{{\delta _1}{\delta _2}}^2(D + 1)(D + 2) - 4{\sigma _{{\delta _1}{\delta _2}}}(D + 1) + 2)}}{{(D - 1){D^2}(D + 1)(D + 2)(D + 3)}}\nonumber\\
&- \frac{{4{\rm{Tr}}\left( {P_{ - ,\ell }^{{i_1}}P_{ - ,\ell }^{{i_1}}P_{ - ,\ell }^{{i_2}}P_{ - ,\ell }^{{i_2}}} \right)\left( {\sigma _{{\delta _1}{\delta _2}}^2(D + 1) - {\sigma _{{\delta _1}{\delta _2}}}\left( {{D^2} + D + 2} \right) + D + 1} \right)}}{{(D - 1){D^2}(D + 1)(D + 2)(D + 3)}}\nonumber\\
&+ \frac{{2{\rm{Tr}}\left( {P_{ - ,\ell }^{{i_1}}P_{ - ,\ell }^{{i_2}}P_{ - ,\ell }^{{i_1}}P_{ - ,\ell }^{{i_2}}} \right)(\sigma _{{\delta _1}{\delta _2}}^2(D + 1)(D + 2) - 4{\sigma _{{\delta _1}{\delta _2}}}(D + 1) + 2)}}{{(D - 1){D^2}(D + 1)(D + 2)(D + 3)}}~.
\end{align}

The first two terms in Eq.~\eqref{eq:intergal10} is given by
\begin{align}
&{\left( {\Delta K_{{\delta _1},{\delta _2}}^{{i_1}{i_2}}} \right)^2} \supset \frac{{\left( {2\sigma _{{\delta _1}{\delta _2}}^2 - 4{\sigma _{{\delta _1}{\delta _2}}}(D + 1) + (D + 1)(D + 2)} \right)}}{{(D - 1){D^2}(D + 1)(D + 2)(D + 3)}}\int {d{V_{ - ,\ell }}{\mathop{\rm Tr}\nolimits} \left( {{{\left( {P_{ - ,\ell }^{{i_1}}} \right)}^2}} \right){\mathop{\rm Tr}\nolimits} \left( {{{\left( {P_{ - ,\ell }^{{i_2}}} \right)}^2}} \right)} \nonumber\\
& + \frac{{2\left( {\sigma _{{\delta _1}{\delta _2}}^2(D + 1)(D + 2) - 4{\sigma _{{\delta _1}{\delta _2}}}(D + 1) + 2} \right)}}{{(D - 1){D^2}(D + 1)(D + 2)(D + 3)}}\int d {V_{ - ,\ell }}{{\mathop{\rm Tr}\nolimits} ^2}\left( {P_{ - ,\ell }^{{i_1}}P_{ - ,\ell }^{{i_2}}} \right)~.
\end{align}

We can represent these two terms as follows:
\begin{align}
\int d {V_{ - ,\ell }}{\mathop{\rm Tr}\nolimits} \left( {\overleftarrow P \left( {P_{ - ,\ell }^{{i_1}}P_{ - ,\ell }^{{i_2}}} \right)\overrightarrow P \left( {P_{ - ,\ell }^{{i_3}}P_{ - ,\ell }^{{i_4}}} \right)} \right)~,
\end{align}
due to the following identities
\begin{align}
\int {d{V_{ - ,\ell }}{\mathop{\rm Tr}\nolimits} \left( {{{\left( {P_{ - ,\ell }^{{i_1}}} \right)}^2}} \right){\mathop{\rm Tr}\nolimits} \left( {{{\left( {P_{ - ,\ell }^{{i_2}}} \right)}^2}} \right)}  &= \int {d{V_{ - ,\ell }}{\mathop{\rm Tr}\nolimits} \left( {\overleftarrow P {{\left( {P_{ - ,\ell }^{{i_1}}} \right)}^2}\overrightarrow P {{\left( {P_{ - ,\ell }^{{i_2}}} \right)}^2}} \right)} ~,\nonumber\\
\int d {V_{ - ,\ell }}{{\mathop{\rm Tr}\nolimits} ^2}\left( {P_{ - ,\ell }^{{i_1}}P_{ - ,\ell }^{{i_2}}} \right) &= \int d {V_{ - ,\ell }}{\mathop{\rm Tr}\nolimits} \left( {\overleftarrow P \left( {P_{ - ,\ell }^{{i_1}}P_{ - ,\ell }^{{i_2}}} \right)\overrightarrow P \left( {P_{ - ,\ell }^{{i_1}}P_{ - ,\ell }^{{i_2}}} \right)} \right)~.
\end{align}

In total, we have the following $16$ terms:
\begin{align}
 &+ {\mathop{\rm Tr}\nolimits} (\overleftarrow P A{B_1}A{B_2}\overrightarrow P A{B_3}A{B_4}) =+ {\mathop{\rm Tr}\nolimits} \left( {\overleftarrow P {X_\ell }V_{ - ,\ell }^\dag {O_{{i_1}}}{V_{ - ,\ell }}{X_\ell }V_{ - ,\ell }^\dag {O_{{i_2}}}{V_{ - ,\ell }}\overrightarrow P {X_\ell }V_{ - ,\ell }^\dag {O_{{i_3}}}{V_{ - ,\ell }}{X_\ell }V_{ - ,\ell }^\dag {O_{{i_4}}}{V_{ - ,\ell }}} \right)~,\nonumber\\
 &- {\mathop{\rm Tr}\nolimits} (\overleftarrow P A{B_1}A{B_2}\overrightarrow P A{B_3}{B_4}A) =- {\mathop{\rm Tr}\nolimits} \left( {{X_\ell }\overleftarrow P {X_\ell }V_{ - ,\ell }^\dag {O_{{i_1}}}{V_{ - ,\ell }}{X_\ell }V_{ - ,\ell }^\dag {O_{{i_2}}}{V_{ - ,\ell }}\overrightarrow P {X_\ell }V_{ - ,\ell }^\dag {O_{{i_3}}}{O_{{i_4}}}{V_{ - ,\ell }}} \right)~,\nonumber\\
 &- {\mathop{\rm Tr}\nolimits} (\overleftarrow P A{B_1}A{B_2}\overrightarrow P {B_3}AA{B_4}) =- {\mathop{\rm Tr}\nolimits} \left( {\overleftarrow P {X_\ell }V_{ - ,\ell }^\dag {O_{{i_1}}}{V_{ - ,\ell }}{X_\ell }V_{ - ,\ell }^\dag {O_{{i_2}}}{V_{ - ,\ell }}\overrightarrow P V_{ - ,\ell }^\dag {O_{{i_3}}}{O_{{i_4}}}{V_{ - ,\ell }}} \right)~,\nonumber\\
 &+ {\mathop{\rm Tr}\nolimits} (\overleftarrow P A{B_1}A{B_2}\overrightarrow P {B_3}A{B_4}A) =+ {\mathop{\rm Tr}\nolimits} \left( {{X_\ell }\overleftarrow P {X_\ell }V_{ - ,\ell }^\dag {O_{{i_1}}}{V_{ - ,\ell }}{X_\ell }V_{ - ,\ell }^\dag {O_{{i_2}}}{V_{ - ,\ell }}\overrightarrow P V_{ - ,\ell }^\dag {O_{{i_3}}}{V_{ - ,\ell }}{X_\ell }V_{ - ,\ell }^\dag {O_{{i_4}}}{V_{ - ,\ell }}} \right)~,\nonumber\\
 &- {\mathop{\rm Tr}\nolimits} (\overleftarrow P A{B_1}{B_2}A\overrightarrow P A{B_3}A{B_4}) =- {\mathop{\rm Tr}\nolimits} \left( {\overleftarrow P {X_\ell }V_{ - ,\ell }^\dag {O_{{i_1}}}{O_{{i_2}}}{V_{ - ,\ell }}{X_\ell }\overrightarrow P {X_\ell }V_{ - ,\ell }^\dag {O_{{i_3}}}{V_{ - ,\ell }}{X_\ell }V_{ - ,\ell }^\dag {O_{{i_4}}}{V_{ - ,\ell }}} \right)~,\nonumber\\
 &+ {\mathop{\rm Tr}\nolimits} (\overleftarrow P A{B_1}{B_2}A\overrightarrow P A{B_3}{B_4}A) =+ {\mathop{\rm Tr}\nolimits} \left( {{X_\ell }\overleftarrow P {X_\ell }V_{ - ,\ell }^\dag {O_{{i_1}}}{O_{{i_2}}}{V_{ - ,\ell }}{X_\ell }\overrightarrow P {X_\ell }V_{ - ,\ell }^\dag {O_{{i_3}}}{O_{{i_4}}}{V_{ - ,\ell }}} \right)~,\nonumber\\
 &+ {\mathop{\rm Tr}\nolimits} (\overleftarrow P A{B_1}{B_2}A\overrightarrow P {B_3}AA{B_4}) =+ {\mathop{\rm Tr}\nolimits} \left( {\overleftarrow P {X_\ell }V_{ - ,\ell }^\dag {O_{{i_1}}}{O_{{i_2}}}{V_{ - ,\ell }}{X_\ell }\overrightarrow P V_{ - ,\ell }^\dag {O_{{i_3}}}{O_{{i_4}}}{V_{ - ,\ell }}} \right)~,\nonumber\\
 &- {\mathop{\rm Tr}\nolimits} (\overleftarrow P A{B_1}{B_2}A\overrightarrow P {B_3}A{B_4}A) =- {\mathop{\rm Tr}\nolimits} \left( {{X_\ell }\overleftarrow P {X_\ell }V_{ - ,\ell }^\dag {O_{{i_1}}}{O_{{i_2}}}{V_{ - ,\ell }}{X_\ell }\overrightarrow P V_{ - ,\ell }^\dag {O_{{i_3}}}{V_{ - ,\ell }}{X_\ell }V_{ - ,\ell }^\dag {O_{{i_4}}}{V_{ - ,\ell }}} \right)~,\nonumber\\
 &- {\mathop{\rm Tr}\nolimits} (\overleftarrow P {B_1}AA{B_2}\overrightarrow P A{B_3}A{B_4}) =- {\mathop{\rm Tr}\nolimits} \left( {\overleftarrow P V_{ - ,\ell }^\dag {O_{{i_1}}}{O_{{i_2}}}{V_{ - ,\ell }}\overrightarrow P {X_\ell }V_{ - ,\ell }^\dag {O_{{i_3}}}{V_{ - ,\ell }}{X_\ell }V_{ - ,\ell }^\dag {O_{{i_4}}}{V_{ - ,\ell }}} \right)~,\nonumber\\
 &+ {\mathop{\rm Tr}\nolimits} (\overleftarrow P {B_1}AA{B_2}\overrightarrow P A{B_3}{B_4}A) =+ {\mathop{\rm Tr}\nolimits} \left( {{X_\ell }\overleftarrow P V_{ - ,\ell }^\dag {O_{{i_1}}}{O_{{i_2}}}{V_{ - ,\ell }}\overrightarrow P {X_\ell }V_{ - ,\ell }^\dag {O_{{i_3}}}{O_{{i_4}}}{V_{ - ,\ell }}} \right)~,\nonumber\\
 &+ {\mathop{\rm Tr}\nolimits} (\overleftarrow P {B_1}AA{B_2}\overrightarrow P {B_3}AA{B_4}) =+ {\mathop{\rm Tr}\nolimits} \left( {\overleftarrow P V_{ - ,\ell }^\dag {O_{{i_1}}}{O_{{i_2}}}{V_{ - ,\ell }}\overrightarrow P V_{ - ,\ell }^\dag {O_{{i_3}}}{O_{{i_4}}}{V_{ - ,\ell }}} \right)~,\nonumber\\
 &- {\mathop{\rm Tr}\nolimits} (\overleftarrow P {B_1}AA{B_2}\overrightarrow P {B_3}A{B_4}A) =- {\mathop{\rm Tr}\nolimits} \left( {{X_\ell }\overleftarrow P V_{ - ,\ell }^\dag {O_{{i_1}}}{O_{{i_2}}}{V_{ - ,\ell }}\overrightarrow P V_{ - ,\ell }^\dag {O_{{i_3}}}{V_{ - ,\ell }}{X_\ell }V_{ - ,\ell }^\dag {O_{{i_4}}}{V_{ - ,\ell }}} \right)~,\nonumber\\
 &+ {\mathop{\rm Tr}\nolimits} (\overleftarrow P {B_1}A{B_2}A\overrightarrow P A{B_3}A{B_4}) =+ {\mathop{\rm Tr}\nolimits} \left( {\overleftarrow P V_{ - ,\ell }^\dag {O_{{i_1}}}{V_{ - ,\ell }}{X_\ell }V_{ - ,\ell }^\dag {O_{{i_2}}}{V_{ - ,\ell }}{X_\ell }\overrightarrow P {X_\ell }V_{ - ,\ell }^\dag {O_{{i_3}}}{V_{ - ,\ell }}{X_\ell }V_{ - ,\ell }^\dag {O_{{i_4}}}{V_{ - ,\ell }}} \right)~,\nonumber\\
 &- {\mathop{\rm Tr}\nolimits} (\overleftarrow P {B_1}A{B_2}A\overrightarrow P A{B_3}{B_4}A) =- {\mathop{\rm Tr}\nolimits} \left( {\overleftarrow P V_{ - ,\ell }^\dag {O_{{i_1}}}{V_{ - ,\ell }}{X_\ell }V_{ - ,\ell }^\dag {O_{{i_2}}}{V_{ - ,\ell }}{X_\ell }\overrightarrow P {X_\ell }V_{ - ,\ell }^\dag {O_{{i_3}}}{O_{{i_4}}}{V_{ - ,\ell }}{X_\ell }} \right)~,\nonumber\\
 &- {\mathop{\rm Tr}\nolimits} (\overleftarrow P {B_1}A{B_2}A\overrightarrow P {B_3}AA{B_4}) =- {\mathop{\rm Tr}\nolimits} \left( {\overleftarrow P V_{ - ,\ell }^\dag {O_{{i_1}}}{V_{ - ,\ell }}{X_\ell }V_{ - ,\ell }^\dag {O_{{i_2}}}{V_{ - ,\ell }}{X_\ell }\overrightarrow P V_{ - ,\ell }^\dag {O_{{i_3}}}{O_{{i_4}}}{V_{ - ,\ell }}} \right)~,\nonumber\\
 &+ {\mathop{\rm Tr}\nolimits} (\overleftarrow P {B_1}A{B_2}A\overrightarrow P {B_3}A{B_4}A) =+ {\mathop{\rm Tr}\nolimits} \left( {\overleftarrow P V_{ - ,\ell }^\dag {O_{{i_1}}}{V_{ - ,\ell }}{X_\ell }V_{ - ,\ell }^\dag {O_{{i_2}}}{V_{ - ,\ell }}{X_\ell }\overrightarrow P V_{ - ,\ell }^\dag {O_{{i_3}}}{V_{ - ,\ell }}{X_\ell }V_{ - ,\ell }^\dag {O_{{i_4}}}{V_{ - ,\ell }}{X_\ell }} \right)~.
\end{align}

To simply the calculation for the general case, we define
\begin{align}
&A \leftarrow {X_\ell }~,\nonumber\\
&{B_1} \leftarrow V_{ - ,\ell }^\dag {O_{{i_1}}}{V_{ - ,\ell }}~,~~~{B_2} \leftarrow V_{ - ,\ell }^\dag {O_{{i_2}}}{V_{ - ,\ell }}~,\nonumber\\
&{B_3} \leftarrow V_{ - ,\ell }^\dag {O_{{i_3}}}{V_{ - ,\ell }}~,~~~{B_4} \leftarrow V_{ - ,\ell }^\dag {O_{{i_4}}}{V_{ - ,\ell }}~.
\end{align}

For the case when
\begin{align}
\int {d{V_{ - ,\ell }}{\mathop{\rm Tr}\nolimits} \left( {{{\left( {P_{ - ,\ell }^{{i_1}}} \right)}^2}} \right){\mathop{\rm Tr}\nolimits} \left( {{{\left( {P_{ - ,\ell }^{{i_2}}} \right)}^2}} \right)}  = \int {d{V_{ - ,\ell }}{\mathop{\rm Tr}\nolimits} \left( {\overleftarrow P {{\left( {P_{ - ,\ell }^{{i_1}}} \right)}^2}\overrightarrow P {{\left( {P_{ - ,\ell }^{{i_2}}} \right)}^2}} \right)} ~,
\end{align}
we set
\begin{align}
&A \leftarrow {X_\ell }~,\nonumber\\
&{B_1} \leftarrow V_{ - ,\ell }^\dag {O_{{i_1}}}{V_{ - ,\ell }}~,~~~{B_2} \leftarrow V_{ - ,\ell }^\dag {O_{{i_1}}}{V_{ - ,\ell }}~,\nonumber\\
&{B_3} \leftarrow V_{ - ,\ell }^\dag {O_{{i_2}}}{V_{ - ,\ell }}~,~~~{B_4} \leftarrow V_{ - ,\ell }^\dag {O_{{i_2}}}{V_{ - ,\ell }}~.
\end{align}

Similarly, for
\begin{align}
\int d {V_{ - ,\ell }}{{\mathop{\rm Tr}\nolimits} ^2}\left( {P_{ - ,\ell }^{{i_1}}P_{ - ,\ell }^{{i_2}}} \right) = \int d {V_{ - ,\ell }}{\mathop{\rm Tr}\nolimits} \left( {\overleftarrow P \left( {P_{ - ,\ell }^{{i_1}}P_{ - ,\ell }^{{i_2}}} \right)\overrightarrow P \left( {P_{ - ,\ell }^{{i_1}}P_{ - ,\ell }^{{i_2}}} \right)} \right)~,
\end{align}
we set,
\begin{align}
&A \leftarrow {X_\ell }~,\nonumber\\
&{B_1} \leftarrow V_{ - ,\ell }^\dag {O_{{i_1}}}{V_{ - ,\ell }}~,~~~{B_2} \leftarrow V_{ - ,\ell }^\dag {O_{{i_2}}}{V_{ - ,\ell }}~,\nonumber\\
&{B_3} \leftarrow V_{ - ,\ell }^\dag {O_{{i_1}}}{V_{ - ,\ell }}~,~~~{B_4} \leftarrow V_{ - ,\ell }^\dag {O_{{i_2}}}{V_{ - ,\ell }}~.
\end{align}
Using \texttt{RTNI}, we obtain 
\begin{itemize}
\item Four 4-design formulas with a positive sign. All of them are equal. 
\item Eight 3-design formulas with a negative sign. Half of them are equal to 
\begin{align}
\frac{{{\rm{Tr}}\left( {{O_{{i_3}}}{O_{{i_4}}}} \right)\left( {{\rm{Tr}}\left( {{O_{{i_1}}}{O_{{i_2}}}} \right) - N{\rm{Tr}}\left( {{O_{{i_1}}}} \right){\rm{Tr}}\left( {{O_{{i_2}}}} \right)} \right)}}{{{D^2} - 1}}~,
\end{align}
and the other half of them are equal to
\begin{align}
\frac{{{\rm{Tr}}\left( {{O_{{i_1}}}{O_{{i_2}}}} \right)\left( {{\rm{Tr}}\left( {{O_{{i_3}}}{O_{{i_4}}}} \right) - N{\rm{Tr}}\left( {{O_{{i_3}}}} \right){\rm{Tr}}\left( {{O_{{i_4}}}} \right)} \right)}}{{{D^2} - 1}}~.
\end{align}
\item Four 2-design formulas with a positive sign. All of them are equal to
\begin{align}
{\mathop{\rm Tr}\nolimits} \left( {{O_{{i_1}}}{O_{{i_2}}}} \right){\mathop{\rm Tr}\nolimits} \left( {{O_{{i_3}}}{O_{{i_4}}}} \right)~.
\end{align}

\end{itemize}

The last two terms in Eq.~\eqref{eq:intergal10} are given by
\begin{align}
&{\left( {\Delta K_{{\delta _1},{\delta _2}}^{{i_1}{i_2}}} \right)^2} \supset  - \frac{{4\left( {\sigma _{{\delta _1}{\delta _2}}^2(D + 1) - {\sigma _{{\delta _1}{\delta _2}}}\left( {{D^2} + D + 2} \right) + D + 1} \right)}}{{(D - 1){D^2}(D + 1)(D + 2)(D + 3)}}\int {d{V_{ - ,\ell }}{\mathop{\rm Tr}\nolimits} \left( {P_{ - ,\ell }^{{i_1}}P_{ - ,\ell }^{{i_1}}P_{ - ,\ell }^{{i_2}}P_{ - ,\ell }^{{i_2}}} \right)} \nonumber\\
&+ \frac{{2\left( {\sigma _{{\delta _1}{\delta _2}}^2(D + 1)(D + 2) - 4{\sigma _{{\delta _1}{\delta _2}}}(D + 1) + 2} \right)}}{{(D - 1){D^2}(D + 1)(D + 2)(D + 3)}}\int {d{V_{ - ,\ell }}{\mathop{\rm Tr}\nolimits} \left( {P_{ - ,\ell }^{{i_1}}P_{ - ,\ell }^{{i_2}}P_{ - ,\ell }^{{i_1}}P_{ - ,\ell }^{{i_2}}} \right)} ~.
\end{align}

Both terms are of the following form:
\begin{align}
\int {d{V_{ - ,\ell }}{\mathop{\rm Tr}\nolimits} \left( {P_{ - ,\ell }^{{i_1}}P_{ - ,\ell }^{{i_2}}P_{ - ,\ell }^{{i_3}}P_{ - ,\ell }^{{i_4}}} \right)} ~.
\end{align}

Therefore, we get the following 16 terms:
\begin{align}
&+ {\rm{Tr}}(A{B_1}A{B_2}A{B_3}A{B_4}) =  + {\rm{Tr}}\left( {{X_\ell }V_{ - ,\ell }^\dag {O_{{i_1}}}{V_{ - ,\ell }}{X_\ell }V_{ - ,\ell }^\dag {O_{{i_2}}}{V_{ - ,\ell }}{X_\ell }V_{ - ,\ell }^\dag {O_{{i_3}}}{V_{ - ,\ell }}{X_\ell }V_{ - ,\ell }^\dag {O_{{i_4}}}{V_{ - ,\ell }}} \right)~,\nonumber\\
&- {\rm{Tr}}(A{B_1}A{B_2}A{B_3}{B_4}A) =  - {\rm{Tr}}\left( {{X_\ell }V_{ - ,\ell }^\dag {O_{{i_2}}}{V_{ - ,\ell }}{X_\ell }V_{ - ,\ell }^\dag {O_{{i_3}}}{O_{{i_4}}}{O_{{i_1}}}{V_{ - ,\ell }}} \right)~,\nonumber\\
&- {\rm{Tr}}(A{B_1}A{B_2}{B_3}AA{B_4}) =  - {\rm{Tr}}\left( {{X_\ell }V_{ - ,\ell }^\dag {O_{{i_1}}}{V_{ - ,\ell }}{X_\ell }V_{ - ,\ell }^\dag {O_{{i_2}}}{O_{{i_3}}}{O_{{i_4}}}{V_{ - ,\ell }}} \right)~,\nonumber\\
&+ {\rm{Tr}}(A{B_1}A{B_2}{B_3}A{B_4}A) =  + {\rm{Tr}}\left( {{X_\ell }V_{ - ,\ell }^\dag {O_{{i_2}}}{O_{{i_3}}}{V_{ - ,\ell }}{X_\ell }V_{ - ,\ell }^\dag {O_{{i_4}}}{O_{{i_1}}}{V_{ - ,\ell }}} \right)~,\nonumber\\
&- {\rm{Tr}}(A{B_1}{B_2}AA{B_3}A{B_4}) =  - {\rm{Tr}}\left( {{X_\ell }V_{ - ,\ell }^\dag {O_{{i_1}}}{O_{{i_2}}}{O_{{i_3}}}{V_{ - ,\ell }}{X_\ell }V_{ - ,\ell }^\dag {O_{{i_4}}}{V_{ - ,\ell }}} \right)~,\nonumber\\
&+ {\rm{Tr}}(A{B_1}{B_2}AA{B_3}{B_4}A) =  + {\rm{Tr}}\left( {{O_{{i_1}}}{O_{{i_2}}}{O_{{i_3}}}{O_{{i_4}}}} \right)~,\nonumber\\
&+ {\rm{Tr}}(A{B_1}{B_2}A{B_3}AA{B_4}) =  + {\rm{Tr}}\left( {{X_\ell }V_{ - ,\ell }^\dag {O_{{i_1}}}{O_{{i_2}}}{V_{ - ,\ell }}{X_\ell }V_{ - ,\ell }^\dag {O_{{i_3}}}{O_{{i_4}}}{V_{ - ,\ell }}} \right)~,\nonumber\\
&- {\rm{Tr}}(A{B_1}{B_2}A{B_3}A{B_4}A) =  - {\rm{Tr}}\left( {{X_\ell }V_{ - ,\ell }^\dag {O_{{i_3}}}{V_{ - ,\ell }}{X_\ell }V_{ - ,\ell }^\dag {O_{{i_4}}}{O_{{i_1}}}{O_{{i_2}}}{V_{ - ,\ell }}} \right)~,\nonumber\\
&- {\rm{Tr}}(A{B_1}A{B_2}A{B_3}A{B_4}) =  - {\rm{Tr}}\left( {{X_\ell }V_{ - ,\ell }^\dag {O_{{i_4}}}{O_{{i_1}}}{O_{{i_2}}}{V_{ - ,\ell }}{X_\ell }V_{ - ,\ell }^\dag {O_{{i_3}}}{V_{ - ,\ell }}} \right)~,\nonumber\\
&+ {\rm{Tr}}(A{B_1}A{B_2}A{B_3}{B_4}A) =  + {\rm{Tr}}\left( {{X_\ell }V_{ - ,\ell }^\dag {O_{{i_1}}}{O_{{i_2}}}{V_{ - ,\ell }}{X_\ell }V_{ - ,\ell }^\dag {O_{{i_3}}}{O_{{i_4}}}{V_{ - ,\ell }}} \right)~,\nonumber\\
&+ {\rm{Tr}}(A{B_1}A{B_2}{B_3}AA{B_4}) =  + {\rm{Tr}}\left( {{O_{{i_1}}}{O_{{i_2}}}{O_{{i_3}}}{O_{{i_4}}}} \right)~,\nonumber\\
&- {\rm{Tr}}(A{B_1}A{B_2}{B_3}A{B_4}A) =  - {\rm{Tr}}\left( {{X_\ell }V_{ - ,\ell }^\dag {O_{{i_1}}}{O_{{i_2}}}{O_{{i_3}}}{V_{ - ,\ell }}{X_\ell }V_{ - ,\ell }^\dag {O_{{i_4}}}{V_{ - ,\ell }}} \right)~,\nonumber\\
&+ {\rm{Tr}}({B_1}A{B_2}AA{B_3}A{B_4}) =  + {\rm{Tr}}\left( {{X_\ell }V_{ - ,\ell }^\dag {O_{{i_4}}}{O_{{i_1}}}{V_{ - ,\ell }}{X_\ell }V_{ - ,\ell }^\dag {O_{{i_2}}}{O_{{i_3}}}{V_{ - ,\ell }}} \right)~,\nonumber\\
&- {\rm{Tr}}({B_1}A{B_2}AA{B_3}{B_4}A) =  - {\rm{Tr}}\left( {{X_\ell }V_{ - ,\ell }^\dag {O_{{i_1}}}{V_{ - ,\ell }}{X_\ell }V_{ - ,\ell }^\dag {O_{{i_2}}}{O_{{i_3}}}{O_{{i_4}}}{V_{ - ,\ell }}} \right)~,\nonumber\\
&- {\rm{Tr}}({B_1}A{B_2}A{B_3}AA{B_4}) =  - {\rm{Tr}}\left( {{X_\ell }V_{ - ,\ell }^\dag {O_{{i_3}}}{O_{{i_4}}}{O_{{i_1}}}{V_{ - ,\ell }}{X_\ell }V_{ - ,\ell }^\dag {O_{{i_2}}}{V_{ - ,\ell }}} \right)~,\nonumber\\
&+ {\rm{Tr}}({B_1}A{B_2}A{B_3}A{B_4}A) =  + {\rm{Tr}}\left( {{X_\ell }V_{ - ,\ell }^\dag {O_{{i_1}}}{V_{ - ,\ell }}{X_\ell }V_{ - ,\ell }^\dag {O_{{i_2}}}{V_{ - ,\ell }}{X_\ell }V_{ - ,\ell }^\dag {O_{{i_3}}}{V_{ - ,\ell }}{X_\ell }V_{ - ,\ell }^\dag {O_{{i_4}}}{V_{ - ,\ell }}} \right)~.
\end{align}
In the above formulas, we set
\begin{align}
&A \leftarrow {X_\ell }~,\nonumber\\
&{B_1} \leftarrow V_{ - ,\ell }^\dag {O_{{i_1}}}{V_{ - ,\ell }}~,~~~{B_2} \leftarrow V_{ - ,\ell }^\dag {O_{{i_2}}}{V_{ - ,\ell }}~,\nonumber\\
&{B_3} \leftarrow V_{ - ,\ell }^\dag {O_{{i_3}}}{V_{ - ,\ell }}~,~~~{B_4} \leftarrow V_{ - ,\ell }^\dag {O_{{i_4}}}{V_{ - ,\ell }}~.
\end{align}
For
\begin{align}
\int d V_{-, \ell} \operatorname{Tr}\left(P_{-, \ell}^{i_{1}} P_{-, \ell}^{i_{1}} P_{-, \ell}^{i_{2}} P_{-, \ell}^{i_{2}}\right) ~,
\end{align}
we set
\begin{align}
&A \leftarrow {X_\ell }~,\nonumber\\
&{B_1} \leftarrow V_{ - ,\ell }^\dag {O_{{i_1}}}{V_{ - ,\ell }}~,~~~{B_2} \leftarrow V_{ - ,\ell }^\dag {O_{{i_1}}}{V_{ - ,\ell }}~,\nonumber\\
&{B_3} \leftarrow V_{ - ,\ell }^\dag {O_{{i_2}}}{V_{ - ,\ell }}~,~~~{B_4} \leftarrow V_{ - ,\ell }^\dag {O_{{i_2}}}{V_{ - ,\ell }}~.
\end{align}
Similarly, for 
\begin{align}
\int d V_{-, \ell} \operatorname{Tr}\left(P_{-, \ell}^{i_{1}} P_{-, \ell}^{i_{2}} P_{-, \ell}^{i_{1}} P_{-, \ell}^{i_{2}}\right)~,
\end{align}
we set,
\begin{align}
&A \leftarrow {X_\ell }~,\nonumber\\
&{B_1} \leftarrow V_{ - ,\ell }^\dag {O_{{i_1}}}{V_{ - ,\ell }}~,~~~{B_2} \leftarrow V_{ - ,\ell }^\dag {O_{{i_2}}}{V_{ - ,\ell }}~,\nonumber\\
&{B_3} \leftarrow V_{ - ,\ell }^\dag {O_{{i_1}}}{V_{ - ,\ell }}~,~~~{B_4} \leftarrow V_{ - ,\ell }^\dag {O_{{i_2}}}{V_{ - ,\ell }}~.
\end{align}
Using \texttt{RTNI}, we obtain 
\begin{itemize}
\item Two 4-design terms with a plus sign. Both of them are equal.
\item Twelve 2-design terms. Eight of them have a plus sign, while four of them have a minus sign.
\item Two constant terms.
\end{itemize}

Therefore, we get
\begin{align}
&{\left( {\Delta K_{{\delta _1},{\delta _2}}^{{i_1}{i_2}}} \right)^2} \supset \sum\limits_\ell  {\int d } {V_{ + ,\ell }}d{V_{ - ,\ell }}\left( {\begin{array}{*{20}{c}}
{{\mathop{\rm Tr}\nolimits} \left( {S_{{\delta _1}{\delta _2}}^\dag V_{ + ,\ell }^\dag P_{ - ,\ell }^{{i_1}}{V_{ + ,\ell }}{S_{{\delta _1}{\delta _2}}}V_{ + ,\ell }^\dag P_{ - ,\ell }^{{i_2}}{V_{ + ,\ell }}} \right)}\\
{{\mathop{\rm Tr}\nolimits} \left( {S_{{\delta _1}{\delta _2}}^\dag V_{ + ,\ell }^\dag P_{ - ,\ell }^{{i_1}}{V_{ + ,\ell }}{S_{{\delta _1}{\delta _2}}}V_{ + ,\ell }^\dag P_{ - ,\ell }^{{i_2}}{V_{ + ,\ell }}} \right)}
\end{array}} \right)\nonumber\\
&= \frac{L}{{{D^4}}}\left( \begin{array}{l}
8\sigma _{{\delta _1}{\delta _2}}^2{\rm{Tr}}^2\left( {{O_{{i_1}}}{O_{{i_2}}}} \right) + 4\sigma _{{\delta _1}{\delta _2}}^2{\rm{Tr}}\left( {{O_{{i_1}}}{O_{{i_2}}}{O_{{i_1}}}{O_{{i_2}}}} \right)\\
 + 8{\sigma _{{\delta _1}{\delta _2}}}{\rm{Tr}}\left( {O_{{i_1}}^2O_{{i_2}}^2} \right) + 4{\rm{Tr}}\left( {O_{{i_1}}^2} \right){\rm{Tr}}\left( {O_{{i_2}}^2} \right)
\end{array} \right) + \mathcal{O}(\frac{L}{{{D^5}}})~.
\end{align}

After combining all four terms in Eq.~\eqref{eq:intergal10}, we get
\begin{align}
{\left( {\Delta K_{{\delta _1},{\delta _2}}^{{i_1}{i_2}}} \right)^2} &=L(L-1)\frac{4{{\left( D{{\sigma }_{{{\delta }_{1}}{{\delta }_{2}}}}-1 \right)}^{2}}}{{{\left( {{D}^{2}}-1 \right)}^{4}}}{{\left( D\operatorname{Tr}\left( {{O}_{{{i}_{1}}}}{{O}_{{{i}_{2}}}} \right)-\operatorname{Tr}\left( {{O}_{{{i}_{1}}}} \right)\operatorname{Tr}\left( {{O}_{{{i}_{2}}}} \right) \right)}^{2}}\nonumber\\
&+ \frac{L}{{{D^4}}}\left( {\begin{array}{*{20}{l}}
{8\sigma _{{\delta _1}{\delta _2}}^2{\rm{T}}{{\rm{r}}^2}\left( {{O_{{i_1}}}{O_{{i_2}}}} \right) + 4\sigma _{{\delta _1}{\delta _2}}^2{\rm{Tr}}\left( {{O_{{i_1}}}{O_{{i_2}}}{O_{{i_1}}}{O_{{i_2}}}} \right)}\\
{ + 8{\sigma _{{\delta _1}{\delta _2}}}{\rm{Tr}}\left( {O_{{i_1}}^2O_{{i_2}}^2} \right) + 4{\rm{Tr}}\left( {O_{{i_1}}^2} \right){\rm{Tr}}\left( {O_{{i_2}}^2} \right)}
\end{array}} \right) \nonumber\\
&- 4{{L}^{2}}\frac{{{\left( D{{\sigma }_{{{\delta }_{1}}{{\delta }_{2}}}}-1 \right)}^{2}}}{{{\left( {{D}^{2}}-1 \right)}^{4}}}{{\left( D\operatorname{Tr}\left( {{O}_{{{i}_{1}}}}{{O}_{{{i}_{2}}}} \right)-\operatorname{Tr}\left( {{O}_{{{i}_{1}}}} \right)\operatorname{Tr}\left( {{O}_{{{i}_{2}}}} \right) \right)}^{2}} + {\cal O}(\frac{L}{{{D^5}}})\nonumber\\
&= \frac{{4L}}{{{D^4}}}\left( \begin{array}{l}
\sigma _{{\delta _1}{\delta _2}}^2{{\mathop{\rm Tr}\nolimits} ^2}\left( {{O_{{i_1}}}{O_{{i_2}}}} \right) + {\sigma _{{\delta _1}{\delta _2}}}{\mathop{\rm Tr}\nolimits} \left( {{O_{{i_1}}}{O_{{i_2}}}{O_{{i_1}}}{O_{{i_2}}}} \right)\\
 + 2{\sigma _{{\delta _1}{\delta _2}}}{\mathop{\rm Tr}\nolimits} \left( {O_{{i_1}}^2O_{{i_2}}^2} \right) + {{\mathop{\rm Tr}\nolimits} ^2}\left( {{O_{{i_1}}}{O_{{i_2}}}} \right)
\end{array} \right)+ {\cal O}(\frac{L}{{{D^5}}})~,
\end{align}
assuming $\sigma_{\delta_1 \delta_2} \approx \mathcal{O}(1)$. 

\subsection{Numerical simulations for a supervised learning task}
\begin{figure}[htp]
\centering
\includegraphics[width=0.48\textwidth]{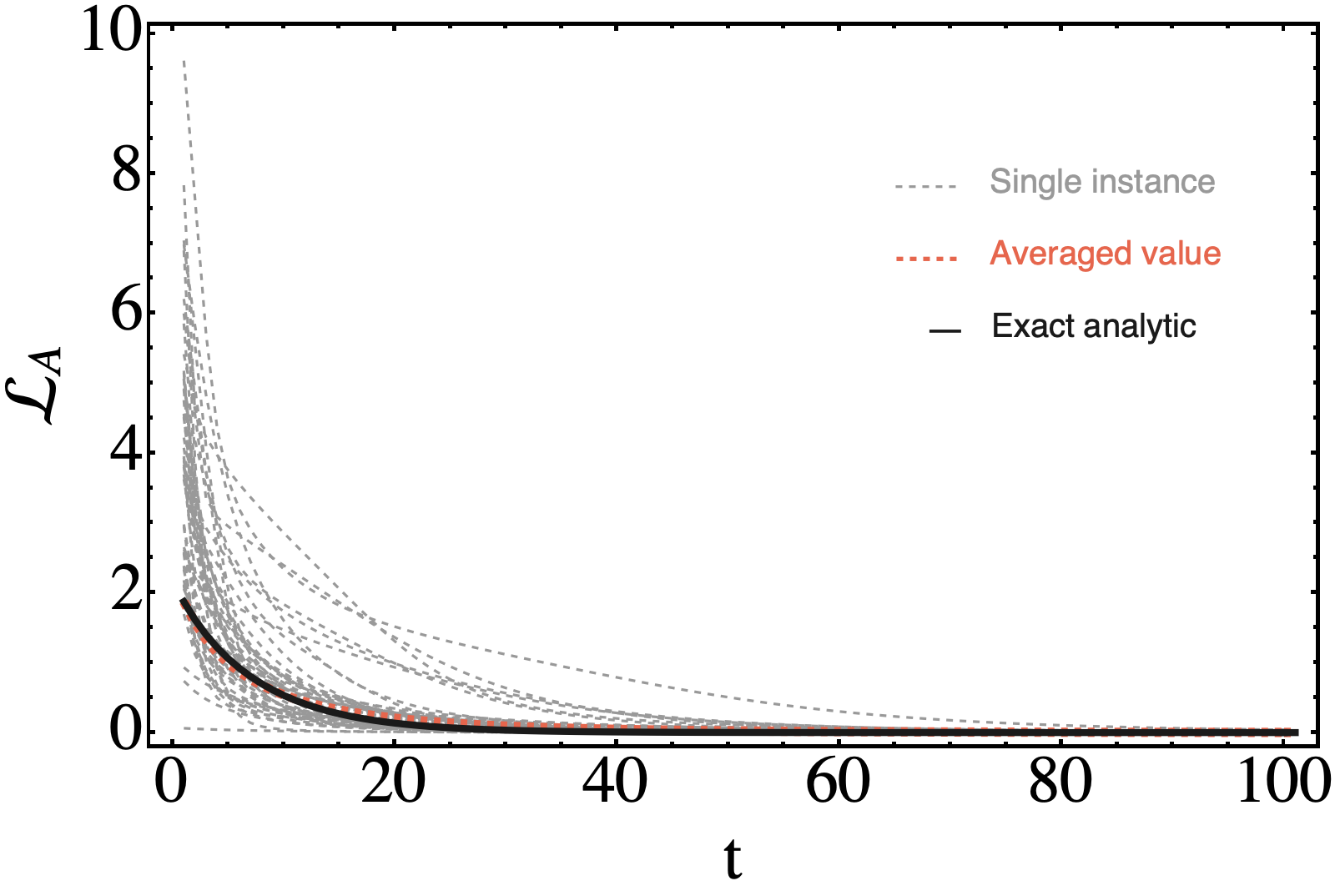}
\caption{The loss function $\mathcal{L}_{\mathcal{A}}$ versus the gradient descent steps $t$ for 2 qubits and $L=64$ for a random ansatz defined in Eq.~\eqref{ansatz}. We randomly assign supervised labels for a data set of size $\abs{\mathcal{A}}=3$. We use random initial angles and perform the gradient descent dynamics with a learning rate $\eta=10^{-3}$ for 100 steps. For 50 different initialization, we plot the dynamics of $\mathcal{L}_{\mathcal{A}}(t)$ (black-dashed curves), and the average theoretical value of the QNTK (black curve), and the numerical average of the QNTK (red-dashed curve). }
\label{fig:classification}
\end{figure}

We extend the results of Figure \ref{fig:error} in the main text in Figure \ref{fig:classification} for a supervised learning task. We randomly assign $-1$ and $+1$ labels to a data set of size $\abs{\mathcal{A}}=3$. We observe the exponential convergence of the training error for 50 randomly selected initial conditions. We select feature maps such that the input data vectors are orthogonal to each other.

\subsection{Classification task over hardware efficient ansatz}
Here, we provide numerical results based on the hardware-efficient ansatz for another supervised learning task. The supervised learning task is performed over a synthetic dataset which has been prepared as follows: For 2 qubits, we generate 20 random data sets in a 2-dimensional coordinate space. Then we attribute a positive parity label for data locating above the $x=y$ line and a negative parity for the data lying below the $x=y$ line. Then we use $\textbf{NeuralNetworkClassifier}$ from Qiskit \cite{Qiskit_ML} which allows the training over the labeled data by performing gradient descent over the squared-loss function. We also generalize the classification task for data in a 3-dimensional coordinate space and analyze the dynamics of the training error as described above.

\begin{figure*}[ht!]
\centering
\includegraphics[width=1\textwidth]{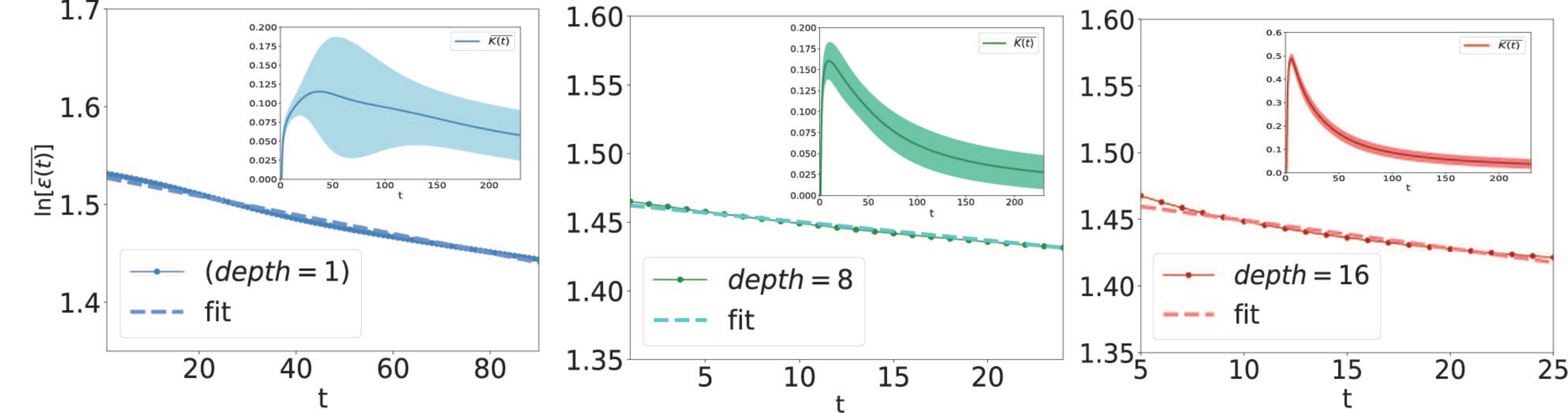}
\includegraphics[width=1\textwidth]{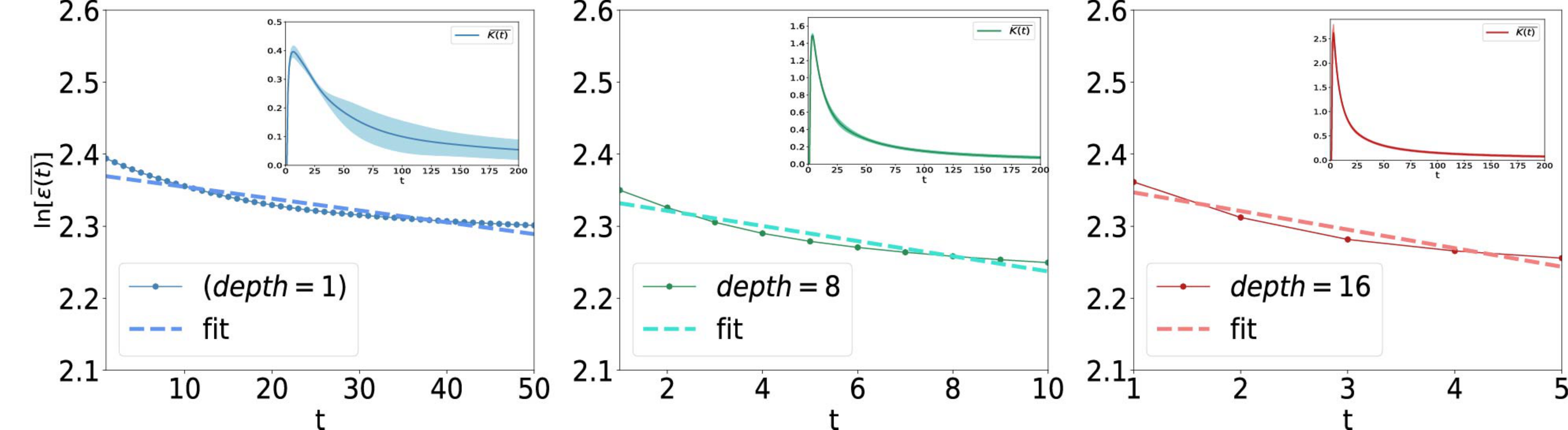}
\caption{We plot the residual error $\varepsilon$ and $K$ obtained from the training dynamics for a classification task for 2 qubits in the top row and 3 qubits in the bottom row. The simulation is performed for a SU(2) hardware-efficient ansatz on IBM Qiskit backend for $\eta=0.01$, and three different depth values $d = 1, 8$ and $16$, depicted respectively from left to right. Here, $\varepsilon$ and $K$ are averaged over 20 different realizations.  Similar to the optimization task, $\varepsilon$ indicates an exponential decay over time as predicted from our analytic results for a supervised-learning task. Moreover, as the depth of the ansatz increases, $\overline{K}$ approaches the analytic value.}
\label{fig:classify_Hardware}
\end{figure*}

In Figure \ref{fig:classify_Hardware}, we plot the residual error $\varepsilon$ as a function of time averaged over 20 different realizations for 2 qubits (top row) and 3 qubits (bottom row). In both cases, we have repeated the numerical simulation for three different values of the circuit depth $d= 1,8$ and 16. In all cases, the residual training error decays exponential, showing a good agreement with our theoretical predictions. In the inset, we plot $\overline{K(t)}$ and its variation over different realizations.  Similar to the optimization task, for a supervised-learning problem, the variance in the QNTK decreases as the depth increases. 
\color{black}

\end{document}